\documentclass[twocolumn, times, 11pt]{aastex63}
\usepackage{hyperref}

\newcommand{\lt}{\ensuremath <}

\revised{\today}

\shorttitle{The Evryscope Fast Transient Engine}
\shortauthors{Corbett et al.}

\begin{document}
\title{The Evryscope Fast Transient Engine: Real-Time Detection for Rapidly Evolving Transients}

\correspondingauthor{Hank Corbett}
\email{htc@unc.edu}

\author[0000-0002-6339-6706]{Hank Corbett}
\affil{University of North Carolina at Chapel Hill, 120 E. Cameron Ave., Chapel Hill, NC 27514, USA}

\author[0000-0001-8544-584X]{Jonathan Carney}
\affil{University of North Carolina at Chapel Hill, 120 E. Cameron Ave., Chapel Hill, NC 27514, USA}

\author[0000-0001-5083-8272 ]{Ramses Gonzalez}
\affil{University of North Carolina at Chapel Hill, 120 E. Cameron Ave., Chapel Hill, NC 27514, USA}

\author[0000-0002-4227-9308]{Octavi Fors}
\affil{Dept. de F\'isica Qu\`antica i Astrof\'isica, Institut de Ci\`encies del Cosmos (ICCUB), Universitat de Barcelona,
IEEC-UB, Mart\'i i Franqu\`es 1, E-08028 Barcelona, Spain}

\author[0000-0001-8105-1042]{Nathan Galliher}
\affil{University of North Carolina at Chapel Hill, 120 E. Cameron Ave., Chapel Hill, NC 27514, USA}

\author[0000-0001-9981-4909]{Amy Glazier}
\affil{University of North Carolina at Chapel Hill, 120 E. Cameron Ave., Chapel Hill, NC 27514, USA}

\author[0000-0002-0583-0949]{Ward S. Howard}
\affil{Department of Astrophysical and Planetary Sciences, University of Colorado, 2000 Colorado Avenue, Boulder, CO 80309, USA}

\author[0000-0001-9380-6457]{Nicholas M. Law}
\affil{University of North Carolina at Chapel Hill, 120 E. Cameron Ave., Chapel Hill, NC 27514, USA}

\author[0000-0001-9171-5236]{Robert Quimby}
\affil{San Diego State University, 5500 Campanile Dr., San Diego, CA 92182, USA}
\affil{Kavli Institute for the Physics and Mathematics of the Universe (WPI), The University of Tokyo Institutes for Advanced Study, The University of Tokyo, Kashiwa, Chiba 277-8583, Japan}

\author[0000-0001-8791-7388]{Jeffrey K. Ratzloff}
\affil{University of North Carolina at Chapel Hill, 120 E. Cameron Ave., Chapel Hill, NC 27514, USA}

\author[0000-0001-9981-4909]{Alan Vasquez Soto}
\affil{University of North Carolina at Chapel Hill, 120 E. Cameron Ave., Chapel Hill, NC 27514, USA}

\begin{abstract}
Astrophysical transients with rapid development on sub-hour timescales are
intrinsically rare. Due to their short durations, events like stellar
superflares, optical flashes from gamma-ray bursts, and shock breakouts from
young supernovae are difficult to identify on timescales that enable
spectroscopic followup. This paper presents the Evryscope Fast Transient Engine
(\textsc{EFTE}), a new data reduction pipeline designed to provide low-latency
transient alerts from the Evryscopes, a North-South pair of ultra-wide-field
telescopes with an instantaneous footprint covering 38\% of the entire sky, and
tools for building long-term light curves  from Evryscope data. \textsc{EFTE}
leverages the optical stability of the Evryscopes by using a simple direct image
subtraction routine suited to continuously monitoring the transient sky at
minute cadence. Candidates are produced within the base Evryscope two-minute
cadence for 98.5\% of images, and internally filtered using \textsc{vetnet}, a
convolutional neural network real-bogus classifier. \textsc{EFTE} provides an
extensible, robust architecture for transient surveys probing similar
timescales, and serves as the software testbed for the real-time analysis
pipelines and public data distribution systems for the Argus Array, a next
generation all-sky observatory with a data rate $62\times$ higher than
Evryscope. 
\end{abstract}


\section{Introduction}  

Optical transients evolving on short, sub-hour timescales are difficult to study
using the multi-wavelength, multi-facility approaches typically used for
longer-lived transients. For the fastest events, including prompt optical
flashes from long gamma-ray bursts (GRBs) \citep{fox_2003, cucchiara_2011,
vestrand_2014, martin_carrillo_2014, troja_2017}, shock breakout in young
supernovae \citep{garnavich_2016, bersten_2018}, and stellar flares
\citep{Howard_2022, pietras_2022, aizawa_2022}, the
duration of the event can be $\leq$ 1 hour, shorter than the base observing
cadence of conventional tiling surveys, such as the Zwicky Transient Facility
\citep[ZTF;][]{ztf_instrument}, Pan-STARRS \citep{panstarrs_instrument}, the
Catalina Sky Survey \citep[CSS;][]{catalina_instrument} and Catalina Real-Time
Transient Survey \citep[CRTS;][]{crts_instrument}, SkyMapper
\citep{skymapper_instrument}, the Asteroid Terrestrial-impact Last Alert System
\citep[ATLAS;][]{atlas_instrument}, the All-Sky Automated Survey for Supernovae
\citep[ASAS-SN;][]{asassn_instrument}, the Dark Energy Survey
\citep[DES;][]{des_program}, the Gravitational Wave Optical Transient
Observatory \citep[GOTO;][]{goto_instrument}, and the Mobile Astronomical System
of Telescope-Robots \citep[MASTER;][]{master_instrument}. Each of these surveys
tile the sky on timescales of days to maximize their likelihood of detecting
supernova-like transients, which evolve over the course of days and months. 

Faster events, occurring on minute-to-hour timescales, are detected in
conventional tiling surveys, but with frequently undersampled light curves.
Tiling surveys are also not typically optimized for minute-scale latency between
detection and reporting, precluding spectroscopic follow-up on timescales
comparable to the lifetime of the transient. As a result, searches for
short-lived events typically require simultaneous coordinated observations of
small sky regions, as in the Deeper-Wider-Faster program
\citep[DWF;][]{dwf_program}. However, previous searches for fast transients in
this regime, by the DWF team \citep{andreoni_2020}, as well as from PanSTARRS
\citep{berger_2013}, iPTF \citep{ho_2018}, PTF \citep{van_roestel_2019},
Tomo-e-Gozen \citep{richmond_2020}, and the Organized Autotelescopes for
Serendipitous Event Survey \citep[OASES;][]{Arimatsu_2021} have only produced
upper-limits on the extragalactic event rate of fast transients, suggesting that
increased areal survey rates are necessary to observe any new populations of
high-speed transients.

\begin{figure*}
	\includegraphics[width=\textwidth]{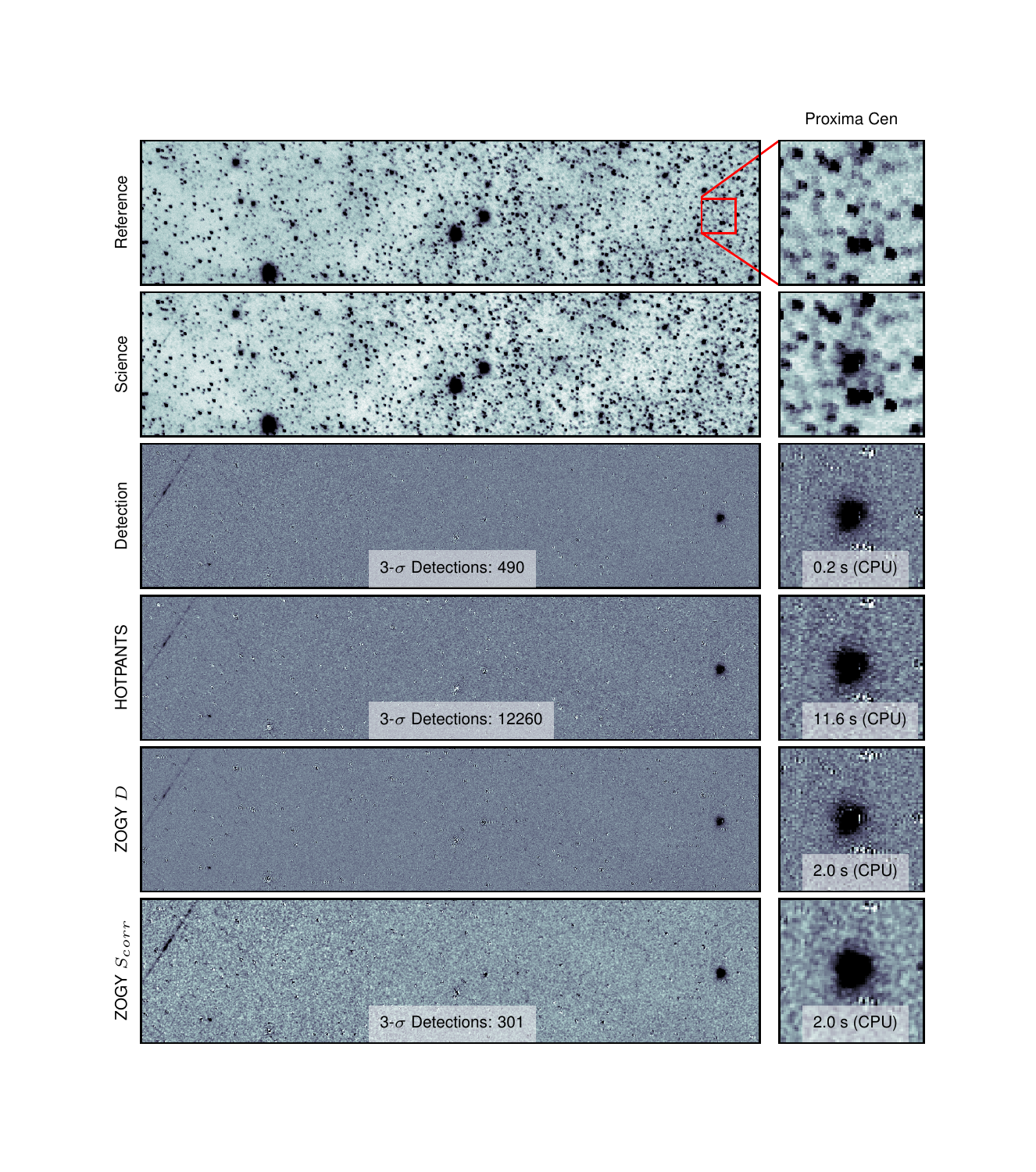}
	\caption{\label{fig:proxima_diff}\textbf{(Left)} A $4^\circ\times1^\circ$
	region near Proxima Centauri from a reference, science, and
	subtraction images from \textsc{EFTE}, the High Order Transform of PSF and
	Template Subtraction algorithm \citep[HOTPANTS;][]{hotpants}, and the Zackay,
	Ofek, and Gal-Yam algorithm \citep[ZOGY;][]{zogy}. For ZOGY, we include both
	the scaled $S_{corr}$ image used for point source detection and the proper difference
	image $D$.	In the subtraction images, a faint satellite streak (left),
	variable star (bottom left), and M-dwarf super flare
	\citep{proxima_superflare} (far right) are successfully recovered.
	The EFTE direct subtraction produces 47\% more 3-$\sigma$ detections
	than ZOGY for this field and can be computed is 10$\times$ faster for this
	image size. \textbf{(Right)} Cutouts from the left images showing the
	$1'\times1'$ region centered on Proxima Cen.}
\end{figure*} 

An alternate approach to probing the dynamic sky at short timescales is to
survey an extreme field of view, typically sacrificing some depth and resolution
relative to conventional tiling surveys in exchange for rapid-cadence
monitoring. This approach enables time-resolved detection of fast optical
transients, and poses unique challenges and opportunities for real-time data
reduction. Rapid localizations of transients across wide fields of view have
recently been used to make spectroscopic observations of flares using the
Ground-based Wide-Angle Camera system (GWAC) with latencies as low as 20 minutes
\citep{wang_2021,xin_2021}.

Galactic transients at minute-to-hour timescales are plentiful, with stellar
flares from M-dwarf stars making up the majority of these events
\citep{kulkarni_2006}. Flares are caused by reconnections in the stellar
magnetic field, producing radiation across the electromagnetic spectrum on
timescales ranging from seconds to hours. Radiation from the largest events,
so-called \emph{superflares}, reach energies $\geq 10^{33}$ erg -- orders of
magnitude greater than the largest solar flares \citep{schaefer_2000}. Flares
are responsible for much of the UV environment of rocky planets orbiting cool
stars \citep{Walkowicz_2008}, potentially providing the bioactive UV flux
necessary for prebiotic chemistry \citep{Ranjan_2017}, or even eroding
Earth-like atmospheres \citep{segura_2010, proxima_superflare}. Spectroscopic
observations taken during the initial stages of a flare can reveal temperature
and emission-line evolution during their most impulsive phases, which is
valuable for constraining fundamental flare physics  as well as potential
impacts of flare activity on exoplanet atmospheres. Such observations are
crucial for assessing the habitability of Earth's closest neighbors: the nearest
exoplanet to Earth, Proxima Centauri b, is subject to frequent high-energy flare
activity \citep{proxima_superflare}.

In addition to time-resolution constraints on survey design, searches for
sub-hour transients like flares require software data pipelines capable of
rapidly identifying candidates for spectroscopic followup and classification.
Examples of bespoke pipelines optimized for minimal latency are presented in
\citep{mary_pipeline, Perrett_2010, Cao_2016, Forster_2016, Kumar_2015}. Such
pipelines are often built around difference image analysis, a method for
isolating sources with variable flux by subtracting an earlier reference image
of the field, complicated by the need to match the seeing-limited point spread
functions (PSFs) of images from multiple epochs. Methods for subtracting images
in the presence of variable PSFs include deconvolution with a matching kernel
\citep{alard_luptin_1998, bramich_2008, hotpants}, which can be computationally
expensive and numerically unstable, and more recently, the statistically optimal
\textsc{ZOGY} method \citep{zogy}, which requires a robust and static model of the image
PSF, and the Saccadic Fast Fourier Transform \textsc{sFFT} method \citep{Hu_2022}. 

In this paper, we present the Evryscope Fast Transient Engine (\textsc{EFTE}), a
real-time discovery pipeline for the Evryscopes. The Evryscopes are a pair of
gigapixel-scale survey instruments which continuously image 38\% of the
celestial sphere at two-minute cadence. \textsc{EFTE} is optimized for
sensitivity to short-duration transients, including stellar flares and
flash-like optical counterparts to multi-messenger or multi-wavelength events.
Using \textsc{EFTE}, we are able to produce transient candidates within survey
cadence, with actionable alerts indexed into our database before the next image
in the sequence. \textsc{EFTE} also provides a processing workflow for batch
processing of Evryscope image data, and forms the basis for the Evryscope
precision photometry pipeline, providing years-long light curves for every star
brighter than $g\sim15$.

One goal for the \textsc{EFTE} pipeline is to minimize the computational
resources necessary for data analysis; a single co-located compute node supports
each Evryscope site. Low resource requirements are particularly necessary when
looking towards next-generation sky surveys, such as the Argus Optical Array
\citep{law_2022,law_2022b}. The upcoming Argus Array Pathfinder instrument,
consisting of thirty-eight 20 cm telescopes, will produce up to 180 TiB of data
per night at 1-second cadence and 6 TiB of data per night at the base 30-second
cadence; maximizing science returns from data-intensive systems like Argus will
require time- and cost-efficient algorithms and pipelines. For Argus, all images
must be reduced within the observing cadence to provide sufficiently low latency
for followup and to avoid a backlog of data, which can require runaway compute
resources for ``catch-up.'' Incoming Argus images are resampled to a pre-defined
HEALPix \citep{gorski_2011} grid using a custom GPU-based code. By parallelizing
direct subtraction based on the EFTE algorithm, the Argus pipelines are able to
reduce each image into transient candidates and compressed images in an average
of of 925 ms. \cite{Corbett_2022} presents a full description of the Argus Array
pipelines and data reduction strategy.

The paper is organized as follows. In Section~\ref{sec:instrument}, we give an
overview of the Evryscope instruments and survey strategy. In
Section~\ref{sec:efte}, we describe the \textsc{EFTE} pipeline and present
algorithms for data analysis and transient discovery in ultra-wide field
systems, including a simple image subtraction method suitable for time-sensitive
searches (see Figure~\ref{fig:proxima_diff}). In Section~\ref{sec:autovet}, we describe the selection metrics and
machine learning (ML) approaches used to select candidates from the event stream. In
Section~\ref{sec:results}, we characterize the photometric, astrometric, and
latency performance of the pipeline, including the expected survey completeness
and characterization of the convolutional neural network (CNN) used for vetting
candidates. In Section~\ref{sec:science}, we summarize early science returns
from the \textsc{EFTE} pipeline, including a characterization of the impact of
satellite glints on rapid-response transient surveys and rapid-response
observations of stellar flares using the Goodman High-Throughput Spectrograph on
the SOAR 4.1 m telescope \citep{clemens_2004}. We summarize, consider
extensibility of the \textsc{EFTE} pipeline to data from other surveys, and
describe next steps towards producing a public event stream in
Section~\ref{sec:summary_conclusions}.

\section{Evryscope Survey Overview}\label{sec:instrument}

\subsection{Instrument Description}

The Evryscopes are a pair of multiplexed wide-field survey telescopes, located
at Cerro Tololo Inter-American Observatory (CTIO) in Chile and Mount Laguna
Observatory (MLO) in California. Each site consists of up to twenty-seven 6.1 cm
aperture camera units, arranged to observe the majority of the sky above an
airmass of $\sim$2 simultaneously. Collectively, the Evryscopes have a
instantaneous field of view of 16,512 sq. degrees (15,929 sq. degrees accounting
for overlaps between adjacent cameras) with a resolution of 13.2 arcseconds per
pixel across a 1.24 gigapixels combined image plane. The telescopes observe at a
constant two-minute cadence with a 97\% duty cycle, and collect an average of
600 GiB of data per night. While the primary Evryscope survey has been conducted
in the Sloan $g$-band, the Northern Hemisphere system is also equipped with a
Sloan $r$ filter for use in future surveys. All of the telescopes at each
site are attached on a single mount, tracking the sky in two-hour increments.

The instruments are fully robotic, operating autonomously based on a local
weather station. Evryscope-South has been in operation since May 2015, and
Evryscope-North began science operations in January 2019. For a full description
of the instrument and Evryscope science programs, see
\cite{evryscope_instrument} and \cite{evryscope_project}. The instrument
parameters are summarized in Table~\ref{tab:evryscope_params}.
\begin{table}
	\centering
	\caption{System properties for the Evryscopes. For further information, see
	\citet{evryscope_instrument}}\label{tab:evryscope_params}
	\begin{tabular}{lcc} 
		\hline
		Property & Evryscope-South & Evryscope-North\\
		\hline
		Field of View (Deduplicated) & 8520 sq. deg & 7409 sq. deg\\
		Field of View (Total) & 8832 sq. deg & 7680 sq. deg\\

		Detector Size &  662.4 MPix & 576 MPix\\
		Cadence & \multicolumn{2}{c}{2 minutes}\\
    	Aperture & \multicolumn{2}{c}{6.1 cm} \\
		Pixel Scale & \multicolumn{2}{c}{13.2 arcsec/pixel}\\
		Data Rate  & \multicolumn{2}{c}{165 Mbps (1.2 GiB/minute)}\\
		\hline
	\end{tabular}
\end{table}

\subsection{Evryscope Observation Strategy}\label{sec:observation_strategy} 

The Evryscopes utilize two distinct strategies for determining the
two-hour-observing fields that are observed over the course of the night:
\begin{enumerate}
    \item Semi-random, with the eastern-most edge of the field placed 30 degrees
    above the horizon to the east at the start of each 2-hour observation. 
    \item Fixed pointings, chosen from 48 overlapping regions, separated by 7.5
    degrees in right ascension. 
\end{enumerate}
In both scenarios, the duration of a single pointing is limited by the time it
takes the westward edge of the field to pass beneath an airmass of $\sim$2. This
timescale (a ``ratchet'') is typically on the order of two hours.  Each
Evryscope tracks continuously at the sidereal rate. Minimal (few arcminute)
drift due to polar alignment is present over the course of a ratchet, but the
visible field between consecutive two-minute exposures is consistent,
point-like, and unstreaked. 

Semi-random pointings (currently used for Evryscope-South) are preferred for
long-term photometric performance, as diverse field positions allow sensor-plane
effects to average out over the duration of the survey. Because of the
commercial-off-the-shelf optics used, individual cameras exhibit up to 50\%
vignetting at the edge of the sensor field of view. Randomized pointings also
minimize the effects of camera-to-camera periodic noise, provide some resilience
against CCD sensor defects, and limit the prevalence of pathological coordinates
that are always located at the edge of a sensor and thus unduly affected by
optical vignetting. The trade-off is that individual fields repeat only on
timescales of months (cross camera) or years (single camera), and only to
few-degree precision. 

Fixed-fields, by contrast, are used for Evryscope-North, and result in a fields
that repeat with arc-minute precision on 1-3 day timescales. This repeatability
is convenient for transient searches, as it allows us to build up an archive of
reference frames to use for image subtraction. For fields above an airmass of 2,
76\% are observed within two days of the previous visit, and 97\% are observed
within a week. Because adjacent fields overlap by $\sim95\%$, a given sky region
will appear in many different pointings, meaning that the field recurrence time
is independent of the observing cadence.

\section{The Evryscope Fast Transient Engine} \label{sec:efte}

The Evryscope Fast Transient Engine (\textsc{EFTE}) is a pipeline for searching
Evryscope images for bright and rapidly changing sources in real-time,
identifying and recording candidates across the full Evryscope FOV within the
two-minute observing cadence. The primary goal of \textsc{EFTE} is to provide a
reliable event stream with sufficiently minimal latency to enable
multi-wavelength followup of events with sub-hour durations. \textsc{EFTE} also
provides useful general-purpose utilities for interacting with and analyzing
Evryscope data, including a quick-look photometry pipeline independent of the
general-purpose precision-photometry pipeline, a custom astrometric solver, and
CCD calibration functions. \textsc{EFTE} is mostly written in Python, with some
compute-intensive routines (stamp extraction and photometry) implemented in C
and wrapped as Python extensions using Cython \citep{behnel2011cython}.

\textsc{EFTE} is a hierarchical + distributed system, with two analysis servers
on-site at MLO and CTIO streaming reduced data products back to a central
PostgreSQL\footnote{http://www.postgresql.org/} relational database on campus at
University of North Carolina at Chapel Hill (UNC-CH). The analysis servers each have dual AMD EPYC processors (36 CPU cores for
Evryscope-North, 48 for Evryscope-South), and 512 GiB of RAM (384 GiB at
Evryscope-North). The asymmetry between the two sites is due to the additional
four years of archival data from Evryscope-South. The central database for
reduced data products is hosted on a 36-core server with 24 TiB of
flash storage, located on-campus at UNC-CH. This server also hosts a backend application for pipeline monitoring,
associating \textsc{EFTE} transients with external alerts, and end-user
reporting via a Slack\footnote{https://slack.com/}-based web-interface.

\begin{figure*}
	\includegraphics{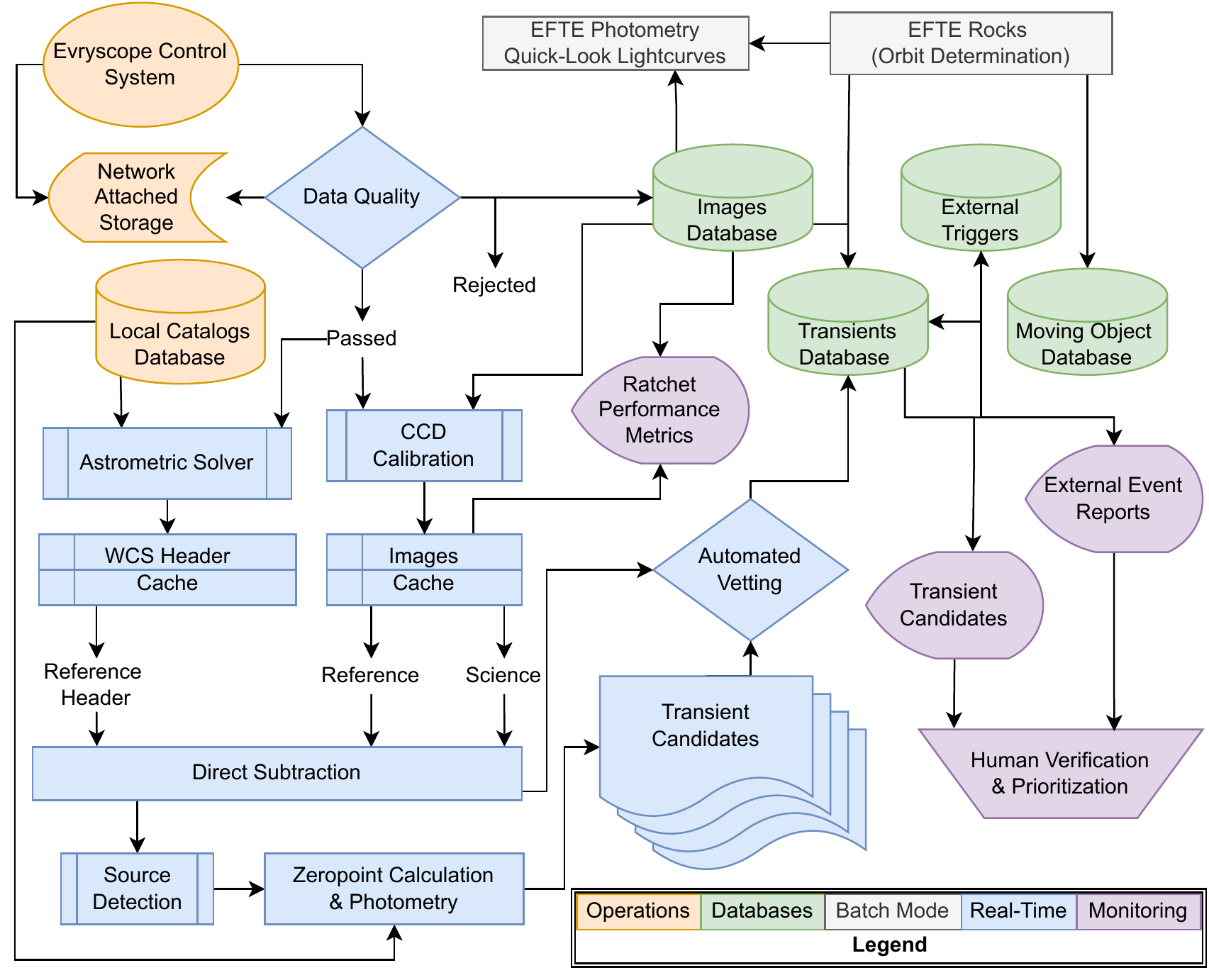}
	\caption{\label{fig:efte_flowchart}Data flow and layout of the
	\textsc{EFTE} pipelines. Operations (orange), and real-time reduction (blue)
	components are independent for each observatory, while pipeline monitoring
	(purple) and shared databases (green) collate data from both Evryscope-North
	and Evryscope-South.}
\end{figure*}

In real-time operation, \textsc{EFTE} instances on each analysis server
communicate with the Evryscope data acquisition system via a TCP socket
connection, receiving notifications for each incoming image once it has been
written to a shared network filesystem. \textsc{EFTE} maintains a per-ratchet,
in-memory database of recent images to be matched for image subtraction,
spawning subprocesses for all analysis tasks. Figure~\ref{fig:efte_flowchart}
shows the primary components of the \textsc{EFTE} pipeline, from the moment that
an image is written to disk to reporting candidates.

\subsection{Image Quality Monitor and CCD
Calibration}\label{sec:data_quality_and_calibration}

Once an exposure is completed, the Evryscope observation daemon sends a TCP
packet to an \textsc{EFTE} instance running on the analysis server.  Upon
receipt of this notification, \textsc{EFTE} will asynchronously record basic metadata
including camera, timing, origin, and instrument configuration to the central
database located at UNC-CH. Before further reduction, the image goes through a
series of general quality assurance steps, including:
\begin{itemize}
    \item verification of the file against the checksum recorded by the
    acquisition system
    \item instrument configuration checks for camera cooling, dome status, and
    exposure type
    \item autocorrelation-based checks for tracking errors and alignment failures
    \item sky-background measurement for saturation and linearity checks.
\end{itemize}
If these conditions are satisfied (as they are for 98\% of images), the image is
converted from ADU to electron units and matched to dark and flat fields for CCD
calibration. 

Dark frames are regularly regenerated using frames taken at the beginning and
ending of each night. Cameras are cooled to a constant $-20^\circ C$ during observing,
but some few-percent level drift in bias level is observed as a function of the
camera external temperature. We believe that this is caused by temperature
gradients across the readout electronics, and make a quadratic correction to
the bias level as a function of the camera electronics temperature, as measured
by an on-board sensor in each camera. Additionally, a small ($<1\%$) linearity
correction is applied per pixel based on a cubic fit to pixel value vs.
exposure time in lab testing. The linearity correction was determined to be
near-identical for all our sensors.

Because of the extreme single-camera field of view, twilight flats contain
significant sky gradient that the Evryscope is unable to compensate for through
diverse pointings because of its fixed camera positions. Instead, we use
photometric flats, calculated based on a 7th order polynomial fit to the
normalized flux offsets of reference stars relative to \emph{g}-band catalog photometry
from the ATLAS All-Sky Stellar Reference Catalog
\citep[ATLAS-REFCAT2;][]{atlas_catalog}. These frames capture the average
vignetting patterns of the individual cameras, which can change sharply at the
edge of the field. Photometric flat fields are stable at the 1\% level over
months-long timescales due to focus stability of the Evryscope Robotilter alignment
system \citep{ratzloff_robotilter}, and are regenerated only when the
instruments are cleaned, which typically requires replacing or removing the
outer optical windows on each camera.

Bad pixels are replaced with the median of the surrounding $3\times3$ pixel
block and then assigned an arbitrarily high uncertainty in the resulting noise
image used for photometry and source detection. Parts of each camera's field of
view, particularly those near the center of the frame, will have undersampled
PSFs. Simple bad-pixel masking (\emph{i.e.,} assigning pixels a \texttt{NaN} or
0 value) will produce sharp artifacts in subsequent analysis requiring pixel
resampling, like the image subtraction described in
Section~\ref{sec:direct_subtraction}.

\subsection{Astrometric Solutions}\label{sec:astrometry}

In parallel with the science-frame calibration steps, the \textsc{EFTE} pipeline
produces an astrometric solution for the image using a custom solver
developed for the highly distorted Evryscope focal plane. Evryscope astrometric
solutions begin with an initial solve based on the center 512$\times$512 pixel
region using a local install of \texttt{astrometry.net} \citep{Lang_Hogg_2010}.
This solution is only used to locate the center of the image. Sources in the
image are then cross-matched against the Tycho-2 catalog \citep{tycho2}, and the
offsets are used to optimize a polynomial distortion solution to 5th order in
each of $x,y$ and radial position on the sensor, plus cross terms. The solution
is then verified against a subset of bright stars from Gaia DR2 \citep{gaia_dr2}
based on crossmatch performance against detections in an even grid of 15
different sensor regions using the following requirements:
\begin{enumerate}
	\itemsep0em
	\item $>$ 80\% recovery in at least 7 regions 
	\item $<$ 50\% recovery in 0 regions 
	\item Uncertain recovery (due to source confusion or non-detections) in no
	more than 2 regions. 
\end{enumerate}

We selected Gaia DR2 for solution verification due to its reference epoch
(J2015.5) coinciding with the beginning of Evryscope observations. Typical RMS
offsets from Gaia DR2 positions are $\sim 4$ arcseconds, or 0.3 pixels. 

The complete solution is written into a world coordinate system (WCS) header
using the TPV convention for distortion polynomials \citep{calabretta2004}. The
TPV representation is an extension to the standard TAN projection, including
additional terms for a general polynomial
correction\footnote{https://fits.gsfc.nasa.gov/registry/tpvwcs/tpv.html}. Due to
atmospheric refraction and tracking errors, the solution must be recalculated
per-image, but the solver is able to start with a pre-computed baseline
distortion solution, averaged over dozens of fields for each camera. We found
that starting with an averaged solution decreased the time required for the final
optimization by a factor of several on average.

This header is archived to network storage, and serialized and passed back into
the in-memory \textsc{EFTE} matching database, where it is associated with the
camera and active field. Finally, a footprint of the image, discretized as a
GeoJSON \citep{geojson} polygon, is stored in the central database, where it is
indexed using PostGIS\footnote{http://postgis.net/} extension to PostgreSQL,
which provides a variety of spatial object types. These footprints support a
variety of use-cases, and allow users to easily query for images containing a
given target, or search for images intersecting with arbitrary sky regions that
can be represented as polygons, such as probability skymaps for gravitational
wave and GRB triggers.

\subsection{Direct Image Subtraction}\label{sec:direct_subtraction}

Like most optical transient surveys, \textsc{EFTE} isolates objects with
changing flux by subtracting each science image from an earlier reference frame
of the same field from each image. We optimize our subtraction algorithm for
speed rather than statistical optimality, electing for per-pixel operations
requiring no additional intermediate data products beyond those produced in the
initial photometric pipeline. Due to the short (sub-hour) timescales of interest
and the dominance of instrumental optical effects on the system PSF, Evryscope images do not
require PSF-matching techniques addressed by standard difference-image analysis
routines, like \textsc{HOTPANTS} \citep{hotpants} or \textsc{ZOGY} \citep{zogy}.
However, \textsc{EFTE} was built for extensibility, and implementations of both
\textsc{HOTPANTS} and \textsc{ZOGY} are included in \textsc{EFTE}.

\begin{figure*}
	\includegraphics{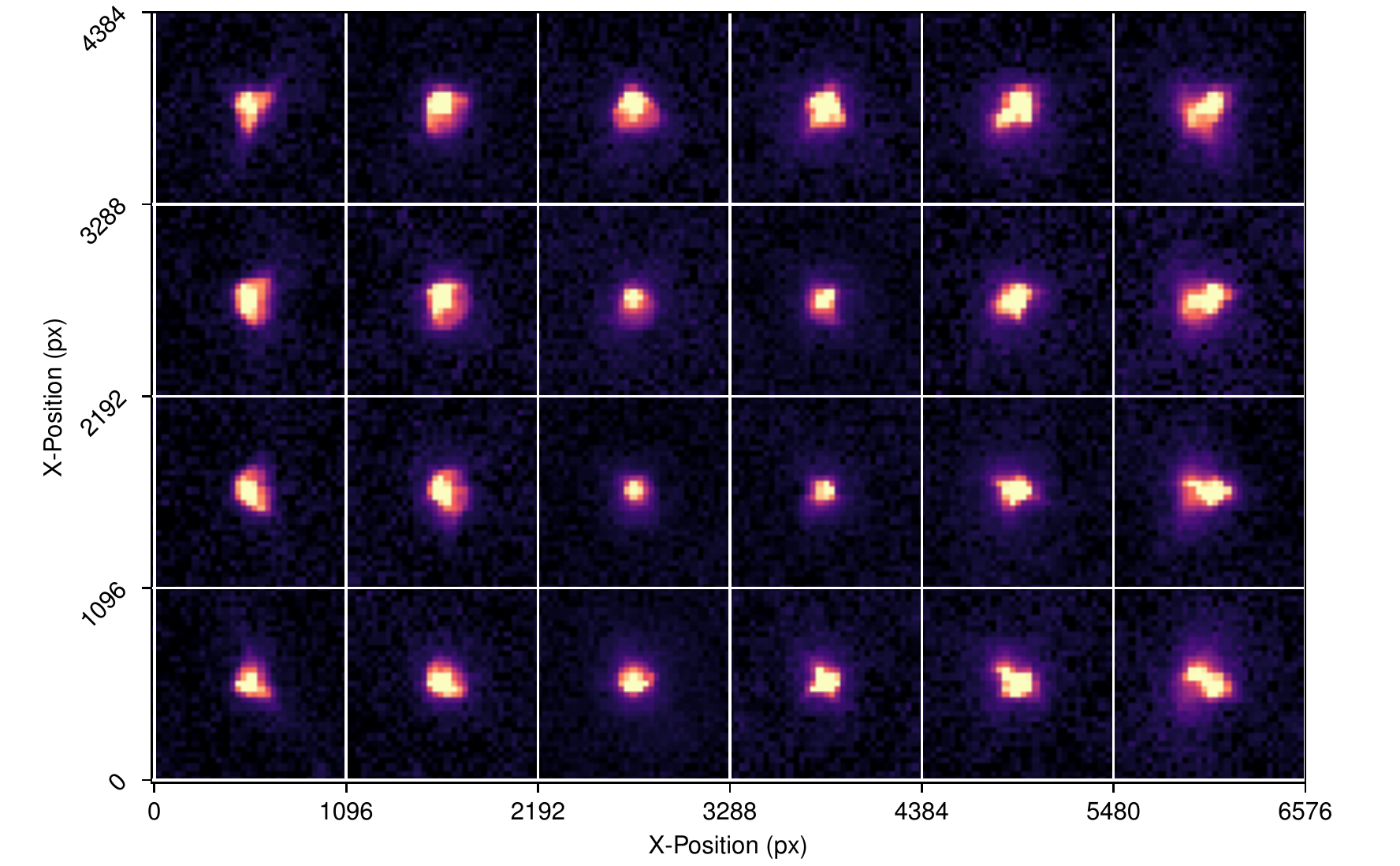} 
\caption{\label{fig:psf_map}Median PSFs
across a $6\times4$ grid of sensor regions. PSF variability as a function of
chip position is evident; however, long-term measurements of Evryscope optical
stability \citep{ratzloff_robotilter} indicate that the PSF is repeatable in
time despite aberrations. }
\end{figure*}

The resolution of Evryscope images is limited by the optical distortions from
the camera lenses and pixel scale, rather than atmospheric effects, under most
practical observing conditions. PSFs vary greatly across the image plane of each
camera, as illustrated in Figure~\ref{fig:psf_map}; however, Evryscope image
quality metrics have been measured to be stable at the few-percent level over
many-month timescales \citep{ratzloff_robotilter}, creating highly repeatable
PSFs for each individual camera.

As a result, we adopt a straightforward algorithm for image subtraction in which
the reference and science images are aligned, matched in flux, and subtracted
directly. The difference of the two images is then weighted by a propagated
uncertainty image to identify significant changes in flux. This approach is
valid only if the following conditions are satisfied for the reference and
science image couplet:
\begin{itemize}
    \item Observed PSFs are dominated by telescope optics and pixel scale, and do not vary
    significantly as a function of observing conditions on the timescale of the
    lag between the images
    \item All sources have near-identical pixel coordinates in both images, offset by no more
    than the PSF-coherence scale, which we define as the pixel distance over which
    spatial PSF variation is less than thermal and atmospheric effects over a 
    few-minute baseline, or to a 1\% maximum change in the
    normalized PSF.  This scale is typically $\sim10$ arcminutes, or $\sim50$
    pixels
    at Evryscope pixel scale
    \item The global flux scaling between the two images is smooth 
\end{itemize}

In the following subsections, we describe the process by which we match image
couplets for subtraction, and the custom method we use to subtract the images
that is optimized for the unique resolution and time domain covered by
\textsc{EFTE}.

\subsubsection{Reference and Science Image Selection}

The primary science targets for the \textsc{EFTE} survey are stellar flares,
which have characteristic optical rise-times of minutes. As a result, there is
minimal benefit from producing reference frames widely spaced in time from our
science images to maximize sensitivity to slowly varying objects. Instead, the
image-matching daemon uses a sliding reference frame, taken from the same
pointing as the science image. The Evryscopes maintain a consistent pointing
over the course of a ratchet, with only few-arcminute drift even at the equator,
meaning that the PSF for a given star is essentially constant during each
two-hour tracking period, up to resampling effects caused by its sub-pixel
position in the 13.2 arcsecond pixels. Additionally, using a reference image
from the same pointing means that the science and reference frames are taken
under near-identical sky conditions, minimizing the amount of flux scaling
necessary. 

In the most aggressive case, we could simply subtract consecutive images to
achieve near-ideal consistency between the new and reference frames. However,
immediate re-use of science images as reference images limits the survey
sensitivity to only transients with a detectable change over the two-minute
image interval, making confirmation images of highly impulsive events unlikely.
In practice, we enforce a short lag, $\Delta t_{D}$, between the reference and
science images. $\Delta t_{D}$ is typically chosen to be 10 minutes.
Over 10 minutes, field drift due to polar alignment error is
consistently less than 10 arcminutes (50 pixels). PSF variability over 50 pixels
typically produces sub-percent subtraction artifacts, and image registration
between the two images can be done with simple transformations with minimal
loss in astrometric precision (see Section~\ref{sec:astrometry_tests}).

\begin{figure}
	\includegraphics[height=5in]{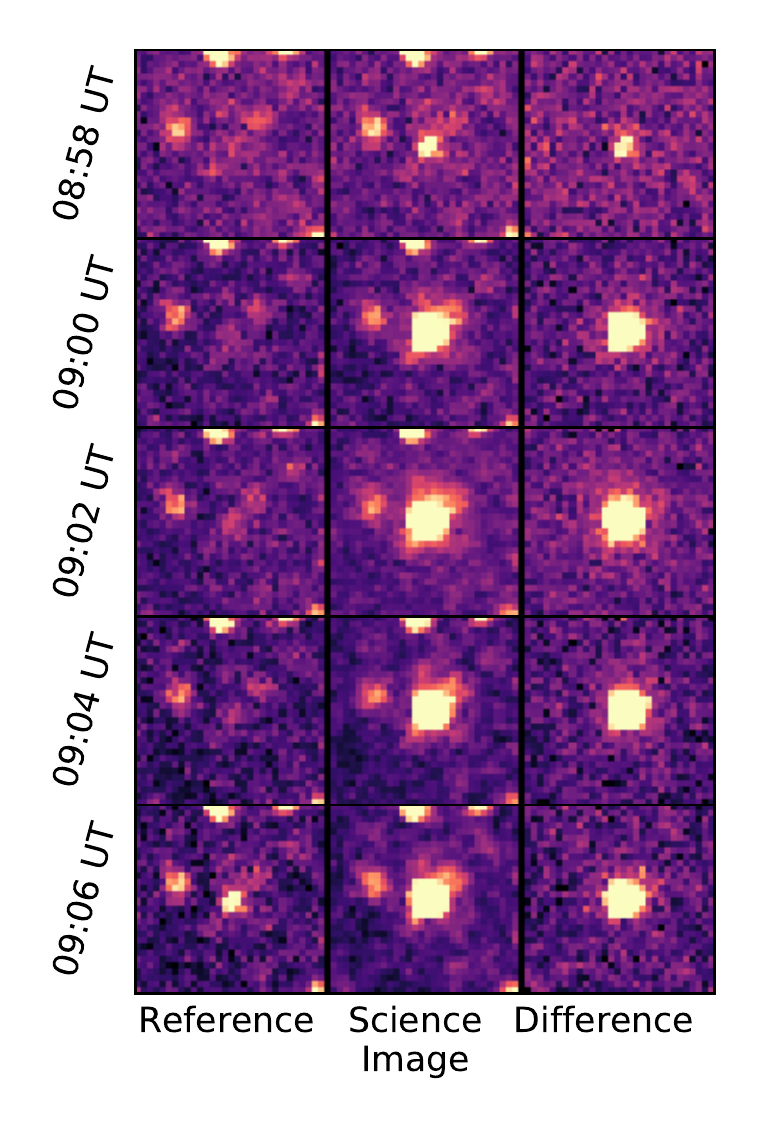}
	\caption{\label{fig:efte_event}EVRT-192099, a 5.5 magnitude 
	flare from a star associated with the 1RXS J174441.6-531551 in the ROSAT
	All-Sky Survey Bright Source Catalogue \citep[][]{fresneau_rosat,
	rosat_survey}. The reference frame from 2019 October 4 at  9:06~UT is the science image from
	8:58~UT, showing the sliding reference frame used by our direct subtraction
	algorithm.}
\end{figure}

Image re-use effects are also evident at $\Delta t_{D} = 10$ minutes, but only
after the potential fourth confirmation image. Figure~\ref{fig:efte_event} shows
this image-reuse effect on a real flare seen on-sky, in which the same image is
both the first science image and last reference image for the transient. While
the amplitude and rise time of this event enabled multiple detections up to 8
minutes after the initial detection, events with shorter rise-times or lower
amplitudes may only be detected in a single epoch. 

Additionally, the sliding reference frame causes photometry in the unscaled
difference image (\emph{i.e.,} the numerator of equation~\ref{eq:det_image}) to be
relative in time. Light curves of \textsc{EFTE} candidates are computed using
forced aperture photometry in the science images, as described in
Section~\ref{sec:photometry}.

Over the course of a ratchet beginning with images A, B, C, D, E and F, the
pipeline will perform the following subtractions: B-A, C-A, D-A, E-A, and F-B.
The rise-time sensitivity of the pipeline increases as a function of the time
delay $\Delta t_{D}$ between the science image and the previous image from the
same pointing chosen as a reference image. $\Delta t_{D}$ is in general a
tunable parameter of the pipeline which could be increased to trade the
viability of the assumptions enumerated above (and thus higher false
positive/false negative alert rates) for increased sensitivity to slower rise
times. Given sufficient computing, multiple instances of \textsc{EFTE} can run
in parallel, enabling sensitivity to different science targets.

\subsubsection{Image Registration}\label{sec:image_registration}

Because of the sliding reference frame selection, drift between the science and
reference images amounts to a maximum of a few pixels during real-time
operations that must be corrected. Additionally, small offsets can have
significant effects on the sampled PSF. As such, the images must be carefully
aligned and resampled to match both in position and PSF. 

For the first image in a ratchet, \textsc{EFTE} must wait for an astrometric
solution. However, the astrometric solution for subsequent images from each
camera can be inferred by alignment to the first image. The effects of image
registration on astrometry performance are addressed in
Section~\ref{sec:astrometry_tests}. Bootstrapping the astrometric solution in
this way reduces delays in the real-time subtraction process due to the
astrometric solver to once per ratchet, on the first image. For image alignment
and resampling, we use WCS-independent asterism-matching, using the Python
\emph{AstroAlign}\footnote{https://github.com/toros-astro/astroalign} package
\citep{astroalign} to calculate a rigid transformation between the two images
and perform quadratic resampling. 

Alternately, in cases where a full WCS solution is available for both images
(e.g., in batch reductions not conducted in real-time) the reference and science
image can be aligned by resampling the images to a common grid using their WCS
solutions using the Astropy-affiliated package \emph{Reproject}. This has the
advantage of allowing for non-rigid transformations and accounting for the
effects of varying per-pixel sky area across the sensor plane. While astrometric
warping due to atmospheric diffraction is negligible for typical $\Delta t_{D}$
values used for real-time reduction, WCS-based resampling is necessary for
longer baselines and inter-night comparisons for fixed fields.

\subsubsection{Flux Scaling}\label{sec:flux_scaling}

Despite the minimal baseline between the reference and science image, we fit a
multiplicative flux scaling factor to the reference image to remove any
discrepancies with respect to the science image due to variations in
transparency and sky brightness, which is particularly important for
observations during twilight conditions. Because each individual Evryscope camera
covers a large sky area, we allow the flux scaling factor to vary across the
image based on the results of forced aperture photometry. 

First, we divide each image into $24\times16$ 274-pixel square regions of equal
pixel area, and select several thousand bright stars from the ATLAS-REFCAT2. We
then calculate the sigma-clipped mean flux ratio between the science and
reference images for stars in each of the 384 image sectors, and interpolate
this back to full resolution using cubic splines. Finally, the flux-matched
reference image $R_m(x,y)$ is calculated from the original, calibrated reference
image $R(x,y)$ and the spatially varying flux ratio $F(x,y)$ as
\begin{equation}
    R_m(x,y) =  R(x,y) *  F(x,y).
\end{equation}

\begin{figure*}
	\includegraphics[width=\textwidth]{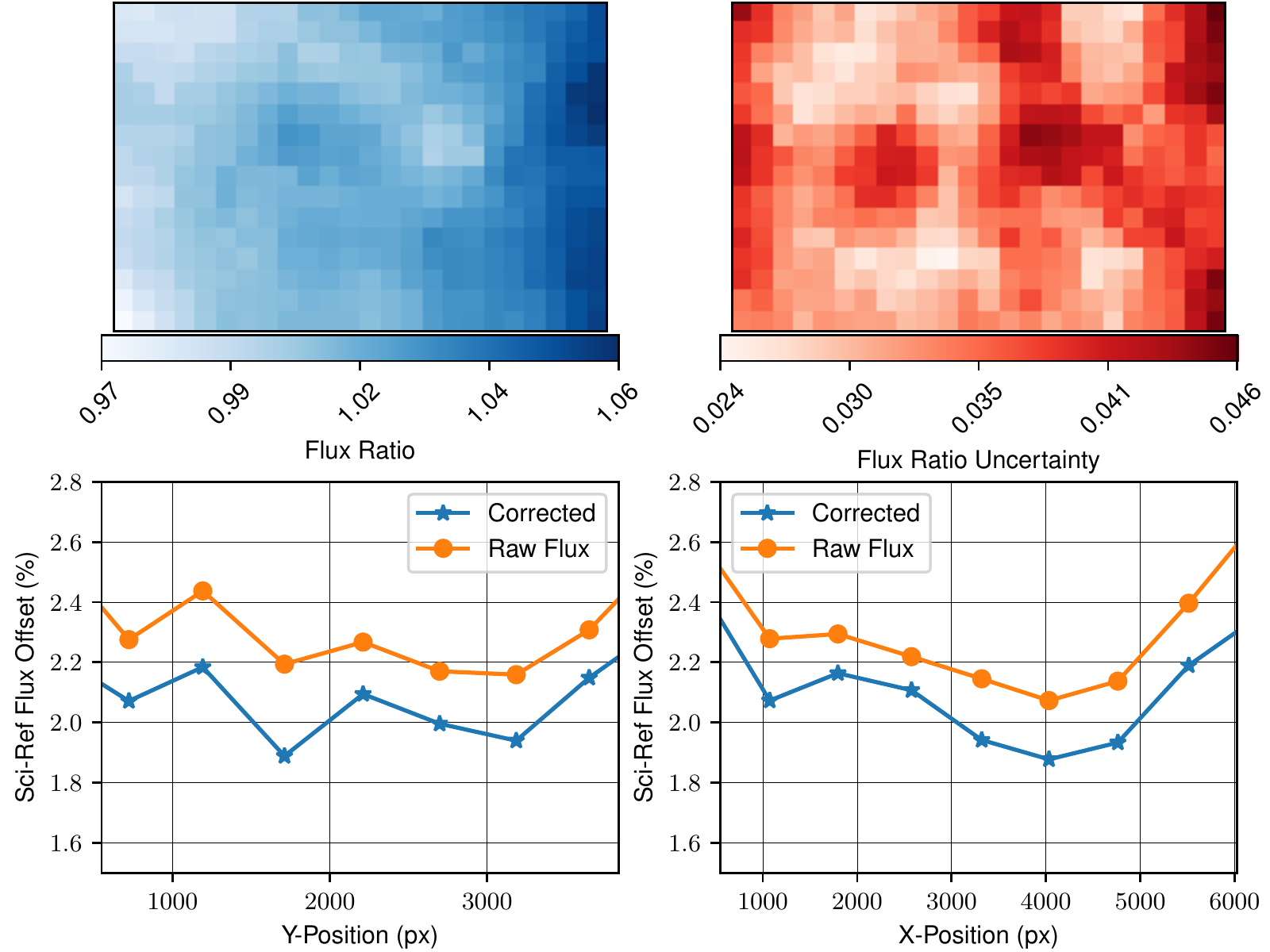}
	\caption{\label{fig:flux_ratio}\textbf{(Top-Left)} Map of the flux ratio
	between a science and reference image couplet, and \textbf{(Top-Right)} the corresponding flux
	ratio error map. The flux ratio is measured based on forced
	aperture photometry of several thousand reference stars, and is interpolated
	across each full-resolution, single-camera image based on the observed flux
	ratio in an equally spaced grid of 384 image sectors.
	\textbf{(Bottom)} Relative aperture flux residuals before and after
	correcting with the interpolated flux ratio as a function of $x$- and
	$y$-position on the sensor. The correction improves the
	flux match between images by $\lt1\%$, which is not significant
	for short ($\Delta t_{D} = 2$ minute) subtraction baselines. Increases in
	the flux offset at the edges of the image are caused by aperture losses due
	to the variable Evryscope PSF.}
\end{figure*}

To calculate the uncertainty in the flux ratio, we calculate the standard
 deviation of the flux ratios for reference stars in each region and then
 interpolate across the full field of the image using cubic
 splines.  The flux ratio uncertainty is propagated forward into the noise
 characterization for the reference image. Figure~\ref{fig:flux_ratio} shows a
 typical low-resolution flux ratio map for a pair of images with $\Delta t_{D} =
 10$ minutes, showing an average 1.7\% change in transparency between the images
 with some internal structure. The magnitude of the scaling is consistent with
 transparency changes due to airmass for a camera placed at the edge of the
 array. Small-scale structure, when present, tends to move smoothly between
 images and is likely caused by high clouds.

Depending on the science program, flux scaling can be skipped during reduction
to minimize latency. For consecutive-image subtraction ($\Delta t_{D} = 2$
minutes), we neglect flux scaling effects, as the uncertainties in the flux
scaling dominate the final noise budget for the image, and the flux scaling is
typically sub-percent under normal observing conditions. The primary driver of
these uncertainties is likely the sub-pixel response function (sPRF),
which is highly local on the Evryscope image sensors, causing the effect to not
average out beyond the 1-3\% level when interpolating across the image plane.
Instead, multiple, slightly offset measurements of the sample star must be
modeled simultaneously, as they are for the precision photometry pipeline and in
coaddition of multiple images of the same field. However, we include the flux
ratio for longer-baseline subtractions, where background variations can dominate
over systematics.

\subsubsection{Error Analysis and the Detection Image}\label{sec:detection_image}

To identify significant changes in the difference image, we need a robust
accounting of the noise sources in each image. For each science and reference
image, we model a spatially varying background based on sigma-clipped and
interpolated mesh using \texttt{sep}, a Python implementation of the core
routines from \texttt{SExtractor} \citep{bertin_1996, Barbary2016}. The standard
deviation of the background $s_B$ is also measured at this step, based on the
sigma-clipped standard deviation. $s_B$ is treated as an empirical measure of
the Gaussian noise contributions to each image, including the readout noise and
dark current uncertainty. We note that this approach can overestimate the noise
due to Poisson contributions from unresolved background star light, and as a result,
the detection significance in the direct subtraction image will tend to be an
underestimate, particularly in crowded fields (\emph{e.g.,} near the galactic plane).

For a given combination of science image $S(x,y)$ and flux-matched reference
image $R_m(x,y)$, both in electron units, the detection image $D(x,y)$ is
defined as
\begin{equation}\label{eq:det_image}
	D(x,y) = \frac{S(x,y) - R_m(x,y)}{\sqrt{s_S^2(x,y) + s_R^2(x,y)}},
\end{equation}
where $s_S^2$ and $s_R^2$ are the total noise images for $S$ and $R_m$, given by
\begin{equation}
	s_S(x,y) = \sqrt{S(x,y) + s_{BS}^2(x,y)},
\end{equation}
and 
\begin{equation}
	s_R(x,y) = R(x,y) \sqrt{\frac{R(x,y) + s_{BR}^2(x,y)}{R^2(x,y)} + \frac{s_F^2(x,y)}{F^2(x,y)}},
\end{equation}
where $s_{BS}(x,y)$ and $s_{BR}(x,y)$ are the measured background standard
deviation maps of the science and reference images, respectively, $F(x,y)$ is
the flux ratio between the two images, and $s_F(x,y)$ is the
spatially varying uncertainty in the flux ratio. $D(x, y)$ image is the simple
difference in the two images, scaled by the combined per-pixel uncertainty. The
detection image has units of standard deviations. We again use \texttt{sep} to mask
the detection image at the desired threshold and identify sources.

Figure~\ref{fig:proxima_diff} shows an example of an image couplet in a crowded
field, along with the resulting direct subtraction image. We also include
subtraction images produced using the \textsc{HOTPANTS} and \textsc{ZOGY}
algorithms \citep{hotpants,zogy}. Direct subtraction produces more false
positives (488) than \textsc{ZOGY} (299 in $S_{corr}$) at the 3-$\sigma$
threshold, but two orders of magnitude fewer than \textsc{HOTPANTS} (12258).
While \textsc{ZOGY} is a computationally efficient approach, direct subtraction
is faster by a factor of ten, largely due to the requirement to calculate a PSF
model for ZOGY. Using the direct detection image, we successfully identify three
astronomical transients with the \textsc{EFTE} pipeline using the vetting
procedures described in Section~\ref{sec:autovet}.

\subsection{Photometric Zeropoints and Forced Photometry}\label{sec:photometry}

To calibrate magnitudes for \textsc{EFTE} transient candidates to a standard
photometric system, we build a spatially varying photometric zeropoint based on
a subset of the ATLAS-REFCAT2, a composite catalog consisting of $griz$ data
from the AAVSO Photometric All-Sky Survey \citep[APASS;][]{apass_dr9},
Pan-STARRS Data Release 1 \citep[][]{panstarrs_dr1}, Skymapper Data Release 1.1
\citep[][]{skymapper_dr1}, Tycho-2 \citep[][]{tycho2, tycho2_griz}, the Yale
Bright Star Catalog \citep[][]{yale_bsc_1991}, GAIA DR2 \citep[][]{gaia_dr2},
plus original data from the ATLAS Pathfinder survey. To ensure that stars used
for determining the zeropoint of the image are well-exposed, but not saturated,
we select a subset of the catalog between $10<{g'}<12$. We also exclude stars
with colors redder than $g-r=1.5$ that might bias the photometry due to
unconstrained chromatic aberrations affecting the PSF. We calculate an
instrumental magnitude for each of the reference stars using a forced aperture
at the catalog position in the background-subtracted and calibrated science
image, typically with an aperture radius of 3 pixels (40 arcseconds). 

To model the variation in photometric zeropoint across the field of view, the
science image is divided into an $8\times12$ grid of square subframes, each of
which subtends 4 square degrees and contains O(100) reference stars. Within each
subframe, we calculate the sigma-clipped median offset between the instrumental
magnitudes calculated via aperture photometry and the catalog values. The
offsets in each region are then smoothly interpolated over the rectangular mesh
of the full-sized using quintic splines to produce a spatially varying
zeropoint $z(x, y)$. The resulting image has units of magnitudes and has as its
values the photometric zeropoint at each pixel, defined such that
\begin{equation}
	m_g = -2.5 \log_{10}{F_{aper}} - z(x, y), 
\end{equation}
where $F_{aper}$ is the measured flux from aperture photometry.

Finally, we calculate magnitudes for each candidate detected as described in
Section~\ref{sec:detection_image}, based on their centroid positions in the
detection image. Centroids are calculated for each candidate by computing their
value-weighted average position (``center of mass''). This process uses a custom
aperture photometry routine, implemented in Cython for the Evryscope precision
photometry pipeline \citep{evryscope_instrument}, on the science image. 

\section{Automated Vetting}\label{sec:autovet}

Despite the optical consistency of Evryscope images chosen for subtraction, the
direct subtraction process produces thousands of false positives per image.
Observed sources of false positives are plentiful from inside the CCD sensors
out to Earth orbit, including:
\begin{itemize}
	\item Cosmic ray muon tracks
	\item Compton recoil electrons from radionuclides in materials at the
	observatory
	\item Optical ghosts
	\item Registration and astrometric errors
	\item Persistent residual charge from bright stars remaining after cycling
	the detectors
	\item Flat-fielding errors
	\item Aircraft strobes
	\item Tumbling satellites and debris \citep{corbett_flashes}
	\item Noise artifacts from both photon and astrometric noise. 
\end{itemize}  

In total, the event rate from these sources can outnumber the real, on-sky rate
of astrophysical transients by orders of magnitude. Human candidate inspection
remains standard, but it is not scalable to surveys producing hundreds of
thousands of candidates per night. As a result, a reliable, efficient, and
automated vetting system for candidates is a core component in any transient
survey producing an actionable event stream that can be delegated to followup
resources. 

Some false-positive sources in the list above can be identified with simple
filters: Bright streaks from satellites involve thousands of pixels, and
residual charge from bright stars can be flagged based on previous astrometric
solutions. Other classes of observed signals can be difficult to identify
from simple metrics in all scenarios. To account for this, we use a combination of data
cuts based on explicit filters and machine learning (ML) methods. 

\subsection{Initial Candidate Filters}

While ML techniques can be comprehensive, simple filters grounded in domain
knowledge can be both more efficient and more easily interpreted. Starting from
an initial deep source extraction ($SNR > 3$ in a minimum of 1 pixel) of the
detection image $D(x,y)$, we implement three first-order quality cuts, removing
candidates which meet any of the following conditions:

\begin{enumerate}
	\itemsep0em 
	\item Centroid within 15 pixels of the edge of the CCD
	\item Ratio of negative-to-positive pixels within a 6-pixel circular
	aperture $>0.4$
	\item More than $750$ pixels above the detection threshold
\end{enumerate}

Detections near the edge of a CCD are typically caused by small amounts of mount
drift between the science and reference images. Large ratios of negative to
positive pixels typically indicate a photon or astrometric noise artifact.
Extended events $>750$ pixels are commonly bright streaks, caused predominantly
by aircraft and LEO satellites. 

We apply an additional filter after the ML vetting described below; we reject
any candidates coming from a subtraction with more than 500 high-confidence
candidates. These failed subtractions rarely occur, and are caused by a doubled
or streaked image due to wind shake at the instrument, or a breakdown of the
assumption of a slow and smoothly varying sky background required for direct image
subtraction, as described in Section~\ref{sec:direct_subtraction}.

With no additional vetting, these simple filters reduce the per-image candidate
count to O($10^2$) using the baseline values stated above; however, these
numbers are readily tunable to the science case and corresponding false positive
tolerance, either by modifying a configuration file for the \textsc{EFTE}
pipeline instance running at each observatory, or by filtering the database
queries used to regularly report candidates to end users. Candidates that pass
the thresholds for these filters at the pipeline-instance level are inserted
into the central database, including small $30\times30$-pixel ``postage stamp''
cutouts around their detection positions.

\subsection{\textsc{vetnet}: Real-Bogus Classification with Convolutional Neural Networks}
\label{sec:convnet_classifier} 

For additional reduction of the \textsc{EFTE} false positive rate, we use an
ML model based on two-dimensional convolutional layers
\citep{lecun-89} with weights conditioned directly on image data. This model is
a binary (``real/bogus'' [RB]) classifier, which assigns each candidate a score
between 0 and 1, where a score of 1 indicates that the candidate is likely real.
RB classifiers have seen long-standing use in transient surveys, starting with
the model built by \citep{bailey_2007} for the Nearby Supernova Factory
\citep{Aldering_2002}. Similar approaches have been used for the Palomar
Transient Factory \citep[PTF;][]{law_2009, bloom_2008}, the Intermediate Palomar
Transient Factory \citep[iPTF;][]{brink_2013}, the Dark Energy Survey
\citep{goldstein_2015}, and most recently, for ZTF \citep{mahabal_2019,
duev_2020} and GOTO \citep{Killestein_2021}.

Deep learning is a type of ML in which ``deep'' stacks of
artificial neural network layers \citep{mcculloch_1943} are used to transform
input data into latent-space encodings that can be mapped to the desired output
quantities. Convolutional Neural Networks (CNNs) \citep{lecun_convnets} are a
sub-class of artificial neural networks that build up a latent space
representation of pixel data using convolutions, which identify
increasingly compressed features of the input as the depth of the network
increases, as opposed to requiring pre-selection of computed - and potentially
sub-optimal - features to represent the data. CNNs have found widespread
use in astronomy for tasks including source detection and deblending
\citep{autosourcid-light, Burke_2019}, in addition to transient real-bogus
vetting \citep{makhlouf_2021, Forster_2016,duev_2020,Killestein_2021}.

In this section, we describe \textsc{vetnet}, a CNN-based vetting algorithm
trained to assign real-bogus probabilites to \textsc{EFTE} candidates directly
from 30$\times$30 pixel cutouts from the reference, science, and direct
subtraction difference images. 

\subsubsection{Training Set and Data Labelling}

Supervised ML classifiers require large datasets of labelled examples to
identify the complex latent associations during training. In general, there are
two options for producing these datasets: simulation or human classification.
Exclusively training on simulated data is risky because the efficacy of the
final model is dependent on how representative the simulations are of real data.
However, human classification is labor-intensive, and prohibitive at the level
of producing thousands or even millions of labelled examples across a
representative sample of the survey.

As a result, we have adopted a compound approach, using both hand-labelled,
on-sky data and simulated events produced via spatially varying PSF injection.
The simulated dataset was used to train intermediate models to pre-screen events
for human labelling, including the prototype CNN used
in \cite{corbett_flashes}. We manually classified the on-sky, moderate-purity
sample produced by the intermediate model to produce a smaller, but
minimally contaminated and representative, data sample for training the
production models.

\subsubsection{Network Architecture}

\begin{figure*}
	\includegraphics[width=\textwidth]{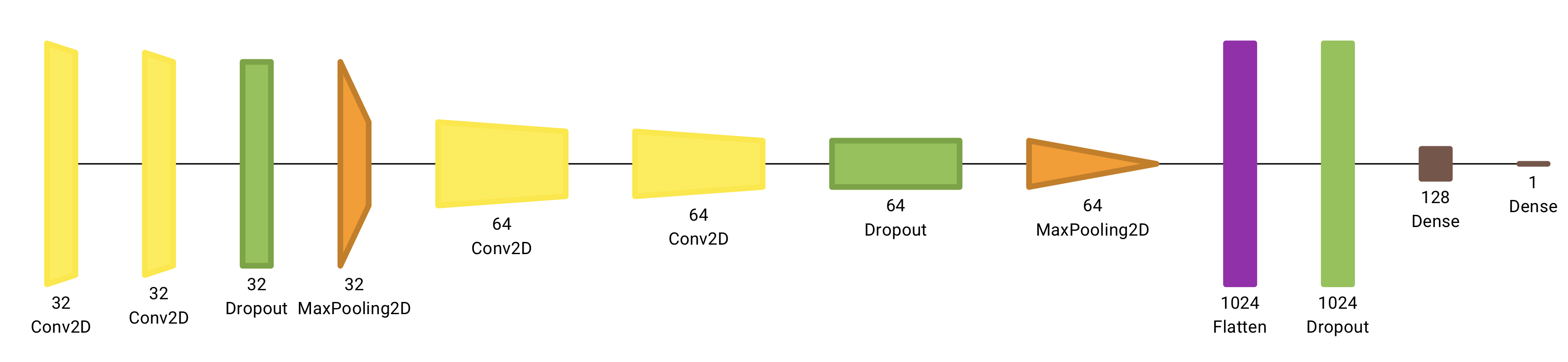}
	\caption{\label{fig:vetnet_arch}Architecture of \textsc{vetnet}, a
	convolutional real-bogus classifier used by \textsc{EFTE}. The inputs to the
	network are a triplet of 30$\times$30 pixel cutouts around the center of
	each candidate, taken from the reference, science, and direct difference
	images. All trainable layers, save for the final dense unit, use ReLU
	activation. Pairs of convolution layers are each followed by max-pooling
	layers and 20\% dropout for regularization. Network visualization generated
	with Net2Vis \citep{net2vis}.} 
\end{figure*}

\textsc{vetnet} uses a sequential, VGGNet-like \citep{vggnet} model with six
trainable layers; four convolutional, and two fully connected output layers.
Each set of convolution layers is subject to 20\% dropout to prevent
overfitting, encouraging the model to build a diverse set of representations
of the data distribution. The dropout fraction at each layer was determined
using the HyperBand band algorithm \citep{hyperband} with a binary cross-entopy
loss function. Further regularization is provided by a pooling layer, which
reduces the dimensionality of each convolution block output by a factor of four.
Outputs of each pooled layer are normalized and re-centered on zero using batch
normalization \citep{ioffe_2015} to improve training performance and model
stability.   
All convolution layers use 3-pixel square kernels and ReLu activation
\citep{relu}, save for the final fully connected node, which has a sigmoid
activation function that produces an output value normalized between 0 and 1.
This output, the \textsc{vetnet} real-bogus (RB) score, can broadly be
interpreted as probability that a given candidate is real.
Figure~\ref{fig:vetnet_arch} depicts the architecture of the model, including
filter depths and the resulting dimensionality. 

\textsc{vetnet} is implemented in Tensorflow \citep{tensorflow2015-whitepaper}, using the
high-level Keras API \citep{chollet2015keras}. 

\subsubsection{Dropout and Model Uncertainty}\label{sec:vetnet_mcdropout} 

CNNs can have an arbitrarily large number of free parameters, and are
accordingly able to overfit training data. As a result, methods of regularizing
the training process and the weights assigned to the convolutional filters are
necessary to maximize performance on actual data at inference time. Dropout
\citep{srivastava14a} is one common technique, in which a tunable fraction of
outputs from a layer are chosen at random and set to zero, preventing them from
contributing to the final network outputs. 

In addition to slowing overfitting, dropout also can be interpreted as an
approximation of Bayesian inference \citep{Gal_2015}. In this framework, each
random sampling of layer outputs can also be considered a sample from the
distribution representing network weights in a fully Bayesian network.
Evaluating a given sample through these different dropout-induced realizations
of the network enables us to similarly approximate the posterior distribution of
the network output. The advantage of this approach, called Monte Carlo (MC)
Dropout, is that the output distribution includes the systematic uncertainty in
the network output due to model selection, distinct from the random uncertainty
produced by the variance of the training set \citep{mcdropout}. To produce an
output from the network, each candidate is processed through multiple
dropout-induced realizations of the network, producing a distribution of
resulting RB scores. We use the median of this distribution as the RB
probability for each source.

Interpretation of MC Dropout is unsettled in the literature (namely,
whether it represents a genuinely Bayesian approximation \citep{folgoc_2021}).
However, it can be used to produce a number that scales with the degree of
consensus within the network and amount of support for a sample within the
training set, and that can be interpreted as a confidence metric. This is
similar to the interpretation of the sigmoid activation of the network as a
whole as a real-bogus probability, despite not representing a normalized
probability density function. We
adopt the entropy-based  
metric from \cite{Killestein_2021} to quantify the network confidence: 
\begin{equation}
	\mathbb{C}=\frac{1}{N}\sum_{i=1}^{N} -p_i\log_2{p_i} - (1-p_i)\log_2{(1-p_i)},
\end{equation}
where $N$ is the number of samples from the posterior distribution and $p_i$ is
the network output for the $i$th sample. The metric $\mathbb{C}$ is the binary
entropy of the Bernoulli process representing real-bogus classification,
averaged across posterior draws, and is bounded on the interval $[0,1]$. In
Section~\ref{sec:convnet_eval}, we demonstrate that $\mathbb{C}$ also matches
the subjective confidence of human vetting.

\begin{figure}
	\includegraphics[width=1\columnwidth]{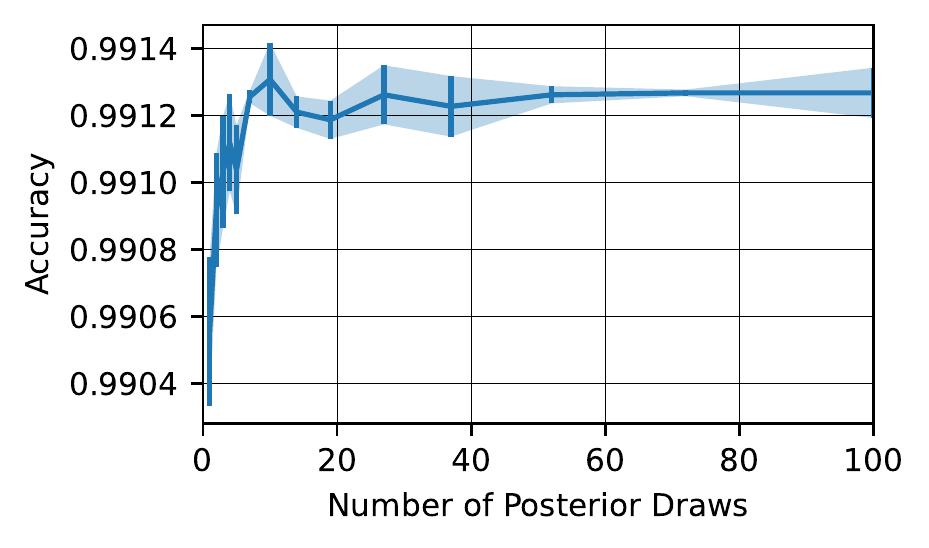}
	\caption{Network accuracy vs. number of forward passes through the network
	for the validation set. Performance converges at 10 samples. Error bars
	represent the standard deviation of the results across 20
	iterations.\label{fig:posterior_draws}}
\end{figure}

The number of forward passes used to approximate the network output posterior
distribution is determined empirically from the validation set.
Figure~\ref{fig:posterior_draws} shows the accuracy of the classifier as a
function of the number of forward passes through the network. The accuracy of
the median RB score converges after 10 inferences, which is consistent with the
findings in \cite{Killestein_2021}, despite the dropout rate here being two
orders of magnitude higher. 

\subsubsection{Training Set and Data Augmentation}\label{sec:vetnet_dataset}

Two datasets were used to train the \textsc{vetnet} classifier; the 435,452
simulated candidate dataset described in Section~\ref{sec:transient_injection},
and a human-annotated sample of on-sky detections containing 31,092 candidates
flagged as probably real by an earlier iteration of \textsc{vetnet} itself
\citep{corbett_flashes}. Unlike the simulated dataset, the on-sky data is
heavily class-imbalanced, with only 9.6\% of examples (2,976) being
human-labelled as real. To account for this class imbalance, we randomly exclude
25,140 of the bogus samples from the on-sky dataset, noting that the simulated
dataset only contains simulated examples of the \emph{real} class. Bogus
examples within the simulated data set are drawn from the same population as the
bogus examples in the on-sky dataset. Our approach for maximizing the return
from this relatively small sample of on-sky data is described in
Section~\ref{sec:vetnet_training}. 

We divided the simulated and annotated on-sky datasets into training,
validation, and testing subsets using an 80:10:10 ratio. We used validation set
for tuning the MC Dropout fractions and the number of posterior draws, and for
monitoring the training process.

To extend the effective size of the dataset, random flips and rotations are
applied to each batch of training samples. As noted by \cite{Killestein_2021}
and \cite{dielman_2015}, rotations (other than in 90 degree increments) require
interpolation and thus distort the data from the pixel grid; however, in our use
case, the data are previously resampled with interpolation by the image
alignment process (see Section~\ref{fig:efte_astrometry}). No data augmentation
is applied during validation or model evaluation, or for training during
fine-tuning with human-annotated data.

\subsubsection{Simulated Data Generation}\label{sec:transient_injection}

We generated a base training set by injecting simulated transients into 300
randomly selected images across the first two years of full Evryscope science
operation at each site. Images were selected uniformly in time, meaning that moon
phase, sky conditions, focus changes due to temperature variations, and
dust-accumulation on the instrument (leading to measurable changes in the
background level and limiting magnitude on few-month timescales) are uniformly
represented. Each image was calibrated as in the pipeline (see
Section~\ref{sec:direct_subtraction}).

For each image, we generated a uniform sample of 5,000 positions within the
image, then deduplicated so that no position was within 50 pixels of any other
position to avoid overlapping transients, resulting in an average of 1,200
injections per image. While this does bias the initial training set against
contemporaneous, spatially coincident events, this is sufficiently rare that we
neglect this scenario. Each injection was
assigned a random magnitude, drawn from a uniform distribution bounded between
the typical saturation limit of g$\sim7$ and the 1.5$\sigma$ detection limit at
the injection position (determined by the photometric zeropoint interpolation
procedure described in Section~\ref{sec:photometry}). 

A second round of injections was done to simulate transients with known
visible progenitors. From the catalog stars within each image, 500 stars
minimally separated by 50 pixels were selected as additional positions for
injection. The variability amplitude was uniformly sampled between 0.25 and 8
magnitudes. The upper limit is set by the maximum contrast visible for a
pre-detected star in an Evryscope image, \emph{i.e.}, a star at the dim-limit of the
survey that reaches the single-exposure saturation limit.

\begin{figure*}
	\includegraphics[width=\textwidth]{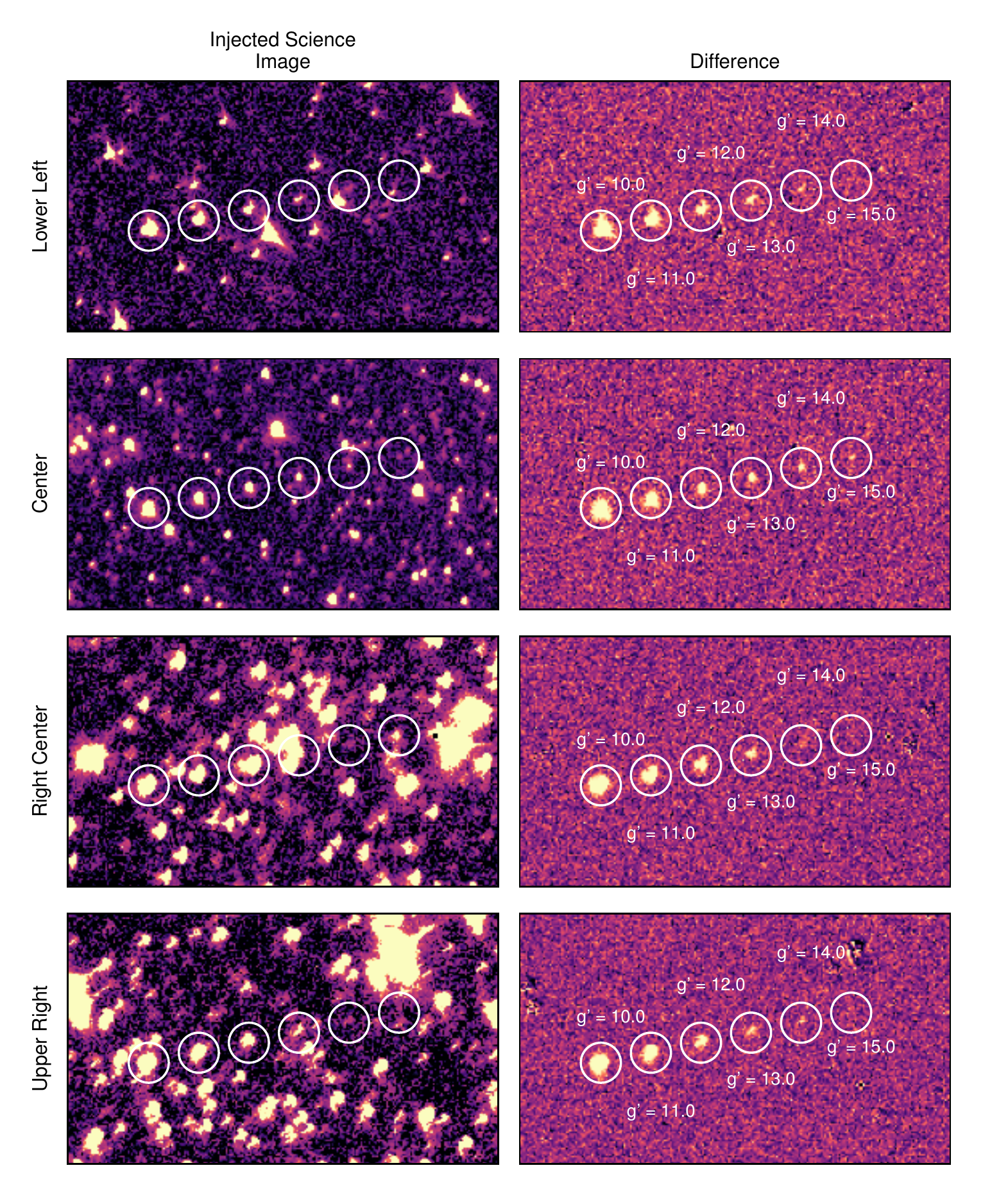}
	\caption{\label{fig:efte_injection} \textbf{(Left)} Examples of simulated transients with
	magnitudes $g' = 10, 11, 12, 13, 14, 15$ at the center and edges of a
	typical mid-galactic latitude Evryscope image. PSFs at each position are
	modeled as a normalized, aligned, and sigma-clipped combination of nearby
	isolated stars, producing morphologically plausible, star-like injections.
	These simulations are used for initial conditioning of our machine-learned
	vetting system. \textbf{(Right)} Difference images for each subimage
	using a reference frame taken two minutes before the injected science image.}
\end{figure*}

Evryscope PSFs are heavily impacted by optical aberrations, and exhibit a wide
variety of morphologies, both between cameras and across the field of view of
individual cameras \citep{ratzloff_robotilter}, making common analytic profiles
(Moffat, Gaussian, Lorentzian) untenable. Further, the coarse pixel scale makes
more complex linear models, such as the ePSF \citep{ePSF} and those used by
\textsc{PSFEx} \citep{psfex} and \texttt{PSFMachine} \citep{hedges_2021}, prone
to poor fits due to aliasing and source confusion. We found that the most robust
method for simulating transients with morphologically plausible, point-like
profiles was to build a model PSF based on nearby, isolated stars. For each
injection position, up to 100 nearby stars having a distance less than 137
pixels and significance of 10$\sigma$ above the local noise are extracted with a
30$\times$30 ``postage stamp'' window. Each stamp is then multiplied by a
smoothly varying (Hanning) window and normalized. The final PSF to be flux
scaled and added into the image is the median of the nearby stamp templates,
weighted by the relative normalized distance from the injection position and
relative flux uncertainty of the template star. Figure~\ref{fig:efte_injection}
shows examples of simulated PSFs using this technique across a typical Evryscope
focal plane for a range of magnitudes, alongside the resulting signal in a
direct subtraction image with a consecutive epoch.

At the end of the transient injection process, the image is ``de-calibrated'' by
adding back in the expected dark current, bias, and background levels, and the
image is converted back into ADU units with pixel values beyond the range of an
unsigned 16-bit integer truncated, matching the histogram of the simulated
images to the distribution expected for science images.

To produce a simulation-augmented dataset, the transient-injected images are
reduced using the \textsc{EFTE} pipeline, and any candidates identified within 2
pixels of an injection location are labelled as real, and all others as bogus.
Despite the large number of injected sources, this process produces an
unbalanced dataset, with artifact detections outnumbering injections at rates up
to 1000-to-1. To balance the dataset, we randomly select a number of bogus
candidates equal to the number of recovered injections for inclusion in the
final data set. This results in a dataset with noisy labels due to both
background transients (likely dominated by short-duration reflections from Earth
satellites \citep{corbett_flashes,nir_2020}) and candidates injected below the
difference image threshold and recovered coincidentally. From visual inspection
of 10,000 injection candidates, uniformly sampled from both known-injections and
predicted artifacts, we estimate label contamination to affect $\leq2.7\%$ of
the candidates in the 435,452 candidate dataset. Deep convolutional models have
been observed in prior work to be robust to many times this level of label noise
\citep{ghosh_2016, rolnick_2018}.

\subsubsection{Staged Training Methodology}\label{sec:vetnet_training}

A common approach for building specialized models with limited training data is
to utilize transfer learning, leveraging the pre-trained feature representations
of existing models built with massive related datasets. Rather than training
an entire model from scratch, which requires a large annotated dataset over the
domain of interest, a pre-trained model can be selectively ``fine-tuned'' over a
representative dataset in a new domain. We adopt a similar approach for making
use of the considerable diversity of observing conditions and sky regions
represented in the simulated dataset described in
Section~\ref{sec:vetnet_dataset}, while minimizing the risk of optimizing for 
properties of the transient injection process (see
Section~\ref{sec:transient_injection}) rather than properties transferable to
on-sky data.

Our training curriculum for \textsc{vetnet} was as follows:
\begin{enumerate}
	\item Train the full model, including all convolutional and fully connected
	layers, on the simulated dataset until convergence to create the synthetic
	base model
	\item Freeze the weights on all convolution layers from the synthetic base,
	and re-train the fully connected layers from scratch using on-sky data
	\item Unfreeze the convolution layers, train at a minimal learning rate
	using on-sky data until convergence to produce the final on-sky model.
\end{enumerate}
We used the Adam optimizer \citep{kingma_2014} with a binary cross-entropy loss
function for all three stages. For the first two training stages, we start with
a maximum learning rate of 0.0003, slowed by a factor of two whenever the loss
on the validation set plateaued for 10 epochs. Scheduling the learning rate in
this way helps the network to converge to a local minimum when near a global
minimum and decreases the oscillation around the minimum of the loss function.
For the final fine-tuning of the entire model, we reduced the initial learning
rate to 0.00001 while maintaining the same scheduled rate decay. 

The synthetic base model converges after $\sim$100 epochs, taking about 6.5
hours when training on an Intel Xeon E5-2695v4 CPU. Fine-tuning with on-sky data converges
after $\sim$150 epochs, but takes less than 10 minutes due to the smaller quantity of
data. 

\begin{figure}
	\includegraphics[width=1\columnwidth]{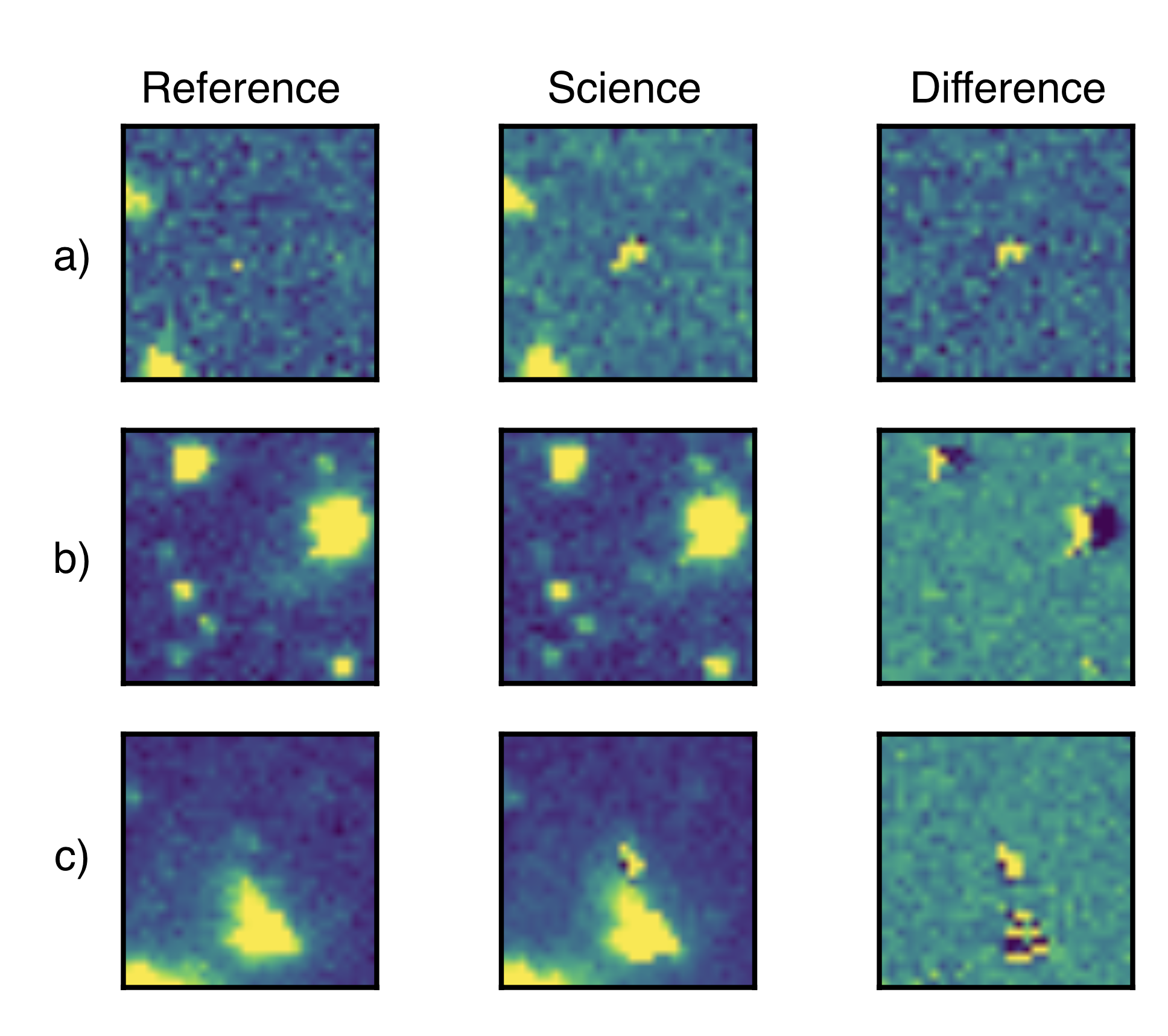}
	\caption{Misclassified training set samples re-labelled based on the
	entropy-based confidence of \textsc{vetnet} predictions. Network
	reclassifications typically affect samples which are difficult to classify
	manually. Samples \textbf{(a)} and \textbf{(c)} were initially
	classified as real by human vetters, but were relabeled as bogus by the
	algorithm. Both have pathological PSFs likely caused by interpolation
	artifacts from resampling near a cosmic ray, particle strike, or unmasked
	bad pixel. Sample \textbf{(b)} is low-significance and off-center, but was
	manually classified as bogus before being confidently relabelled as real by
	\textsc{vetnet}. \label{fig:relabels}}
\end{figure}

In the final stage, we use a single-iteration of the semi-supervised relabelling
routine suggested by \cite{Killestein_2021}; samples in the training set which
are classified differently by the network than by human labelers are flipped to
the model classification in cases where the model confidence $\mathbb{C}$ is
greater than the median. On review, these samples are generally either difficult
to classify by hand, with low-significance peaks relative to the surrounding
noise, areas of the sensor plane with pathological PSFs, or likely errors
in the original labelling process. Figure~\ref{fig:relabels} shows three
representative examples that are re-labelled by \textsc{vetnet} during this
process. In total, less than 4.7\% of samples are changed in the training set
after updating the labels. We note that this is comparable to the label noise in
the synthetic dataset (2.7\%). 

\subsection{Candidate Crossmatching and Source Association}

\textsc{EFTE} candidates from both sites and their corresponding metadata are
stored in a relational database. On insert, candidates are associated with
previous candidates at the same position, which collectively form an ``event,''
via an insert trigger within the database. If a candidate has no antecedent, a
new event is created, and an additional trigger crossmatches the new event's
position with a variety of externally produced reference catalogs.  At time of
writing, these reference catalogs include the International Variable Star Index
\citep[VSX;][]{vsx_catalog}, the Galaxy List for the Advanced Detector Era
\citep[GLADE;][]{glade_catalog}, ATLAS-REFCAT2, and the ASAS-SN Catalog of
Variable Stars \citep{asassn_variables}. Stellar sources are crossmatched with a
radius of 26 arcseconds (corresponding to 2 Evryscope pixels and the worst-case
astrometric performance for \textsc{EFTE} detections - see
Figure~\ref{fig:efte_astrometry}), and galactic sources from GLADE are
crossmatched with a 1 arcminute radius. 

To accelerate the in-database crossmatching and candidate queries, all
candidates, events, and reference catalogs are indexed using the Quad Tree Cube
(Q3C) pixelization scheme,\footnote{See:
\href{https://ascl.net/1905.008}{https://ascl.net/1905.008} \citep{q3c_ascl}} a
PostgreSQL extension for efficient spherical crossmatching and radial queries
\citep{q3c_paper}. Sky areas, such as the on-sky footprints of images or
probability contour regions for multi-messenger transient events, are indexed
using PostGIS with a custom non-geodetic projection. This projection does not
include the WGS-84 \citep{kumar1988world} reference ellipsoid, and represents
the right ascension and declination in the standard  barycentric celestial
reference system in all \textsc{EFTE} application code.

Candidates can also be crossmatched against external triggers received by
\textsc{EFTE} via automated circulars from the NASA Gamma-ray Coordinates
Network/Transient Astronomy Network (GCN). Alerts are inserted in the central
database by an automated ingest microservice, where they are indexed by position
either using Q3C for tightly localized triggers, or as PostGIS polygons for
events that are distributed as polygon skymaps, like LIGO/Virgo skymaps
\citep{ligo_instrument} or GRB alerts from the Fermi Gamma Burst Monitor
\citep[Fermi GBM;][]{fermi_gbm_instrument}.

\section{Pipeline Performance Evaluation}\label{sec:results}

\subsection{Photometric Solutions}\label{sec:photometry_tests}

\begin{figure}
	\includegraphics[width=\columnwidth]{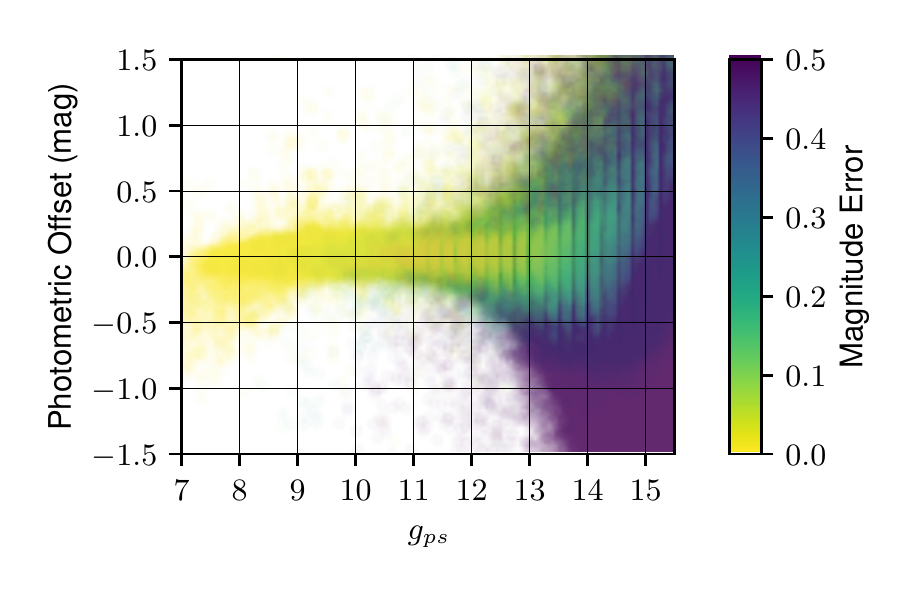}
	\caption{\label{fig:phot_perf}Photometric calibration offsets between the
	ATLAS All-Sky Photometric Reference Catalog and \textsc{EFTE}. Median RMS
	offset in the  region $8 < m_g < 14.5$ is 0.06 magnitude. Anomalously bright
	and high-precision measurements (upper right) are due to source confusion
	and blending. Under-reporting of magnitudes due to saturation is evident for
	stars brighter than $g' = 8$.}
\end{figure}

To evaluate the performance of the photometric calibration using a
smoothly varying zeropoint, as described in Section~\ref{sec:photometry}, we
compare single-epoch forced photometry from 3,217,215 catalog stars from the
ATLAS-REFCAT2 across 500 randomly selected images from the 2018 observation
year. The images were required to pass the quality assurance metrics described
in Section~\ref{sec:data_quality_and_calibration}, but were not otherwise
filtered for sky or instrumental conditions. Figure~\ref{fig:phot_perf} gives
the distribution of photometric offsets and offset RMS as a function of
magnitude in individual images. The resulting photometry is calibrated to the
reference catalog with an RMS offset of 0.05 magnitude between $8 < m_g < 14.0$,
measured using 5 iterations of a 5-$\sigma$ clip to remove outliers due to
single-epoch failures.

These numbers likely represent an upper limit on photometric RMS for isolated
and dim events, as the distribution is dominated by source confusion beyond $g'
= 14$ (e.g., dim catalog stars with a brighter star near or within the 6-pixel
aperture), causing anomalously bright and high-precision measurements of dim
catalog sources. Sources brighter than $g' = 9$ are occasionally saturated when
they appear near the center of the image, though typically, sources as bright as
$g' = 8$ are well-calibrated and linear. There is a noise floor around $\sim5\%$
for single-epoch detections from Evryscope due to variation in the sub-pixel
response across the image plane. These effects are modeled in data products from
the Evryscope precision photometry pipeline \citep{evryscope_instrument}, but
are prominent in raw single-epoch bright-star photometry from \textsc{EFTE}.

Additional color and airmass terms can be applied to light curves  of EFTE
photometry as needed, using the equation
\begin{equation}
	g_{EVR} = g_{PS} + A + B(g_{PS} - r_{PS}) + k_1X + k_2X(g_{PS} - r_{PS}), 
\end{equation}
where $g_{EVR}$ is the magnitude in Evryscope $g$-band, $g_{PS}$ and $r_{PS}$
are the PanSTARRS magnitudes from the ATLAS reference catalog, $X$ is the
airmass of the star, and $A$, $B$, $k_1$, $k_2$ are fitted photometric
conversion factors between the Evryscope and PanSTARRS bandpasses. Based
on fits to forced-aperture light curves using a robust estimator
\citep{fischler_bolles_1981}, the photometric conversion terms are $A = 0.037
\pm 0.002$, $B = -0.051 \pm 0.004$, $k_1 = 0.021 \pm 0.002$, and $k_2 = -0.051
\pm 0.003$. The light curves used to fit these parameters were chosen from a
random sample of 25,000 Northern Hemisphere stars, which we then filtered based
on a quality metric which includes source variability relative to nearby stars,
saturation, and the shape of the aperture flux growth curve, leaving a cleaner
sample of 10,671 lightcurves with an average of 17,140 epochs. 

\begin{figure*}
	\gridline{\fig{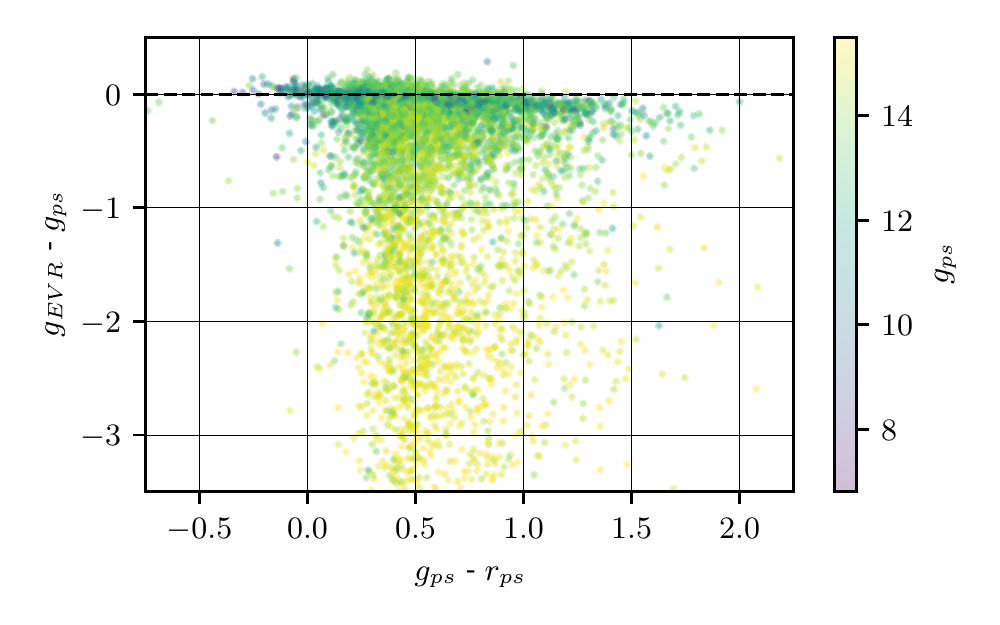}{0.5\textwidth}{(a)}
			  \fig{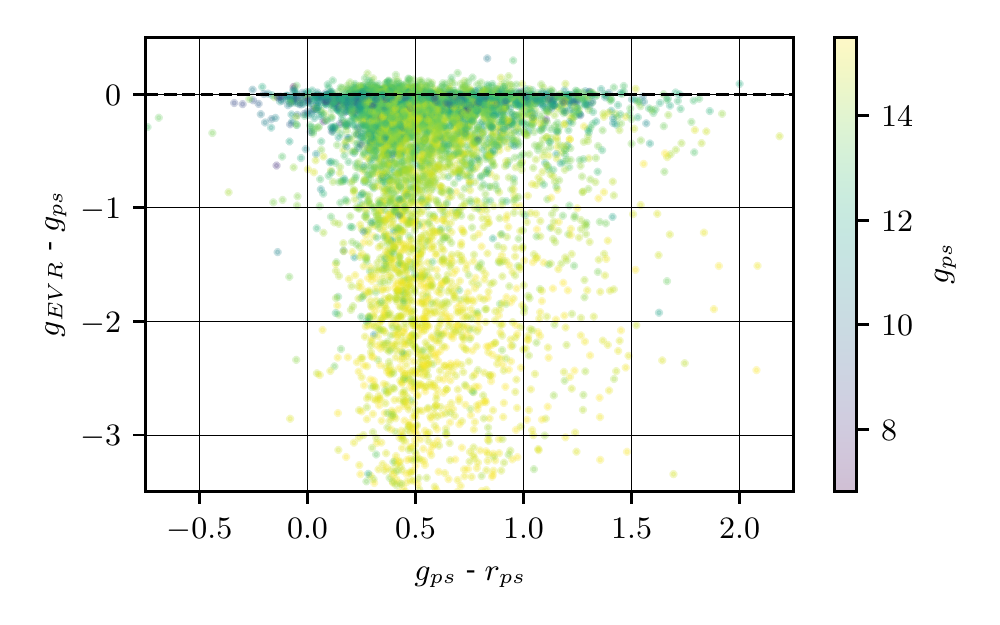}{0.5\textwidth}{(b)}}
	\gridline{\fig{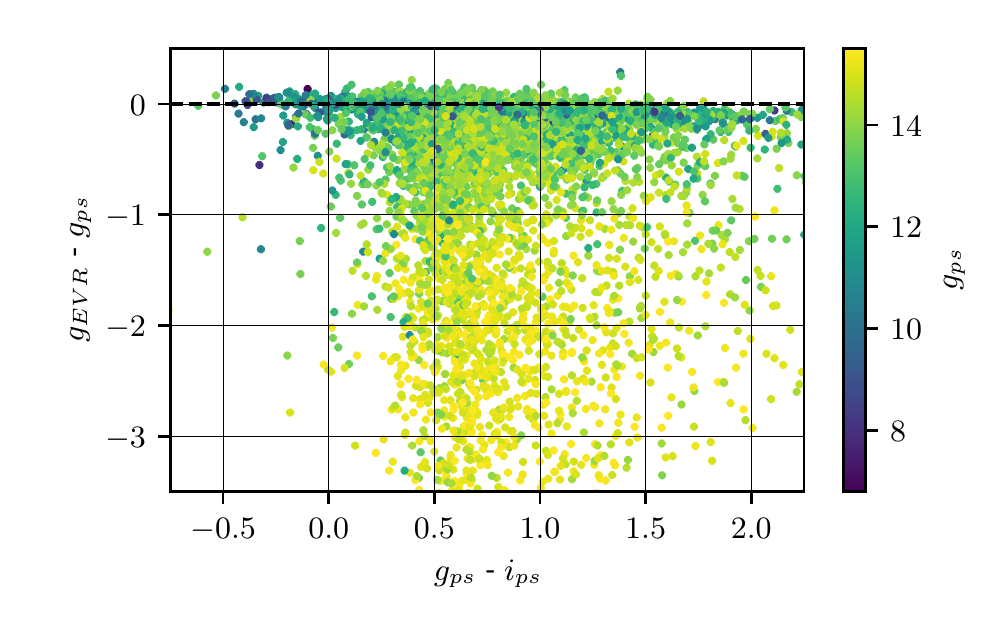}{0.5\textwidth}{(c)}
			  \fig{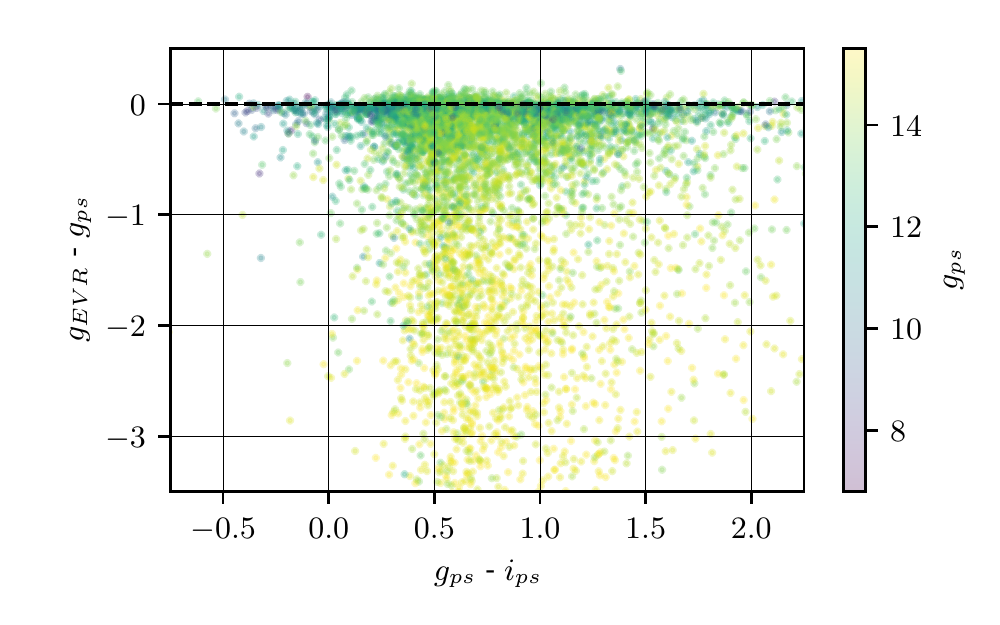}{0.5\textwidth}{(d)}}
	\gridline{\fig{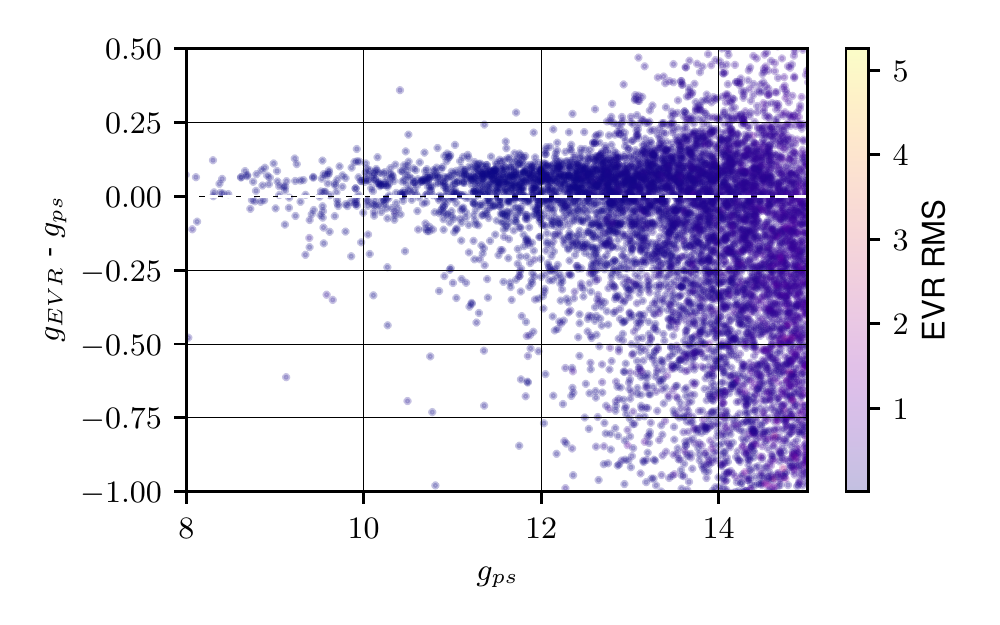}{0.5\textwidth}{(e)}
			  \fig{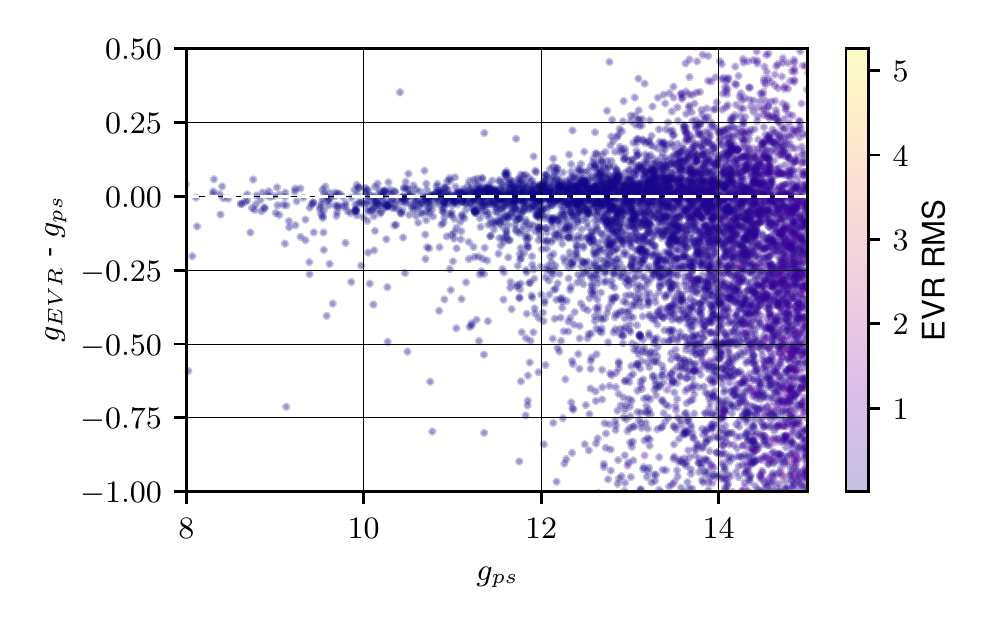}{0.5\textwidth}{(f)}}
	\caption{\textbf{(a)}, \textbf{(b)} Photometric offsets between Evryscope
	and PanSTARRS g-band as a function of $g-r$ before and after calibration
	with the color and airmass terms above, respectively. Large offsets below
	the linear trend are caused by blended sources and low-SNR
	detections that were not filtered based on the light curve quality metric
	described in Section~\ref{sec:photometry_tests}. \textbf{(c)}, \textbf{(d)}
	Same as above, but with $g-i$ colors in place of $g-r$ to demonstrate
	perfomance over a wider variety of colors. No calibration fits are made as a
	function of $i$-band colors. \textbf{(e)}, \textbf{(f)} Photometric
	calibration performance for many-epoch light curves as a function of
	magnitude. The sigma-clipped RMS photometric offset decreases from 0.16 mags
	to 0.06 mags for sources between $8 < m_g < 14.5$ with application of color
	terms. \label{fig:photometric_terms}}
\end{figure*}

Figure~\ref{fig:photometric_terms}
shows the photometric offset as a function of $g-r$ and $g-i$ colors before and after
applying the calibration offset, as well as the resulting impact on the
long-term photometric accuracy of light curves. Application of the
color and airmass correction brings the RMS calibration accuracy of long-term
light curves  from 0.16 mags to 0.06 mags, in line with the single-epoch
measurements above. 

\begin{figure}
	\includegraphics[width=\columnwidth]{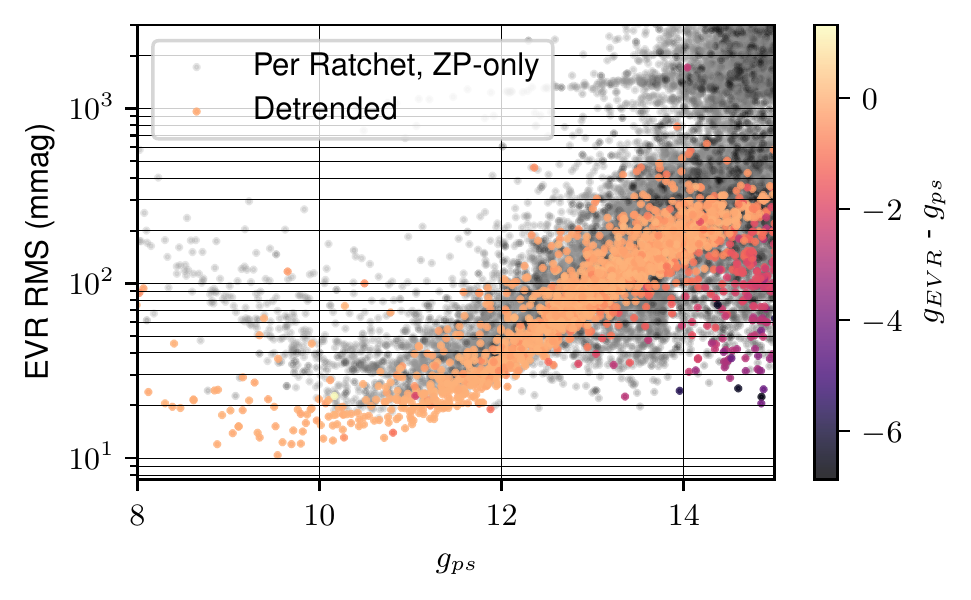}
	\caption{\label{fig:phot_precision}Measured RMS of 10671 randomly selected,
	long-term Evryscope light curves. Performance in the detrended lightcurves ranges from 20 mmag at the
	bright end to 20\% for dim sources. RMS for raw light curves is an average
	of the measured RMS in each pointing, neglecting zerpoint offsets between ratchets.}
\end{figure}

Public EFTE data products, including both transient alerts and long-term
photometric light curves , do not include color and airmass terms. For
light curves , calibration for photometric precision, rather than accuracy, is
prioritized. Evryscope light curves  are first decorrelated from subpixel PSF
variations, and then detrended using a customized version of
the SysREM algorithm \citep{tamuz_2005} to correct for systematics, ultimately
producing light curves  that are self-consistent at the $\leq$20 mmag level at the
bright end of Evryscope's operating range, as shown in
Figure~\ref{fig:phot_precision}.

\subsection{Astrometric Localization}\label{sec:astrometry_tests}

The custom astrometry routines developed for the Evryscopes are capable of
providing 1-2\arcsec (0.08-0.15 pixel) RMS astrometry over the field of view of
the Evryscopes in single images; however, astrometric localizations from
\textsc{EFTE} also depend on the quality of the alignment procedure used for
real-time reduction, and must therefore be characterized separately. In addition
to the sub-pixel scatter induced by photon noise, \textsc{EFTE} localizations
depend on the consistency of the pointing between consecutive images and the
accuracy of the rigid transformation calculated to align each image with the
previous image for which a WCS solution is available.

We evaluated the quality of this alignment routine by performing source
detection in individual science images aligned to a previous target image in the
same pointing. We chose target images with a typical $\Delta t_{D}$ of $10$
minutes, with samples of $\Delta t_{D} < 10$ minutes representative of what
would occur in the first few images in a ratchet. 

\begin{figure}
	\includegraphics[width=\columnwidth]{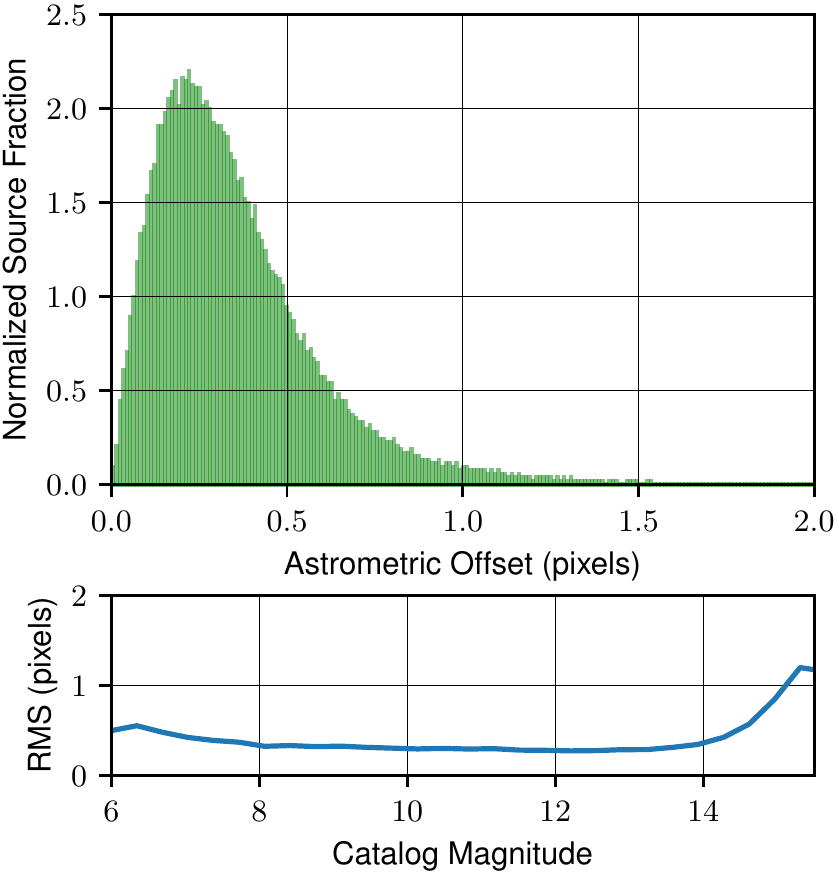}
	\caption{\label{fig:efte_astrometry}(\textbf{top}) Astrometric localization
	performance for the \textsc{EFTE} pipeline, and the RMS localization scatter
	as a function of magnitude (\textbf{bottom}). Performance is sub-pixel at
	the 99th percentile, with a typical RMS scatter of 7\arcsec, excepting stars
	brighter than 7th magnitude, which are typically saturated, and dimmer than
	14th magnitude, where source confusion dominates in the source extraction.}
\end{figure}

The detected sources were then cross-matched with sources in the ATLAS-REFCAT2
\citep{atlas_catalog}. As in Section~\ref{sec:photometry_tests}, 500 science
images for testing were randomly selected from the 2018 observing data set,
across all weather and Moon conditions. Figure~\ref{fig:efte_astrometry} shows a
histogram of the offsets between the catalog positions and the recovered
positions in the aligned science image with a re-used WCS solution.  Astrometric
performance was sub-pixel for 99\% of detected sources, with an RMS scatter less
than 4 arcseconds between 8th and 14th magnitude. As for the photometry,
precision is limited by saturation effects at the bright end, and by source
confusion for sources dimmer than 14.5. In all cases, the localization is
accurate to within 2 pixels.

\subsection{\textsc{vetnet} Model Evaluation}\label{sec:convnet_eval} 

We evaluated the \textsc{vetnet} real-bogus model using both the held-back test set
described in Section~\ref{sec:vetnet_dataset}, and an injection-recovery program
over a sample of randomly selected images. 

\subsubsection{On-Sky Test Set}

\begin{figure*}
	\gridline{\fig{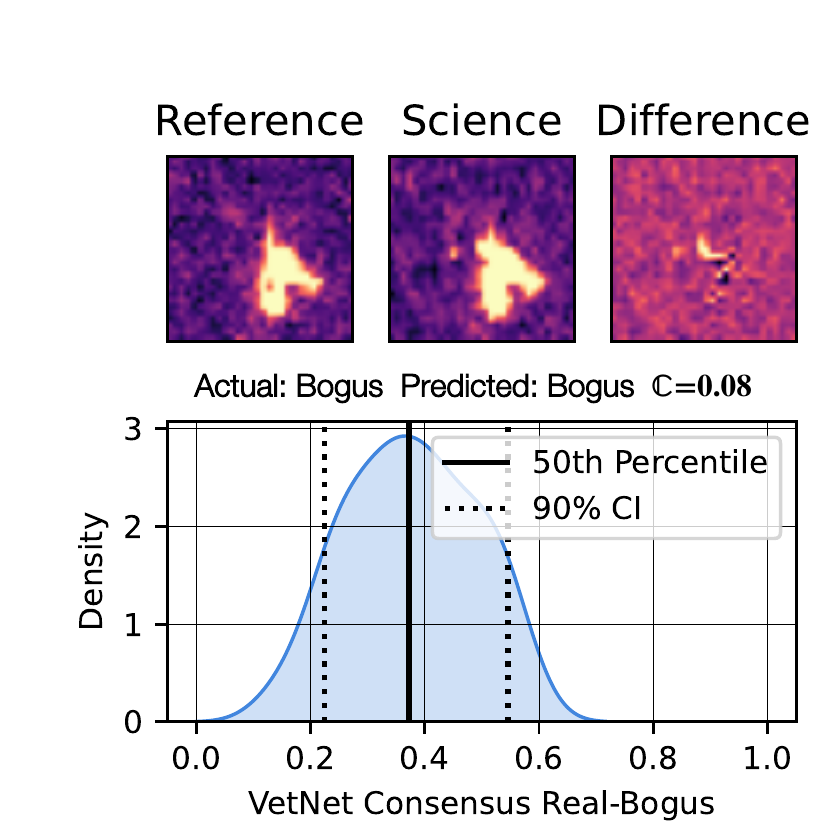}{0.3\textwidth}{(a)}
			  \fig{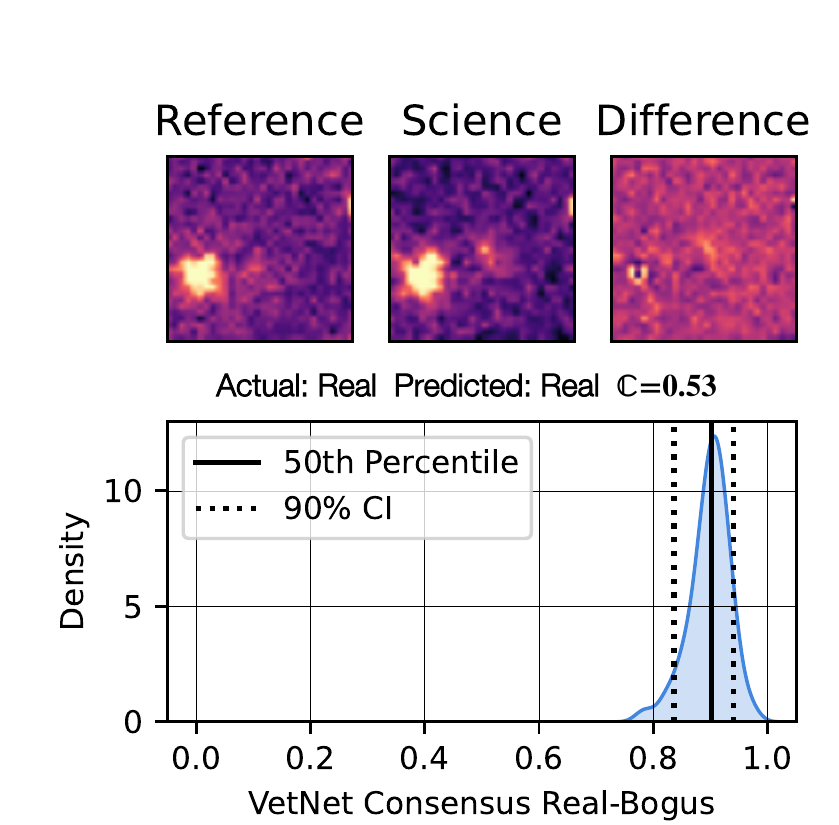}{0.3\textwidth}{(b)}
			  \fig{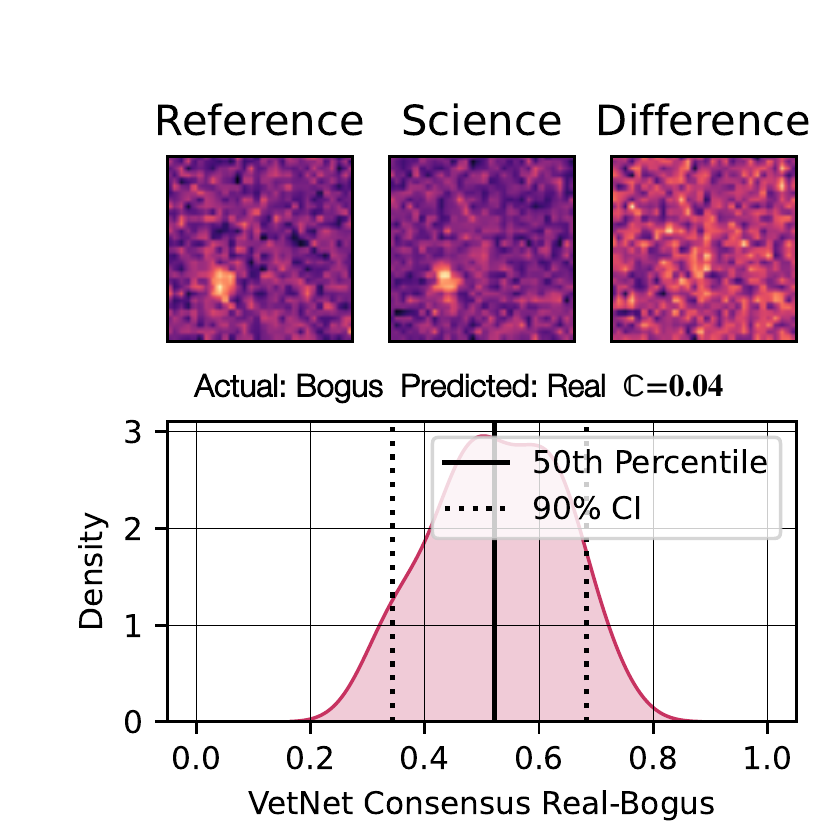}{0.3\textwidth}{(c)}}
	\vspace{-0.8cm}
	\gridline{\fig{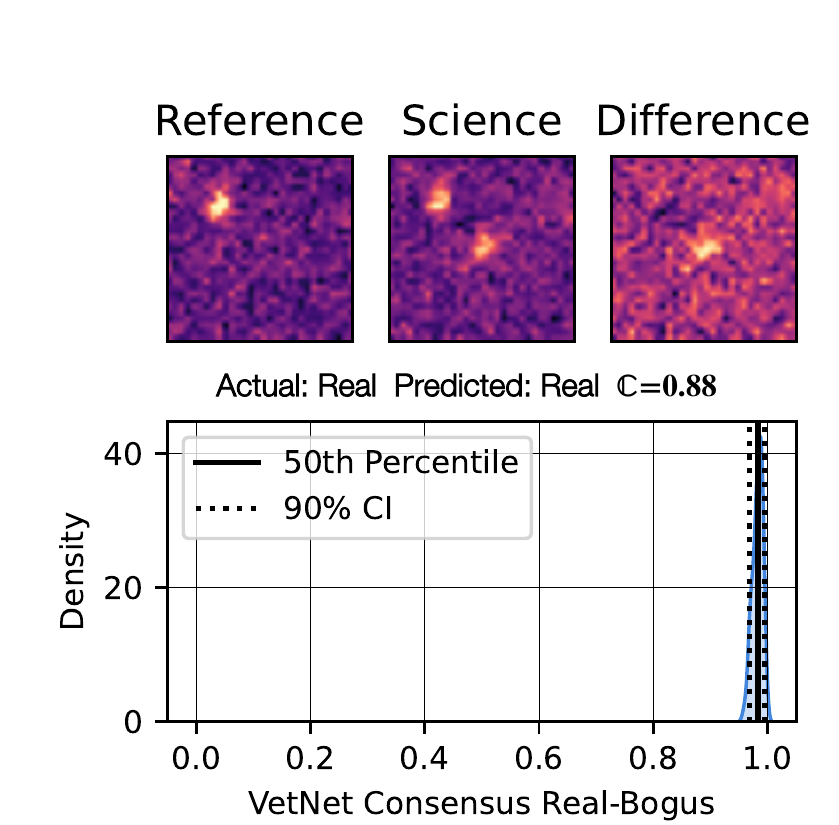}{0.3\textwidth}{(d)}
		      \fig{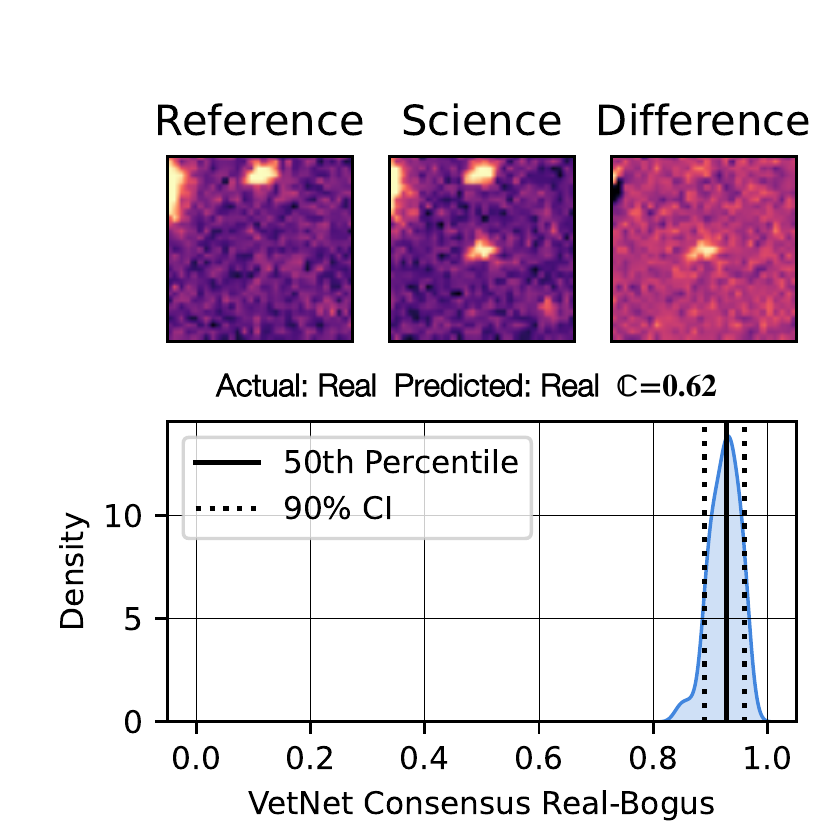}{0.3\textwidth}{(e)}
			  \fig{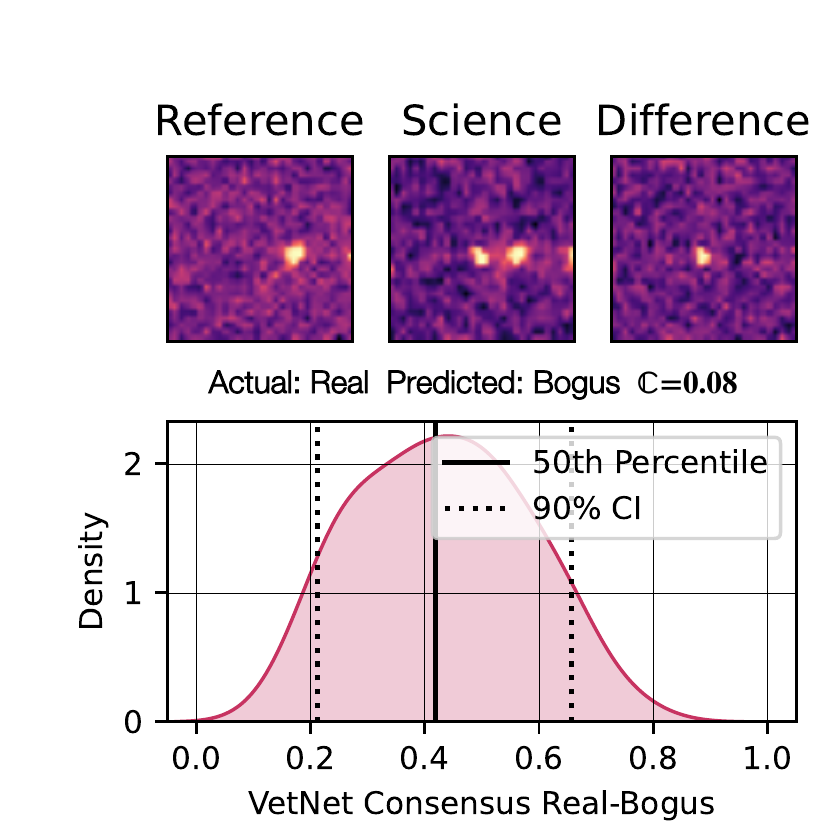}{0.3\textwidth}{(f)}}
	\vspace{-0.8cm}

	\gridline{\fig{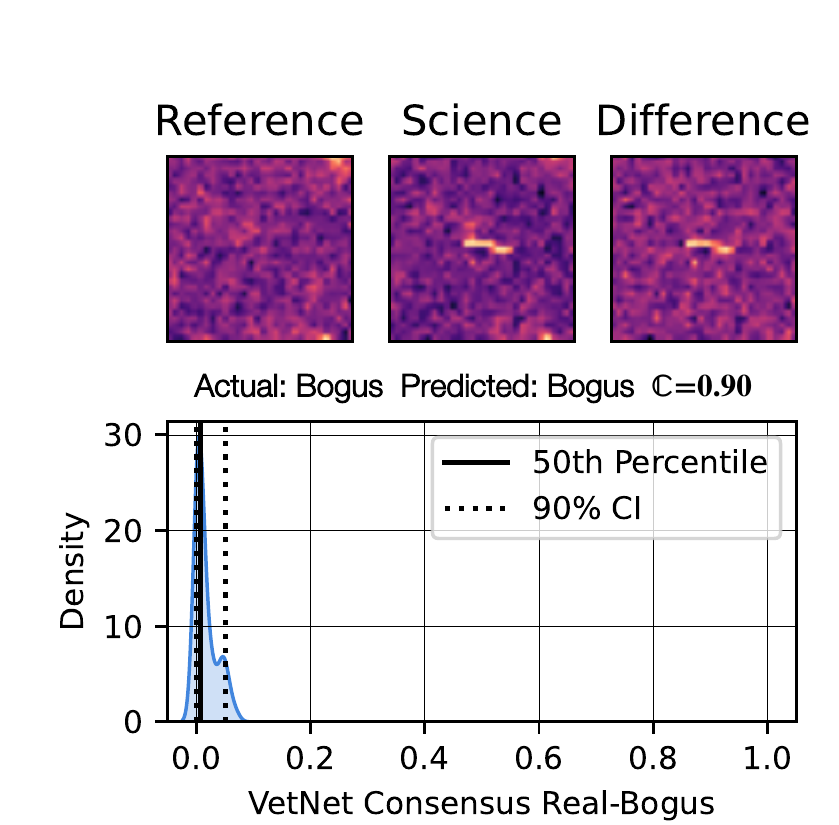}{0.3\textwidth}{(g)}
		      \fig{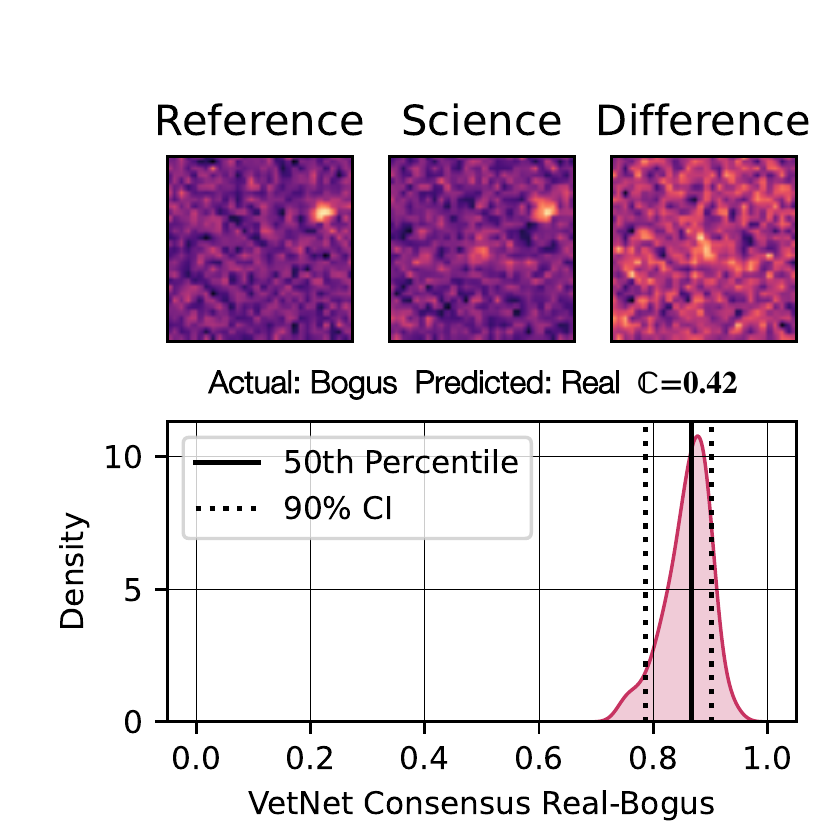}{0.3\textwidth}{(h)}
			  \fig{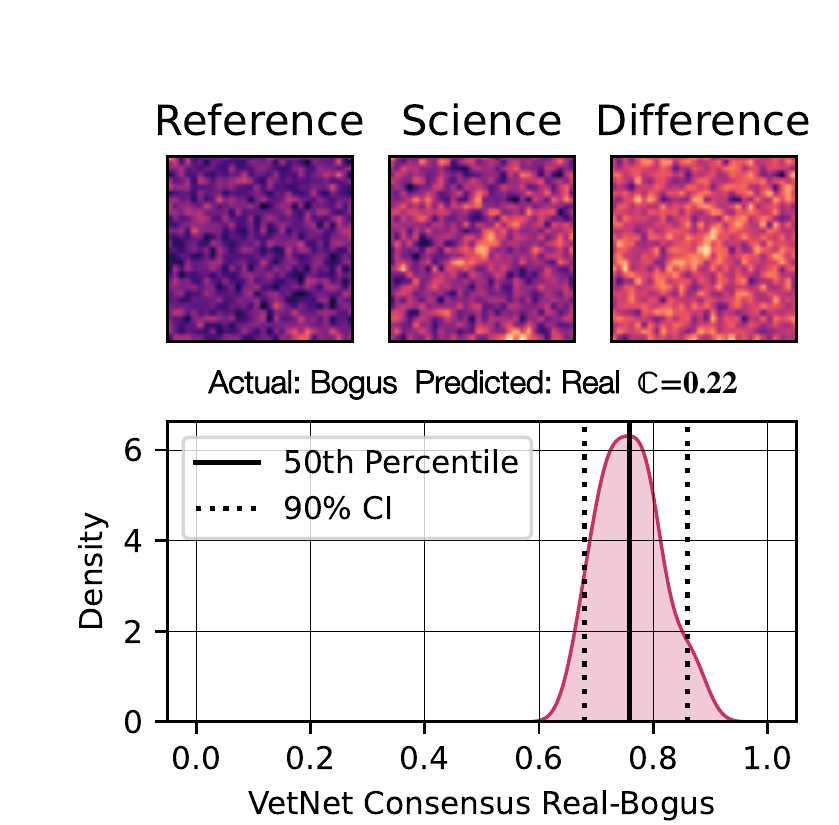}{0.3\textwidth}{(i)}}
	\vspace{-0.8cm}

	\gridline{\fig{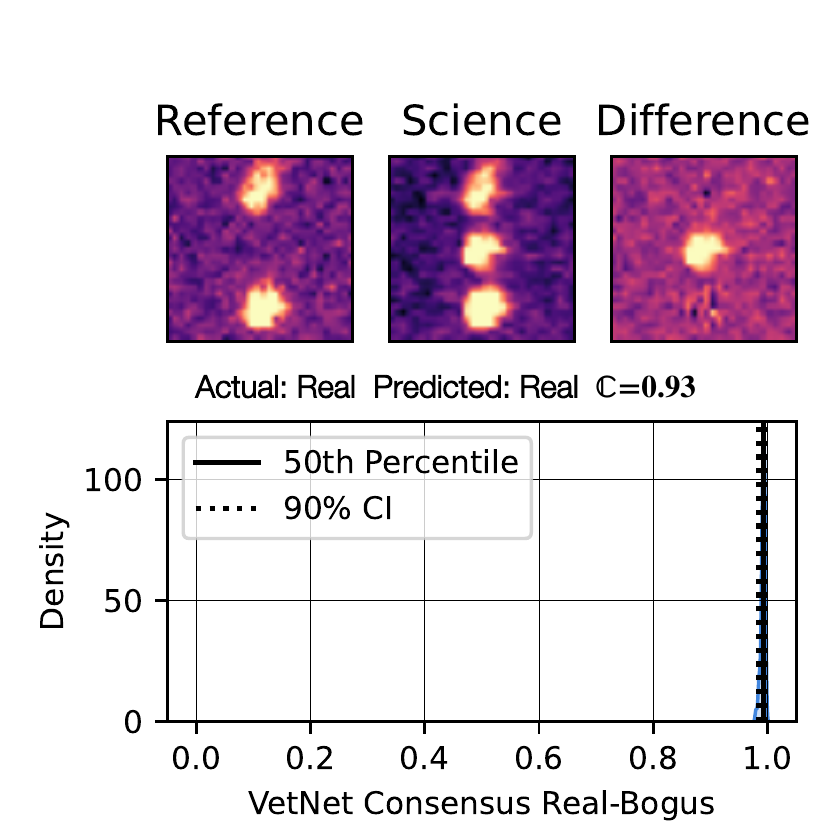}{0.3\textwidth}{(j)}
              \fig{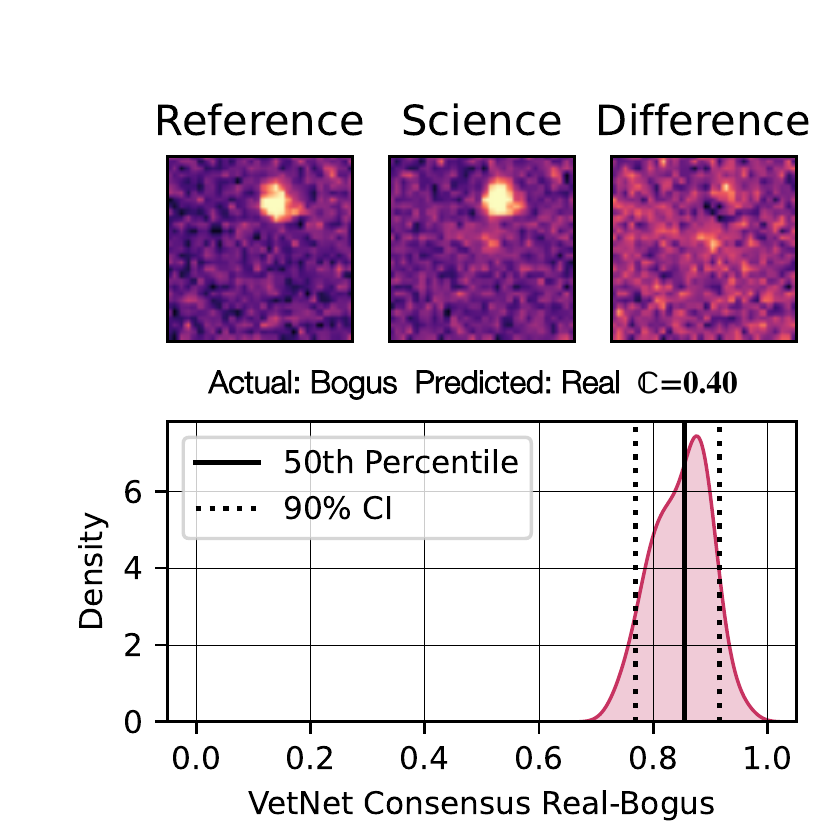}{0.3\textwidth}{(k)}
			  \fig{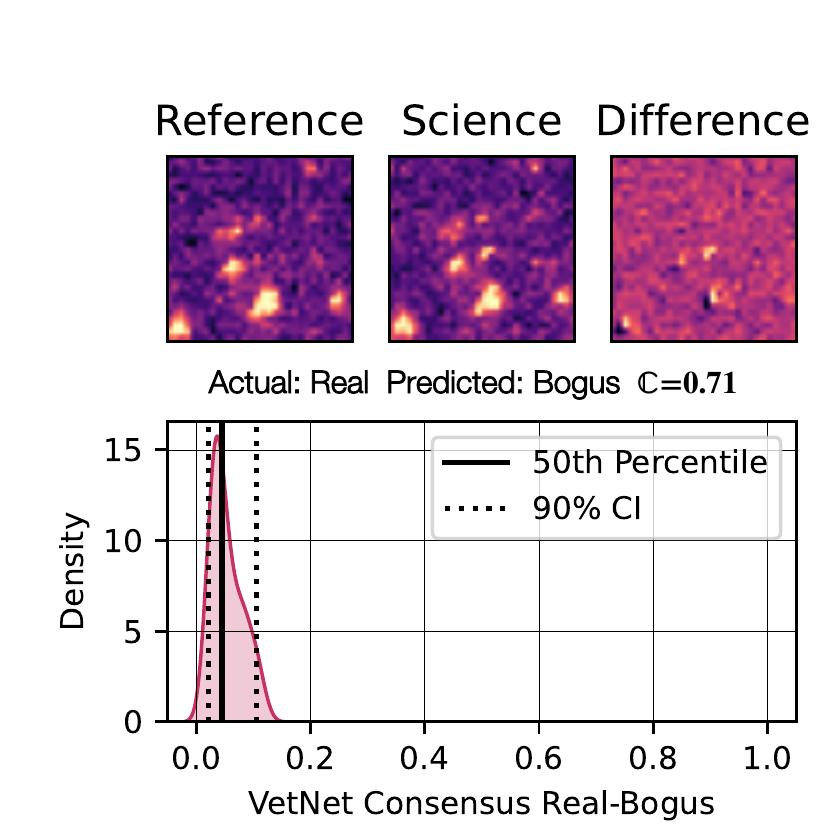}{0.3\textwidth}{(l)}}
	\vspace{-0.2cm}

	\caption{Sample on-sky candidates taken from the \textsc{vetnet} test set.
	``Actual'' and ``Pred'' values represent sky truth class determined by a
	human inspector and the predicted class by the network, respectively.
	Histograms are an approximation of the normalized Bayesian posterior distribution for
	the probability of the candidate representing a real astrophysical event,
	quantified using the entropy-based $\mathbb{C}$ metric from
	\cite{Killestein_2021}. In cases where the sky truth and the network
	prediction disagree, $\mathbb{C}$ is typically $<0.5$, or extenuating
	circumstances exist, such as the anomolous PSFs in panels \textbf{(l)} and
	\textbf{(f)}, or the potential human misclassifications in \textbf{(h)},
	\textbf{(i)}, and \textbf{(k)}. \label{fig:vetnet_samples}}
\end{figure*}

Figure~\ref{fig:vetnet_samples} shows postage stamp cutouts and classification
histograms from the held-back test set of on-sky transients, divided evenly
between cases where \textsc{vetnet} classifcations and the human-assigned labels agreed
and cases where they disagreed. In both categories, the entropy-based confidence
score scales with subjective appraisal of the candidates; candidates (h), (k),
(i), and (c) are faint borderline detections, assigned accordingly low confidence
scores. Candidate (g) is a linear particle collision. Notably, candidates (f)
and (l) have anomolously sharp PSFs that were both counted as real by human
labellers, but were assigned bogus scores by the network, suggesting that they
are morphologically more similar to cosmics. Even in cases where the labels are
consistent, the confidence drops in areas with pathological PSFs, as in example
(b), but particularly in example (a). 

\begin{figure}
	\includegraphics[width=\columnwidth]{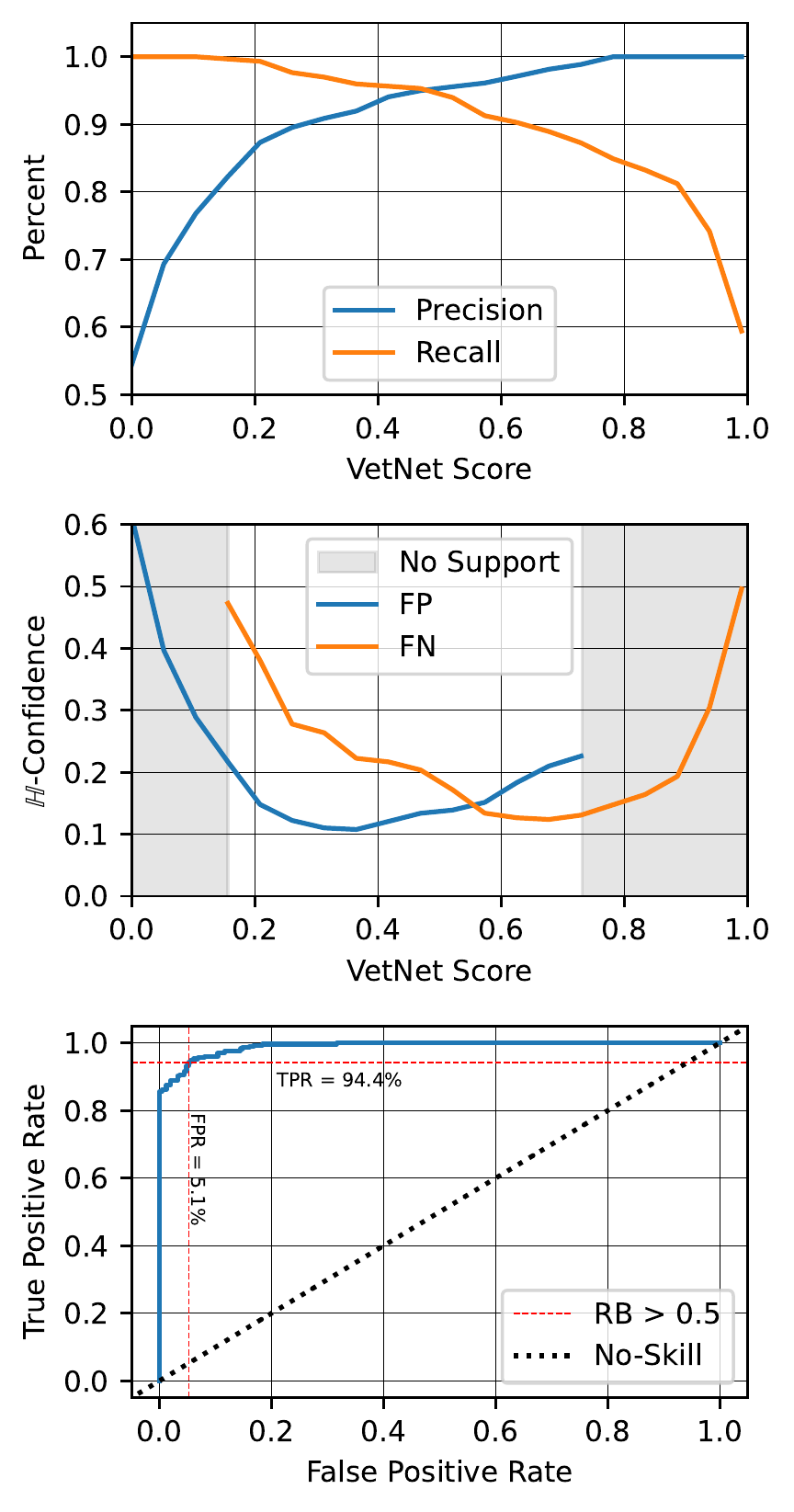}
	\caption{Performance of \textsc{vetnet} on the on-sky test set.
	(\textbf{top}) Precision and recall as a function of the \textsc{vetnet}
	real-bogus score. At the $RB = 0.5$ threshold, the observed precision and
	recall are 95.4\% and 94.4\% respectively. (\textbf{middle}) Entropy-based
	confidence scores for false-positives and false-negatives as a function of
	\textsc{vetnet} real-bogus score threshold. Shaded regions indicate score
	regions where no false positives or false negatives occur within the on-sky
	test set. (\textbf{bottom}) ROC curve for \textsc{vetnet}. The area under
	the ROC curve (the ROC-AUC metric) is 0.99, representing the probability
	that a random real candidate will receive a higher real-bogus score than a
	random bogus candidate. No-skill line indicates the expected performance
	curve for a random classifier. \label{fig:vetnet_prec_recall}}
\end{figure}

Figure~\ref{fig:vetnet_prec_recall} shows the performance of the model on the
on-sky test set. The magnitude-integrated precision and recall at a \textsc{vetnet} score
threshold of 0.5 are 95.4\% and 94.4\% respectively, with a false positive rate
of 5.1\%. Depending on the science case and false-alarm tolerance of follow-up resources,
these numbers can be tuned using a combination of the \textsc{vetnet} RB score and the
$\mathbb{C}$ rating; for instance, a sub-percent FPR is measured above a RB
threshold of 0.7. 

\subsection{Candidate Production Latency}\label{sec:latency_tests}

To enable rapid followup, \textsc{EFTE} must produce candidates on timescales
comparable to the earliest and most impulsive phases of the astrophysical events
of interest, ideally within the base cadence of the survey. For Evryscope, this
means adding candidates to an actionable event stream within two minutes of the
end of each exposure. We consider the candidate production latency as our figure
of merit for speed, defined here as the time delay between the shutter close
time for the image and candidates being fully inserted into the central
\textsc{EFTE} database in Chapel Hill, with all automated vetting and
in-database source association and deduplication actions complete.

\begin{figure*}
	\gridline{\fig{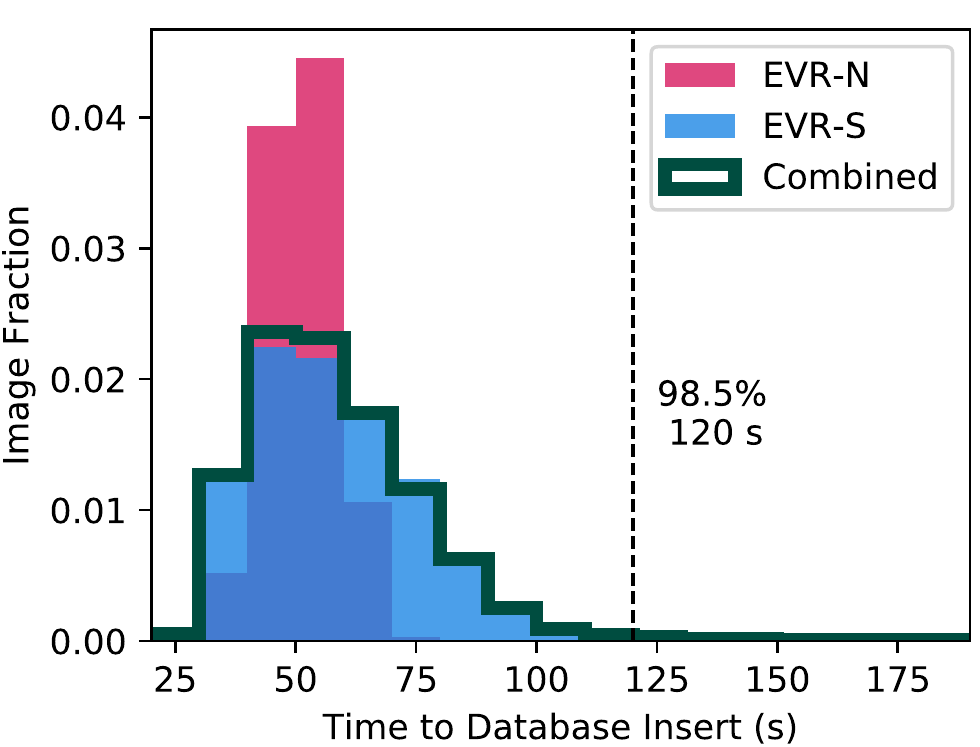}{0.5\textwidth}{(a)}
			  \fig{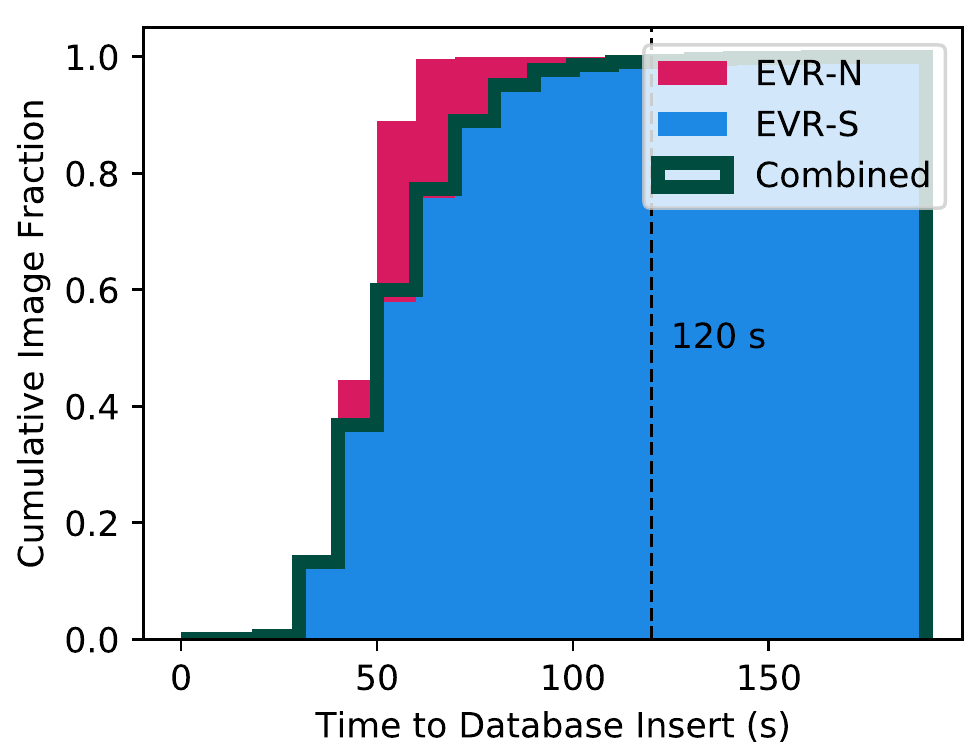}{0.5\textwidth}{(b)}}
	\caption{\label{fig:efte_latency}Cumulative histogram (\textbf{a}) and
	distribution (\textbf{b}) of time delay between exposure and insertion of
	vetted candidates into the remote database between 25 November 2019 and 1
	January 2020. 98.5\% of images are fully reduced into lists of transient
	candidates within 120 seconds, before the next image is complete.}
\end{figure*}

Figure~\ref{fig:efte_latency} presents histograms of candidate production
latency for both Evryscope-North and Evryscope-South during early on-sky testing
of \textsc{EFTE} between 25 November 2019 and 1 January 2020. Some variation is
seen between Evryscope-North and Evryscope-South, which we attribute to a
combination of the difference in on-site compute hardware specifications, camera
counts, and varying network connectivity to each observatory.
Cumulatively between both sites, \textsc{EFTE} is able to meet the sub-cadence
latency requirement for 98.5\% of images.

\subsection{Injection-Recovery Testing for Completeness}
To estimate the expected completeness of the survey, we selected 800 images from
the 2021 Evryscope-North data set at random and injected simulated sources using the
routine described in Section~\ref{sec:transient_injection}, and evaluated the
recovery probability as a function of magnitude using the routine described in
\cite{corbett_flashes}. The ratio of variables (injected with a minimum
contrast of 0.25 mag) to transients without a counterpart in the reference image
was 1:6. In total, 960,000 transients were added to the images. 

\begin{figure}
	\includegraphics[width=\columnwidth]{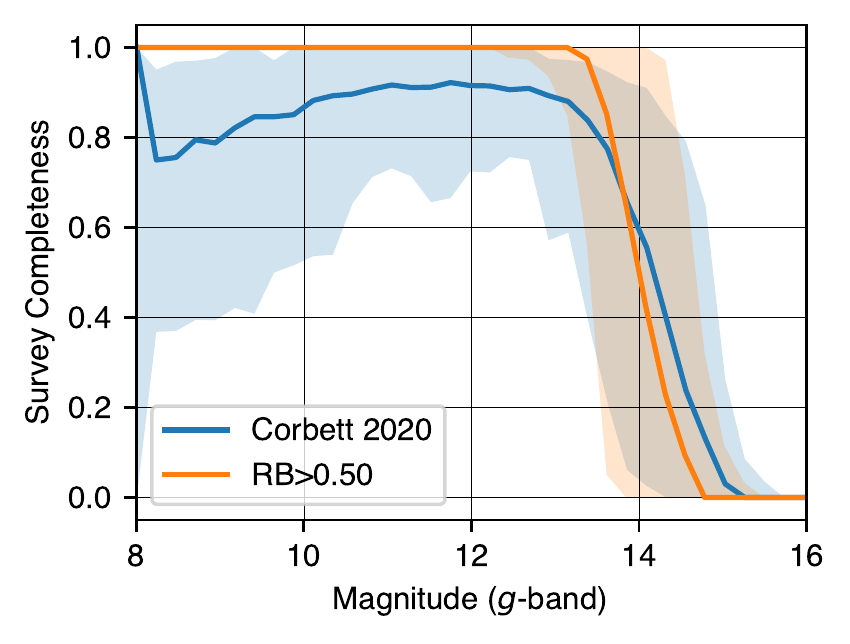}
	\caption{\label{fig:survey_complete}Survey completeness as a function of
	magnitude for both the current EFTE with a threshold \textsc{vetnet}
	real-bogus score of 0.5, and the previous version described in
	\cite{corbett_flashes}. Completeness is measured on a synthetic sample of
	injected transient and variable sources. Recovery probability for sources
	brighter than $m_g=13.2$ is 99.9\%, and rapidly falls to 50\% at $m_g=14$.
	Shaded regions represent the 90\% confidence intervals of each curve, based
	on the percentiles of the per-image recovery functions.}
\end{figure}

Figure~\ref{fig:survey_complete} shows the fraction of simulated transients
recovered from the test set, and the corresponding recovery fraction from
\cite{corbett_flashes}, which used an earlier version of the \textsc{vetnet}
model. We note that dropping the \textsc{vetnet}-RB score threshold to 0.0 has a
marginal effect on the dim end recovery curve, indicating that the decreased
depth (50\% at $m_g=14$ instead of 50\% at $m_g=14.2$) is a property of the
slightly different image sample rather than of the updated \textsc{vetnet}
model. 
Sources brighter than $m_g=13.2$ are successfully recovered in all images. 

\section{Science Results from \textsc{EFTE}}\label{sec:science}

\subsection{Rapid Follow-Up of Stellar Flares with SOAR}\label{sec:soar_followup}

\textsc{EFTE}'s latency is fast enough for flare candidates to be observed by
other telescopes in the minutes immediately following the flare's detection. The
Southern Astrophysical Research (SOAR) telescope is a 4.1-m telescope located at
Cerro Pachon in Chile, which hosts the Goodman High-Throughput Spectrograph
\citep{clemens_2004}. The entirety of Evryscope-South's field of view is within
SOAR's observable area, and a band of Evryscope-North's field of view below
declinations of 10 degrees is accessible to SOAR. Ongoing studies pair SOAR and
Goodman with the \textsc{EFTE} alert stream to acquire spectra of stellar flares
within minutes of the flare's detection, allowing the spectral evolution of
flares to be characterized during the most impulsive phases of the flare with
time resolution limited by the exposure time necessary to obtain a spectrum with
adequate signal to noise ratio (typically sub-minute for \textsc{EFTE}-detected
stellar flares). 

\subsubsection{EVRT-3586872, a $\Delta{m}=4.2$ Flare from a mid-M Dwarf}

In an exposure beginning at 5:39:56 UTC on 15 February 2020, \textsc{EFTE}
detected a new source from Evryscope South, which was then confirmed in the two
consecutive images, at magnitudes 12.7 and 12.8 respectively, indicating that
this source was both astrophysical in nature and potentially detected near its
peak. The source, assigned the identifier EVRT-3586872, crossmatched to a star
in the ATLAS reference catalog with a red color ($g-r=1.186$), suggesting a
possible M-dwarf origin. The offset between the catalog star,
\textsc{2MASS08593584-2340201}, and the \textsc{EFTE} detection is 4.0
arcseconds, well within the expected astrometric error. The star is also
cataloged in \cite{atlas_variables} as an irregular sinusoidal variable,
consistent with an M-dwarf rotational signature. 

\begin{figure*}
	\includegraphics[width=\textwidth]{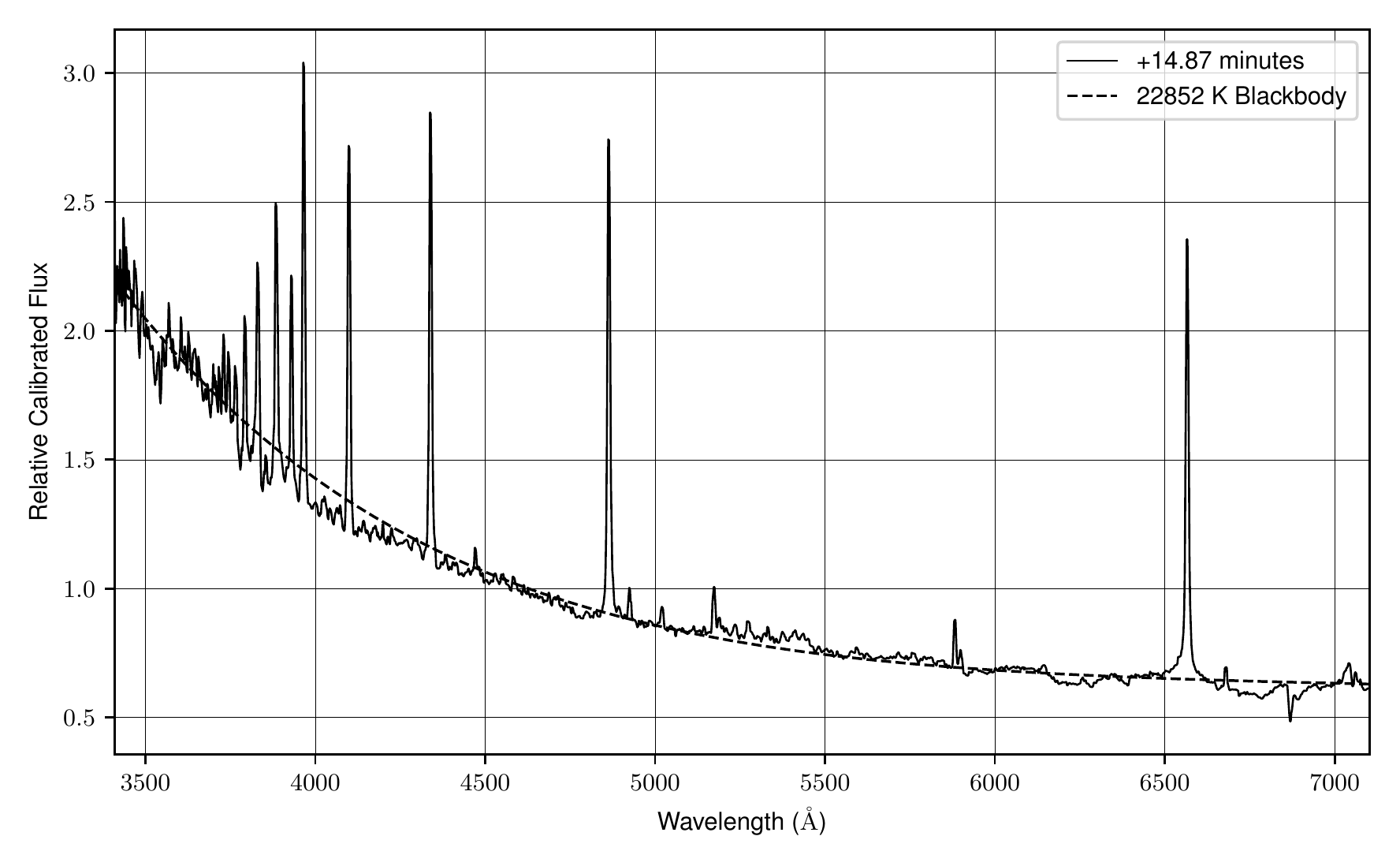}
	\caption{Spectrum of EVRT-3586872 at +14.87 minutes from the end of the
	first Evryscope detection of a flare candidate. The spectrum has been
	flux calibrated and normalized. The overlaid curve is a two-component blackbody
	spectrum consisting of a fitted 22852 K flare continuum and a 3256 K quiescent
	thermal spectrum. \label{fig:soar_spec}} 
\end{figure*}

Upon receiving a notification of consecutive detections of a cataloged red
source via our web interface, we worked with SOAR staff to slew to its location,
and began observing the target 14.9 minutes after the end of the first Evryscope
detection image. We used the 400 line grating in the M1 configuration, covering
approximately the wavelength range from 300 to 705 nm.
Figure~\ref{fig:soar_spec} shows a spectrum of the flare extracted from
a 120-second exposure +15 minutes after the initial trigger, flux calibrated
to the spectrum of LTT2445 \citep{hamuy_92, hamuy_94}. 

We fit a two-component scaled blackbody model to the flare spectrum.
This includes both a fixed contribution from the quiescent star, based on the
Bayesian estimate of $T_{eff}$ from \texttt{StarHorse2} \citep{starhorse2}, and
thermal flare emission with a best-fit temperature of 22,852 K. At almost 23,000
K, this temperature is larger than typically assumed in flare models
\citep{osten_2015}, but consistent with temperatures recently inferred from
broad-band light curves from TESS and Evryscope \citep{howard_2020}. We also
note this measurement is subject to known systematics, including Balmer
continuum emission features \citep{kowalski_2013} and increasing uncertainty as
true temperature increases, due to the optical bandpass primarily sampling the
Rayleigh–Jeans tail of the spectrum at temperatures beyond $10^4$ K
\citep{arcavi_2022}. Analysis of results from the \textsc{EFTE}-SOAR followup
program is ongoing.

\subsection{\textsc{EFTE} Light Curves}\label{sec:lightcurves}

In addition to transient alerts, \textsc{EFTE} enables users to produce
long-term light curves for targets not included in the input catalog used for the
Evryscope high-precision forced photometry pipeline
\citep{evryscope_instrument}. \textsc{EFTE} light curves have been included in
multiple publications, both by the Evryscope team and external collaborators.
Publications using \textsc{EFTE} light curves include analyses of the galactic
novae V1674 Her \citep{Quimby_2021} and V906 Car \citep{Wee_2020}, measurements
of the rotation periods of TESS exoplanet hosts, including one example with a
2.1 mmag amplitude measured from \textsc{EFTE} photometry \citep{Howard_2021},
and long-term monitoring of a mysterious dust-emitting object orbiting the star TIC
400799224 \citep{Powell_2021}.

\subsection{Satellite Glint Foregrounds for Fast Transient
Surveys}\label{sec:fast_flashes}

Image contamination by Earth artificial satellites takes two forms: streaks,
with uniform illumination over extended trajectories, and glints, which appear
as short-duration flashes. These two morphologies are frequently degenerate, and
depend on the structure and orbit of the reflector. Short durations relative to
their motion on the sky and sharp contrast with their associated streaks have
led to glints being mistaken for astrophysical events
\citep{schaefer_1987,Maley_1987,Maley_1991, Rast_1991, shamir_2006}. During the
first six months of \textsc{EFTE} operations, we identified 1,415,722 likely
satellite glints, and modeled an all-sky event rate of $1800^{+600}_{-280}$
sky$^{-1}$ hour$^{-1}$, peaking at $m_g = 13.0$ \citep{corbett_flashes}. This
rate is orders of magnitude higher than the combined rate of public alerts from
all active all-sky fast-timescale transient searches, including neutrino,
gravitational-wave, gamma-ray, and radio observatories. A subsequent study,
using the Weizmann Fast Astronomical Survey Telescope
\citep[W-FAST;][]{nir_2020} revealed that this event rate increases sharply with
depth, reporting an event rate of $9100^{+3000}_{-2000}$ sky$^{-1}$ hour$^{-1}$
for $9 < M_{BP} < 11$ around the equator, 2.3$\times$ the value we measured for
${m_g}<9$. As the majority of the events observed by Evryscope appear to be
generated in low Earth orbit (LEO), we  expect the event rate for satellite
glints to correlate with the rapidly-growning number of LEO
satellites.\footnote{https://www.ucsusa.org/resources/satellite-database}

\section{Summary and Conclusions}\label{sec:summary_conclusions}

In this paper, we presented the Evryscope Fast Transient Engine (\textsc{EFTE}),
the real-time transient discovery pipeline for the Evryscopes. The pipeline is a
fully custom data analysis tool, suited to the unique parameter space inhabited
by the Evryscopes and capable of identifying transient candidates in real-time,
with alerts available for each image within the two-minute cadence of the
Evryscopes for 98.5\% of images. To accomplish this, we reduce the complex image
subtraction process adopted by seeing-limited surveys to a simple direct
subtraction of near-consecutive images. Astrometric performance for transient
alerts is sub-pixel at the 99th percentile, and photometric performance is
within 0.06 mag RMS of the ATLAS-REFCAT2 catalog for reference stars within the
$8 < m_g < 14.5$ sensitivity range of the survey. Using a convolutional
real-bogus classifier, we are able to recover 99.9\% of sources brighter than
$m_g = 13.2$ with a false positive rate of 5.1\%.
 
 While \textsc{EFTE} is specifically adapted to Evryscope data, the
infrastructure, algorithms, and ML models, were developed to enable portability
to instruments with similar survey strategies, such as the NASA Transiting
Exoplanet Survey Satellite \citep[TESS;][]{tess_instrument}, or with stringent
data throughput and reduction latency requirements. Core algorithms from EFTE
have been adapted for usage in the pipeline of the Argus Optical Array, a
5-meter class multiplexed 55 GPix synoptic survey instrument currently in
development \citep{law_2022, law_2022b}. Argus will observe a field of view
equivalent to Evryscope in alternating 1- and 30-second cadences, which produce
up to 4.3 PiB and 145 TiB of raw data per night, respectively. To support this
data rate, Argus data will be reduced in real time, producing low-latency
transient alerts, photometry, and calibrated image data for distribution and
storage. In \cite{corbett_hdps}, we describe the Argus pipeline and data
products, and demonstrate the direct subtraction algorithm described in
Section~\ref{sec:direct_subtraction} on the Argus Array Technology Demonstrator
\citep{corbett_at2d}. The \textsc{VETNET} model described in
Section~\ref{sec:autovet} was optimized for direct image subtraction and
Evryscope data; however, the framework for performing this optimization and for
staged model training using both on-sky and simulated data is similarly portable
to Argus.

A public alert stream from \textsc{EFTE} is in development, based on the
evolving community standard, in use for the Zwicky Alert Distribution System
\citep{zads_paper} and planned for the Rubin Observatory's Legacy Survey of
Space and Time (LSST)\footnote{https://dmtn-093.lsst.io}, of serialized alert
packets distributed via Apache Kafka \citep{Kreps2011KafkaA}. Alerts will be
available via the Arizona-NOIRLab Temporal Analysis and Response to Events
System \citep[ANTARES;][]{antares_paper}. Details of the alert distribution
system and alert schema contents will be addressed in future work.

\section*{Acknowledgements}

The Evryscope was constructed under National Science Foundation ATI grant
AST-1407589, with operating costs from National Science Foundation CAREER grant
1555175. Current operations are supported by AAG-2009645. HC was supported by
the National Science Foundation Graduate Research Fellowship (Grant No.
DGE-1144081), AAG-2009645, AST-2007853, MSIP-2034381, and the North Carolina
Space Grant. OF acknowledges the support by the Spanish Ministerio de Ciencia e Innovaci\'{o}n
(MICINN) under grant PID2019-105510GB-C31 and through the ``Center of Excellence
Mar\'{i}a de Maeztu 2020-2023'' award to the ICCUB (CEX2019-000918-M). The
authors thank Sergio Pizarro, Rodrigo Hernández, and Juan Espinoza for
assistance with SOAR observations of EVRT-3586872, and E. Goeke and A. Jordan
for discussion on transient vetting. This research made use of
Astropy,\footnote{\href{http://www.astropy.org}{http://www.astropy.org}} a
community-developed core Python package for Astronomy \citep{astropy_2013,
astropy_2018}, and
SciPy,\footnote{\href{http://www.scipy.org}{http://www.scipy.org}} a core Python
package for general scientific computing tasks.

\bibliography{evr_fast_transient_engine}{}

\begin{thebibliography}{}
\expandafter\ifx\csname natexlab\endcsname\relax\def\natexlab#1{#1}\fi
\providecommand{\url}[1]{\href{#1}{#1}}
\providecommand{\dodoi}[1]{doi:~\href{http://doi.org/#1}{\nolinkurl{#1}}}
\providecommand{\doeprint}[1]{\href{http://ascl.net/#1}{\nolinkurl{http://ascl.net/#1}}}
\providecommand{\doarXiv}[1]{\href{https://arxiv.org/abs/#1}{\nolinkurl{https://arxiv.org/abs/#1}}}

\bibitem[{Abadi {et~al.}(2015)Abadi, Agarwal, Barham, Brevdo, Chen, Citro,
  Corrado, Davis, Dean, Devin, Ghemawat, Goodfellow, Harp, Irving, Isard, Jia,
  Jozefowicz, Kaiser, Kudlur, Levenberg, Man\'{e}, Monga, Moore, Murray, Olah,
  Schuster, Shlens, Steiner, Sutskever, Talwar, Tucker, Vanhoucke, Vasudevan,
  Vi\'{e}gas, Vinyals, Warden, Wattenberg, Wicke, Yu, \&
  Zheng}]{tensorflow2015-whitepaper}
Abadi, M., Agarwal, A., Barham, P., {et~al.} 2015, {TensorFlow}: Large-Scale
  Machine Learning on Heterogeneous Systems.
\newblock \url{https://www.tensorflow.org/}

\bibitem[{{Abbott} {et~al.}(2009){Abbott}, {Abbott}, {Adhikari}, {Ajith},
  {Allen}, {Allen}, {Amin}, {Anderson}, {Anderson}, {Arain}, {Araya}, {Armand
  ula}, {Armor}, {Aso}, {Aston}, {Aufmuth}, {Aulbert}, {Babak}, {Baker},
  {Ballmer}, {Barker}, {Barker}, {Barr}, {Barriga}, {Barsotti}, {Barton},
  {Bartos}, {Bassiri}, {Bastarrika}, {Behnke}, {Benacquista}, {Betzwieser},
  {Beyersdorf}, {Bilenko}, {Billingsley}, {Biswas}, {Black}, {Blackburn},
  {Blackburn}, {Blair}, {Bland}, {Bodiya}, {Bogue}, {Bork}, {Boschi}, {Bose},
  {Brady}, {Braginsky}, {Brau}, {Bridges}, {Brinkmann}, {Brooks}, {Brown},
  {Brummit}, {Brunet}, {Bullington}, {Buonanno}, {Burmeister}, {Byer},
  {Cadonati}, {Camp}, {Cannizzo}, {Cannon}, {Cao}, {Cardenas}, {Caride},
  {Castaldi}, {Caudill}, {Cavagli{\`a}}, {Cepeda}, {Chalermsongsak},
  {Chalkley}, {Charlton}, {Chatterji}, {Chelkowski}, {Chen}, {Christensen},
  {Chung}, {Clark}, {Clark}, {Clayton}, {Cokelaer}, {Colacino}, {Conte},
  {Cook}, {Corbitt}, {Cornish}, {Coward}, {Coyne}, {Creighton}, {Creighton},
  {Cruise}, {Culter}, {Cumming}, {Cunningham}, {Danilishin}, {Danzmann},
  {Daudert}, {Davies}, {Daw}, {DeBra}, {Degallaix}, {Dergachev}, {Desai},
  {DeSalvo}, {Dhurandhar}, {D{\'\i}az}, {Dietz}, {Donovan}, {Dooley}, {Doomes},
  {Drever}, {Dueck}, {Duke}, {Dumas}, {Dwyer}, {Echols}, {Edgar}, {Effler},
  {Ehrens}, {Espinoza}, {Etzel}, {Evans}, {Evans}, {Fairhurst}, {Faltas},
  {Fan}, {Fazi}, {Fehrmenn}, {Finn}, {Flasch}, {Foley}, {Forrest},
  {Fotopoulos}, {Franzen}, {Frede}, {Frei}, {Frei}, {Freise}, {Frey}, {Fricke},
  {Fritschel}, {Frolov}, {Fyffe}, {Galdi}, {Garofoli}, {Gholami}, {Giaime},
  {Giampanis}, {Giardina}, {Goda}, {Goetz}, {Goggin}, {Gonz{\'a}lez},
  {Gorodetsky}, {Go{\ss}ler}, {Gouaty}, {Grant}, {Gras}, {Gray}, {Gray},
  {Greenhalgh}, {Gretarsson}, {Grimaldi}, {Grosso}, {Grote}, {Grunewald},
  {Guenther}, {Gustafson}, {Gustafson}, {Hage}, {Hallam}, {Hammer}, {Hammond},
  {Hanna}, {Hanson}, {Harms}, {Harry}, {Harry}, {Harstad}, {Haughian},
  {Hayama}, {Heefner}, {Heng}, {Heptonstall}, {Hewitson}, {Hild}, {Hirose},
  {Hoak}, {Hodge}, {Holt}, {Hosken}, {Hough}, {Hoyland}, {Hughey}, {Huttner},
  {Ingram}, {Isogai}, {Ito}, {Ivanov}, {Johnson}, {Johnson}, {Jones}, {Jones},
  {Jones}, {Ju}, {Kalmus}, {Kalogera}, {Kandhasamy}, {Kanner}, {Kasprzyk},
  {Katsavounidis}, {Kawabe}, {Kawamura}, {Kawazoe}, {Kells}, {Keppel},
  {Khalaidovski}, {Khalili}, {Khan}, {Khazanov}, {King}, {Kissel}, {Klimenko},
  {Kokeyama}, {Kondrashov}, {Kopparapu}, {Korand a}, {Kozak}, {Krishnan},
  {Kumar}, {Kwee}, {Lam}, {Landry}, {Lantz}, {Lazzarini}, {Lei}, {Lei},
  {Leindecker}, {Leonor}, {Li}, {Lin}, {Lindquist}, {Littenberg}, {Lockerbie},
  {Lodhia}, {Longo}, {Lormand}, {Lu}, {Lubi{\'n}ski}, {Lucianetti}, {L{\"u}ck},
  {Machenschalk}, {MacInnis}, {Mageswaran}, {Mailand}, {Mandel}, {Mandic},
  {M{\'a}rka}, {M{\'a}rka}, {Markosyan}, {Markowitz}, {Maros}, {Martin},
  {Martin}, {Marx}, {Mason}, {Matichard}, {Matone}, {Matzner}, {Mavalvala},
  {McCarthy}, {McClelland}, {McGuire}, {McHugh}, {McIntyre}, {McKechan},
  {McKenzie}, {Mehmet}, {Melatos}, {Melissinos}, {Men{\'e}ndez}, {Mendell},
  {Mercer}, {Meshkov}, {Messenger}, {Meyer}, {Miller}, {Minelli}, {Mino},
  {Mitrofanov}, {Mitselmakher}, {Mittleman}, {Miyakawa}, {Moe}, {Mohanty},
  {Mohapatra}, {Moreno}, {Morioka}, {Mors}, {Mossavi}, {Mow Lowry}, {Mueller},
  {M{\"u}ller-Ebhardt}, {Muhammad}, {Mukherjee}, {Mukhopadhyay}, {Mullavey},
  {Munch}, {Murray}, {Myers}, {Myers}, {Nash}, {Nelson}, {Newton}, {Nishizawa},
  {Numata}, {O'Dell}, {O'Reilly}, {O'Shaughnessy}, {Ochsner}, {Ogin},
  {Ottaway}, {Ottens}, {Overmier}, {Owen}, {Pan}, {Pankow}, {Papa},
  {Parameshwaraiah}, {Patel}, {Pedraza}, {Penn}, {Perraca}, {Pierro}, {Pinto},
  {Pitkin}, {Pletsch}, {Plissi}, {Postiglione}, {Principe}, {Prix},
  {Prokhorov}, {Punken}, {Quetschke}, {Raab}, {Rabeling}, {Radkins}, {Raffai},
  {Raics}, {Rainer}, {Rakhmanov}, {Raymond}, {Reed}, {Reed}, {Rehbein}, {Reid},
  {Reitze}, {Riesen}, {Riles}, {Rivera}, {Roberts}, {Robertson}, {Robinson},
  {Robinson}, {Roddy}, {R{\"o}ver}, {Rollins}, {Romano}, {Romie}, {Rowan},
  {R{\"u}diger}, {Russell}, {Ryan}, {Sakata}, {de la Jordana}, {Sandberg},
  {Sannibale}, {Santamar{\'\i}a}, {Saraf}, {Sarin}, {Sathyaprakash}, {Sato},
  {Satterthwaite}, {Saulson}, {Savage}, {Savov}, {Scanlan}, {Schilling},
  {Schnabel}, {Schofield}, {Schulz}, {Schutz}, {Schwinberg}, {Scott}, {Scott},
  {Searle}, {Sears}, {Seifert}, {Sellers}, {Sengupta}, {Sergeev}, {Shapiro},
  {Shawhan}, {Shoemaker}, {Sibley}, {Siemens}, {Sigg}, {Sinha}, {Sintes},
  {Slagmolen}, {Slutsky}, {Smith}, {Smith}, {Smith}, {Somiya}, {Sorazu},
  {Stein}, {Stein}, {Steplewski}, {Stochino}, {Stone}, {Strain}, {Strigin},
  {Stroeer}, {Stuver}, {Summerscales}, {Sun}, {Sung}, {Sutton}, {Szokoly},
  {Talukder}, {Tang}, {Tanner}, {Tarabrin}, {Taylor}, {Taylor}, {Thacker},
  {Thorne}, {Th{\"u}ring}, {Tokmakov}, {Torres}, {Torrie}, {Traylor}, {Trias},
  {Ugolini}, {Ulmen}, {Urbanek}, {Vahlbruch}, {Vallisneri}, {van den Broeck},
  {van der Sluys}, {van Veggel}, {Vass}, {Vaulin}, {Vecchio}, {Veitch},
  {Veitch}, {Veltkamp}, {Villar}, {Vorvick}, {Vyachanin}, {Waldman}, {Wallace},
  {Ward}, {Weidner}, {Weinert}, {Weinstein}, {Weiss}, {Wen}, {Wen}, {Wette},
  {Whelan}, {Whitcomb}, {Whiting}, {Wilkinson}, {Willems}, {Williams},
  {Williams}, {Willke}, {Wilmut}, {Winkelmann}, {Winkler}, {Wipf}, {Wiseman},
  {Woan}, {Wooley}, {Worden}, {Wu}, {Yakushin}, {Yamamoto}, {Yan}, {Yoshida},
  {Zanolin}, {Zhang}, {Zhang}, {Zhao}, {Zotov}, {Zucker}, {M{\"u}hlen}, \&
  {Zweizig}}]{ligo_instrument}
{Abbott}, B.~P., {Abbott}, R., {Adhikari}, R., {et~al.} 2009, Reports on
  Progress in Physics, 72, 076901, \dodoi{10.1088/0034-4885/72/7/076901}

\bibitem[{Agarap(2018)}]{relu}
Agarap, A.~F. 2018, arXiv preprint arXiv:1803.08375

\bibitem[{{Aizawa} {et~al.}(2022){Aizawa}, {Kawana}, {Kashiyama}, {Ohsawa},
  {Kawahara}, {Naokawa}, {Tajiri}, {Arima}, {Jiang}, {Hartwig}, {Fujisawa},
  {Shigeyama}, {Arimatsu}, {Doi}, {Kasuga}, {Kobayashi}, {Kondo}, {Mori},
  {Okumura}, {Takita}, \& {Sako}}]{aizawa_2022}
{Aizawa}, M., {Kawana}, K., {Kashiyama}, K., {et~al.} 2022, \pasj, 74, 1069,
  \dodoi{10.1093/pasj/psac056}

\bibitem[{{Alard} \& {Lupton}(1998)}]{alard_luptin_1998}
{Alard}, C., \& {Lupton}, R.~H. 1998, \apj, 503, 325, \dodoi{10.1086/305984}

\bibitem[{{Aldering} {et~al.}(2002){Aldering}, {Adam}, {Antilogus}, {Astier},
  {Bacon}, {Bongard}, {Bonnaud}, {Copin}, {Hardin}, {Henault}, {Howell},
  {Lemonnier}, {Levy}, {Loken}, {Nugent}, {Pain}, {Pecontal}, {Pecontal},
  {Perlmutter}, {Quimby}, {Schahmaneche}, {Smadja}, \&
  {Wood-Vasey}}]{Aldering_2002}
{Aldering}, G., {Adam}, G., {Antilogus}, P., {et~al.} 2002, in Society of
  Photo-Optical Instrumentation Engineers (SPIE) Conference Series, Vol. 4836,
  Survey and Other Telescope Technologies and Discoveries, ed. J.~A. {Tyson} \&
  S.~{Wolff}, 61--72, \dodoi{10.1117/12.458107}

\bibitem[{{Anders} {et~al.}(2022){Anders}, {Khalatyan}, {Queiroz}, {Chiappini},
  {Ard{\`e}vol}, {Casamiquela}, {Figueras}, {Jim{\'e}nez-Arranz}, {Jordi},
  {Mongui{\'o}}, {Romero-G{\'o}mez}, {Altamirano}, {Antoja}, {Assaad},
  {Cantat-Gaudin}, {Castro-Ginard}, {Enke}, {Girardi}, {Guiglion}, {Khan},
  {Luri}, {Miglio}, {Minchev}, {Ramos}, {Santiago}, \&
  {Steinmetz}}]{starhorse2}
{Anders}, F., {Khalatyan}, A., {Queiroz}, A.~B.~A., {et~al.} 2022, \aap, 658,
  A91, \dodoi{10.1051/0004-6361/202142369}

\bibitem[{{Anderson} \& {King}(2000)}]{ePSF}
{Anderson}, J., \& {King}, I.~R. 2000, \pasp, 112, 1360, \dodoi{10.1086/316632}

\bibitem[{{Andreoni} {et~al.}(2017){Andreoni}, {Jacobs}, {Hegarty},
  {Pritchard}, {Cooke}, \& {Ryder}}]{mary_pipeline}
{Andreoni}, I., {Jacobs}, C., {Hegarty}, S., {et~al.} 2017, \pasa, 34, e037,
  \dodoi{10.1017/pasa.2017.33}

\bibitem[{{Andreoni} {et~al.}(2020{\natexlab{a}}){Andreoni}, {Cooke}, {Webb},
  {Rest}, {Pritchard}, {Caleb}, {Chang}, {Farah}, {Lien}, {M{\"o}ller},
  {Ravasio}, {Abbott}, {Bhandari}, {Cucchiara}, {Flynn}, {Jankowski}, {Keane},
  {Moriya}, {Onken}, {Parthasarathy}, {Price}, {Petroff}, {Ryder}, {Vohl}, \&
  {Wolf}}]{dwf_program}
{Andreoni}, I., {Cooke}, J., {Webb}, S., {et~al.} 2020{\natexlab{a}}, \mnras,
  491, 5852, \dodoi{10.1093/mnras/stz3381}

\bibitem[{{Andreoni} {et~al.}(2020{\natexlab{b}}){Andreoni}, {Cooke}, {Webb},
  {Rest}, {Pritchard}, {Caleb}, {Chang}, {Farah}, {Lien}, {M{\"o}ller},
  {Ravasio}, {Abbott}, {Bhandari}, {Cucchiara}, {Flynn}, {Jankowski}, {Keane},
  {Moriya}, {Onken}, {Parthasarathy}, {Price}, {Petroff}, {Ryder}, {Vohl}, \&
  {Wolf}}]{andreoni_2020}
---. 2020{\natexlab{b}}, \mnras, 491, 5852, \dodoi{10.1093/mnras/stz3381}

\bibitem[{{Arcavi}(2022)}]{arcavi_2022}
{Arcavi}, I. 2022, \apj, 937, 75, \dodoi{10.3847/1538-4357/ac90c0}

\bibitem[{{Arimatsu} {et~al.}(2021){Arimatsu}, {Tsumura}, {Usui}, {Ootsubo}, \&
  {Watanabe}}]{Arimatsu_2021}
{Arimatsu}, K., {Tsumura}, K., {Usui}, F., {Ootsubo}, T., \& {Watanabe}, J.-i.
  2021, \aj, 161, 135, \dodoi{10.3847/1538-3881/abd94d}

\bibitem[{{Astropy Collaboration} {et~al.}(2013){Astropy Collaboration},
  {Robitaille}, {Tollerud}, {Greenfield}, {Droettboom}, {Bray}, {Aldcroft},
  {Davis}, {Ginsburg}, {Price-Whelan}, {Kerzendorf}, {Conley}, {Crighton},
  {Barbary}, {Muna}, {Ferguson}, {Grollier}, {Parikh}, {Nair}, {Unther},
  {Deil}, {Woillez}, {Conseil}, {Kramer}, {Turner}, {Singer}, {Fox}, {Weaver},
  {Zabalza}, {Edwards}, {Azalee Bostroem}, {Burke}, {Casey}, {Crawford},
  {Dencheva}, {Ely}, {Jenness}, {Labrie}, {Lim}, {Pierfederici}, {Pontzen},
  {Ptak}, {Refsdal}, {Servillat}, \& {Streicher}}]{astropy_2013}
{Astropy Collaboration}, {Robitaille}, T.~P., {Tollerud}, E.~J., {et~al.} 2013,
  \aap, 558, A33, \dodoi{10.1051/0004-6361/201322068}

\bibitem[{{Astropy Collaboration} {et~al.}(2018){Astropy Collaboration},
  {Price-Whelan}, {Sip{\H{o}}cz}, {G{\"u}nther}, {Lim}, {Crawford}, {Conseil},
  {Shupe}, {Craig}, {Dencheva}, {Ginsburg}, {Vand erPlas}, {Bradley},
  {P{\'e}rez-Su{\'a}rez}, {de Val-Borro}, {Aldcroft}, {Cruz}, {Robitaille},
  {Tollerud}, {Ardelean}, {Babej}, {Bach}, {Bachetti}, {Bakanov}, {Bamford},
  {Barentsen}, {Barmby}, {Baumbach}, {Berry}, {Biscani}, {Boquien}, {Bostroem},
  {Bouma}, {Brammer}, {Bray}, {Breytenbach}, {Buddelmeijer}, {Burke},
  {Calderone}, {Cano Rodr{\'\i}guez}, {Cara}, {Cardoso}, {Cheedella}, {Copin},
  {Corrales}, {Crichton}, {D'Avella}, {Deil}, {Depagne}, {Dietrich}, {Donath},
  {Droettboom}, {Earl}, {Erben}, {Fabbro}, {Ferreira}, {Finethy}, {Fox},
  {Garrison}, {Gibbons}, {Goldstein}, {Gommers}, {Greco}, {Greenfield},
  {Groener}, {Grollier}, {Hagen}, {Hirst}, {Homeier}, {Horton}, {Hosseinzadeh},
  {Hu}, {Hunkeler}, {Ivezi{\'c}}, {Jain}, {Jenness}, {Kanarek}, {Kendrew},
  {Kern}, {Kerzendorf}, {Khvalko}, {King}, {Kirkby}, {Kulkarni}, {Kumar},
  {Lee}, {Lenz}, {Littlefair}, {Ma}, {Macleod}, {Mastropietro}, {McCully},
  {Montagnac}, {Morris}, {Mueller}, {Mumford}, {Muna}, {Murphy}, {Nelson},
  {Nguyen}, {Ninan}, {N{\"o}the}, {Ogaz}, {Oh}, {Parejko}, {Parley}, {Pascual},
  {Patil}, {Patil}, {Plunkett}, {Prochaska}, {Rastogi}, {Reddy Janga},
  {Sabater}, {Sakurikar}, {Seifert}, {Sherbert}, {Sherwood-Taylor}, {Shih},
  {Sick}, {Silbiger}, {Singanamalla}, {Singer}, {Sladen}, {Sooley},
  {Sornarajah}, {Streicher}, {Teuben}, {Thomas}, {Tremblay}, {Turner},
  {Terr{\'o}n}, {van Kerkwijk}, {de la Vega}, {Watkins}, {Weaver}, {Whitmore},
  {Woillez}, {Zabalza}, \& {Astropy Contributors}}]{astropy_2018}
{Astropy Collaboration}, {Price-Whelan}, A.~M., {Sip{\H{o}}cz}, B.~M., {et~al.}
  2018, \aj, 156, 123, \dodoi{10.3847/1538-3881/aabc4f}

\bibitem[{{Bailey} {et~al.}(2008){Bailey}, {Aragon}, {Romano}, {Thomas},
  {Weaver}, \& {Wong}}]{bailey_2007}
{Bailey}, S., {Aragon}, C., {Romano}, R., {et~al.} 2008, Astronomische
  Nachrichten, 329, 292, \dodoi{10.1002/asna.200710932}

\bibitem[{Barbary(2016)}]{Barbary2016}
Barbary, K. 2016, Journal of Open Source Software, 1, 58,
  \dodoi{10.21105/joss.00058}

\bibitem[{{Becker}(2015)}]{hotpants}
{Becker}, A. 2015, {HOTPANTS: High Order Transform of PSF ANd Template
  Subtraction}.
\newblock \doeprint{1504.004}

\bibitem[{Behnel {et~al.}(2011)Behnel, Bradshaw, Citro, Dalcin, Seljebotn, \&
  Smith}]{behnel2011cython}
Behnel, S., Bradshaw, R., Citro, C., {et~al.} 2011, Computing in Science \&
  Engineering, 13, 31

\bibitem[{{Bellm} {et~al.}(2019){Bellm}, {Kulkarni}, {Graham}, {Dekany},
  {Smith}, {Riddle}, {Masci}, {Helou}, {Prince}, {Adams}, {Barbarino},
  {Barlow}, {Bauer}, {Beck}, {Belicki}, {Biswas}, {Blagorodnova}, {Bodewits},
  {Bolin}, {Brinnel}, {Brooke}, {Bue}, {Bulla}, {Burruss}, {Cenko}, {Chang},
  {Connolly}, {Coughlin}, {Cromer}, {Cunningham}, {De}, {Delacroix}, {Desai},
  {Duev}, {Eadie}, {Farnham}, {Feeney}, {Feindt}, {Flynn}, {Franckowiak},
  {Frederick}, {Fremling}, {Gal-Yam}, {Gezari}, {Giomi}, {Goldstein},
  {Golkhou}, {Goobar}, {Groom}, {Hacopians}, {Hale}, {Henning}, {Ho}, {Hover},
  {Howell}, {Hung}, {Huppenkothen}, {Imel}, {Ip}, {Ivezi{\'c}}, {Jackson},
  {Jones}, {Juric}, {Kasliwal}, {Kaspi}, {Kaye}, {Kelley}, {Kowalski},
  {Kramer}, {Kupfer}, {Landry}, {Laher}, {Lee}, {Lin}, {Lin}, {Lunnan},
  {Giomi}, {Mahabal}, {Mao}, {Miller}, {Monkewitz}, {Murphy}, {Ngeow},
  {Nordin}, {Nugent}, {Ofek}, {Patterson}, {Penprase}, {Porter}, {Rauch},
  {Rebbapragada}, {Reiley}, {Rigault}, {Rodriguez}, {van Roestel}, {Rusholme},
  {van Santen}, {Schulze}, {Shupe}, {Singer}, {Soumagnac}, {Stein}, {Surace},
  {Sollerman}, {Szkody}, {Taddia}, {Terek}, {Van Sistine}, {van Velzen},
  {Vestrand}, {Walters}, {Ward}, {Ye}, {Yu}, {Yan}, \&
  {Zolkower}}]{ztf_instrument}
{Bellm}, E.~C., {Kulkarni}, S.~R., {Graham}, M.~J., {et~al.} 2019, \pasp, 131,
  018002, \dodoi{10.1088/1538-3873/aaecbe}

\bibitem[{{Berger} {et~al.}(2013){Berger}, {Leibler}, {Chornock}, {Rest},
  {Foley}, {Soderberg}, {Price}, {Burgett}, {Chambers}, {Flewelling}, {Huber},
  {Magnier}, {Metcalfe}, {Stubbs}, \& {Tonry}}]{berger_2013}
{Berger}, E., {Leibler}, C.~N., {Chornock}, R., {et~al.} 2013, \apj, 779, 18,
  \dodoi{10.1088/0004-637X/779/1/18}

\bibitem[{Beroiz {et~al.}(2020)Beroiz, Cabral, \& Sanchez}]{astroalign}
Beroiz, M., Cabral, J., \& Sanchez, B. 2020, Astronomy and Computing, 32,
  100384, \dodoi{https://doi.org/10.1016/j.ascom.2020.100384}

\bibitem[{{Bersten} {et~al.}(2018){Bersten}, {Folatelli}, {Garc{\'\i}a}, {van
  Dyk}, {Benvenuto}, {Orellana}, {Buso}, {S{\'a}nchez}, {Tanaka}, {Maeda},
  {Filippenko}, {Zheng}, {Brink}, {Cenko}, {de Jaeger}, {Kumar}, {Moriya},
  {Nomoto}, {Perley}, {Shivvers}, \& {Smith}}]{bersten_2018}
{Bersten}, M.~C., {Folatelli}, G., {Garc{\'\i}a}, F., {et~al.} 2018, \nat, 554,
  497, \dodoi{10.1038/nature25151}

\bibitem[{{Bertin}(2011)}]{psfex}
{Bertin}, E. 2011, in Astronomical Society of the Pacific Conference Series,
  Vol. 442, Astronomical Data Analysis Software and Systems XX, ed. I.~N.
  {Evans}, A.~{Accomazzi}, D.~J. {Mink}, \& A.~H. {Rots}, 435

\bibitem[{{Bertin} \& {Arnouts}(1996)}]{bertin_1996}
{Bertin}, E., \& {Arnouts}, S. 1996, \aaps, 117, 393,
  \dodoi{10.1051/aas:1996164}

\bibitem[{{Bhat} {et~al.}(2009){Bhat}, {Meegan}, {Lichti}, {Briggs},
  {Connaughton}, {Diehl}, {Fishman}, {Greiner}, {Kippen}, {Kouveliotou},
  {Paciesas}, {Preece}, \& {von Kienlin}}]{fermi_gbm_instrument}
{Bhat}, P.~N., {Meegan}, C.~A., {Lichti}, G.~G., {et~al.} 2009, in American
  Institute of Physics Conference Series, Vol. 1133, American Institute of
  Physics Conference Series, ed. C.~{Meegan}, C.~{Kouveliotou}, \&
  N.~{Gehrels}, 34--36, \dodoi{10.1063/1.3155916}

\bibitem[{{Bloom} {et~al.}(2008){Bloom}, {Starr}, {Butler}, {Nugent},
  {Rischard}, {Eads}, \& {Poznanski}}]{bloom_2008}
{Bloom}, J.~S., {Starr}, D.~L., {Butler}, N.~R., {et~al.} 2008, Astronomische
  Nachrichten, 329, 284, \dodoi{10.1002/asna.200710957}

\bibitem[{Bramich(2008)}]{bramich_2008}
Bramich, D. 2008, Monthly Notices of the Royal Astronomical Society, 386,
  \dodoi{10.1111/j.1745-3933.2008.00464.x}

\bibitem[{{Brink} {et~al.}(2013){Brink}, {Richards}, {Poznanski}, {Bloom},
  {Rice}, {Negahban}, \& {Wainwright}}]{brink_2013}
{Brink}, H., {Richards}, J.~W., {Poznanski}, D., {et~al.} 2013, \mnras, 435,
  1047, \dodoi{10.1093/mnras/stt1306}

\bibitem[{{Burke} {et~al.}(2019){Burke}, {Aleo}, {Chen}, {Liu}, {Peterson},
  {Sembroski}, \& {Lin}}]{Burke_2019}
{Burke}, C.~J., {Aleo}, P.~D., {Chen}, Y.-C., {et~al.} 2019, \mnras, 490, 3952,
  \dodoi{10.1093/mnras/stz2845}

\bibitem[{{Butler} {et~al.}(2016){Butler}, {Daly}, {Doyle}, {Gillies}, {Hagen},
  \& {Schaub}}]{geojson}
{Butler}, H., {Daly}, M., {Doyle}, A., {et~al.} 2016, {The GeoJSON Format},
  \dodoi{10.17487/RFC7946}

\bibitem[{Bäuerle {et~al.}(2021)Bäuerle, van Onzenoodt, \&
  Ropinski}]{net2vis}
Bäuerle, A., van Onzenoodt, C., \& Ropinski, T. 2021, IEEE Transactions on
  Visualization and Computer Graphics, 27, 2980,
  \dodoi{10.1109/TVCG.2021.3057483}

\bibitem[{{Calabretta} {et~al.}(2004){Calabretta}, {Valdes}, {Greisen}, \&
  {Allen}}]{calabretta2004}
{Calabretta}, M.~R., {Valdes}, F., {Greisen}, E.~W., \& {Allen}, S.~L. 2004, in
  Astronomical Society of the Pacific Conference Series, Vol. 314, Astronomical
  Data Analysis Software and Systems (ADASS) XIII, ed. F.~{Ochsenbein}, M.~G.
  {Allen}, \& D.~{Egret}, 551

\bibitem[{Cao {et~al.}(2016)Cao, Nugent, \& Kasliwal}]{Cao_2016}
Cao, Y., Nugent, P.~E., \& Kasliwal, M.~M. 2016, arXiv,
  \dodoi{10.1088/1538-3873/128/969/114502}

\bibitem[{Chollet {et~al.}(2015)}]{chollet2015keras}
Chollet, F., {et~al.} 2015, Keras,  GitHub.
\newblock \url{https://github.com/fchollet/keras}

\bibitem[{{Clemens} {et~al.}(2004){Clemens}, {Crain}, \&
  {Anderson}}]{clemens_2004}
{Clemens}, J.~C., {Crain}, J.~A., \& {Anderson}, R. 2004, in Society of
  Photo-Optical Instrumentation Engineers (SPIE) Conference Series, Vol. 5492,
  Ground-based Instrumentation for Astronomy, ed. A.~F.~M. {Moorwood} \&
  M.~{Iye}, 331--340, \dodoi{10.1117/12.550069}

\bibitem[{Corbett {et~al.}(2022{\natexlab{a}})Corbett, Soto, Machia, Galliher,
  Gonzalez, \& Law}]{corbett_hdps}
Corbett, H., Soto, A.~V., Machia, L., {et~al.} 2022{\natexlab{a}}, in Software
  and Cyberinfrastructure for Astronomy VII, ed. J.~Ibsen \& G.~Chiozzi, Vol.
  12189, International Society for Optics and Photonics (SPIE), 1218910,
  \dodoi{10.1117/12.2629533}

\bibitem[{Corbett {et~al.}(2022{\natexlab{b}})Corbett, Soto, Machia, Galliher,
  Gonzalez, Walters, \& Law}]{corbett_at2d}
Corbett, H., Soto, A.~V., Machia, L., {et~al.} 2022{\natexlab{b}}, in
  Ground-based and Airborne Telescopes IX, ed. H.~K. Marshall, J.~Spyromilio,
  \& T.~Usuda, Vol. 12182, International Society for Optics and Photonics
  (SPIE), 121824D, \dodoi{10.1117/12.2629489}

\bibitem[{{Corbett} {et~al.}(2022){Corbett}, {Vasquez Soto}, {Machia},
  {Galliher}, {Gonzalez}, \& {Law}}]{Corbett_2022}
{Corbett}, H., {Vasquez Soto}, A., {Machia}, L., {et~al.} 2022, in Society of
  Photo-Optical Instrumentation Engineers (SPIE) Conference Series, Vol. 12189,
  Society of Photo-Optical Instrumentation Engineers (SPIE) Conference Series,
  1218910, \dodoi{10.1117/12.2629533}

\bibitem[{{Corbett} {et~al.}(2020){Corbett}, {Law}, {Soto}, {Howard},
  {Glazier}, {Gonzalez}, {Ratzloff}, {Galliher}, {Fors}, \&
  {Quimby}}]{corbett_flashes}
{Corbett}, H., {Law}, N.~M., {Soto}, A.~V., {et~al.} 2020, \apjl, 903, L27,
  \dodoi{10.3847/2041-8213/abbee5}

\bibitem[{{Cucchiara} {et~al.}(2011){Cucchiara}, {Cenko}, {Bloom}, {Melandri},
  {Morgan}, {Kobayashi}, {Smith}, {Perley}, {Li}, {Hora}, {da Silva},
  {Prochaska}, {Milne}, {Butler}, {Cobb}, {Worseck}, {Mundell}, {Steele},
  {Filippenko}, {Fumagalli}, {Klein}, {Stephens}, {Bluck}, \&
  {Mason}}]{cucchiara_2011}
{Cucchiara}, A., {Cenko}, S.~B., {Bloom}, J.~S., {et~al.} 2011, \apj, 743, 154,
  \dodoi{10.1088/0004-637X/743/2/154}

\bibitem[{D\'alya {et~al.}(2018)D\'alya, Galgóczi, Dobos, Frei, Heng, Macas,
  Messenger, Raffai, \& de Souza}]{glade_catalog}
D\'alya, G., Galgóczi, G., Dobos, L., {et~al.} 2018, Monthly Notices of the
  Royal Astronomical Society, 479, 2374, \dodoi{10.1093/mnras/sty1703}

\bibitem[{{Dark Energy Survey Collaboration} {et~al.}(2016){Dark Energy Survey
  Collaboration}, Abbott, Abdalla, Aleksić, Allam, Amara, Bacon, Balbinot,
  Banerji, Bechtol, Benoit-Lévy, Bernstein, Bertin, Blazek, Bonnett, Bridle,
  Brooks, Brunner, Buckley-Geer, Burke, Caminha, Capozzi, Carlsen,
  Carnero-Rosell, Carollo, Carrasco-Kind, Carretero, Castander, Clerkin,
  Collett, Conselice, Crocce, Cunha, D'Andrea, da~Costa, Davis, Desai, Diehl,
  Dietrich, Dodelson, Doel, Drlica-Wagner, Estrada, Etherington, Evrard,
  Fabbri, Finley, Flaugher, Foley, Fosalba, Frieman, García-Bellido,
  Gaztanaga, Gerdes, Giannantonio, Goldstein, Gruen, Gruendl, Guarnieri,
  Gutierrez, Hartley, Honscheid, Jain, James, Jeltema, Jouvel, Kessler, King,
  Kirk, Kron, Kuehn, Kuropatkin, Lahav, Li, Lima, Lin, Maia, Makler, Manera,
  Maraston, Marshall, Martini, McMahon, Melchior, Merson, Miller, Miquel, Mohr,
  Morice-Atkinson, Naidoo, Neilsen, Nichol, Nord, Ogando, Ostrovski, Palmese,
  Papadopoulos, Peiris, Peoples, Percival, Plazas, Reed, Refregier, Romer,
  Roodman, Ross, Rozo, Rykoff, Sadeh, Sako, Sánchez, Sanchez, Santiago,
  Scarpine, Schubnell, Sevilla-Noarbe, Sheldon, Smith, Smith, Soares-Santos,
  Sobreira, Soumagnac, Suchyta, Sullivan, Swanson, Tarle, Thaler, Thomas,
  Thomas, Tucker, Vieira, Vikram, Walker, Wechsler, Weller, Wester, Whiteway,
  Wilcox, Yanny, Zhang, \& Zuntz}]{des_program}
{Dark Energy Survey Collaboration}, Abbott, T., Abdalla, F.~B., {et~al.} 2016,
  Monthly Notices of the Royal Astronomical Society, 460, 1270,
  \dodoi{10.1093/mnras/stw641}

\bibitem[{{Dieleman} {et~al.}(2015){Dieleman}, {Willett}, \&
  {Dambre}}]{dielman_2015}
{Dieleman}, S., {Willett}, K.~W., \& {Dambre}, J. 2015, \mnras, 450, 1441,
  \dodoi{10.1093/mnras/stv632}

\bibitem[{{Drake} {et~al.}(2009){Drake}, {Djorgovski}, {Mahabal}, {Beshore},
  {Larson}, {Graham}, {Williams}, {Christensen}, {Catelan}, {Boattini},
  {Gibbs}, {Hill}, \& {Kowalski}}]{crts_instrument}
{Drake}, A.~J., {Djorgovski}, S.~G., {Mahabal}, A., {et~al.} 2009, \apj, 696,
  870, \dodoi{10.1088/0004-637X/696/1/870}

\bibitem[{{Duev} {et~al.}(2019){Duev}, {Mahabal}, {Masci}, {Graham},
  {Rusholme}, {Walters}, {Karmarkar}, {Frederick}, {Kasliwal}, {Rebbapragada},
  \& {Ward}}]{duev_2020}
{Duev}, D.~A., {Mahabal}, A., {Masci}, F.~J., {et~al.} 2019, \mnras, 489, 3582,
  \dodoi{10.1093/mnras/stz2357}

\bibitem[{{Dyer} {et~al.}(2018){Dyer}, {Dhillon}, {Littlefair}, {Steeghs},
  {Ulaczyk}, {Chote}, {Galloway}, \& {Rol}}]{goto_instrument}
{Dyer}, M.~J., {Dhillon}, V.~S., {Littlefair}, S., {et~al.} 2018, in Society of
  Photo-Optical Instrumentation Engineers (SPIE) Conference Series, Vol. 10704,
  Observatory Operations: Strategies, Processes, and Systems VII, 107040C,
  \dodoi{10.1117/12.2311865}

\bibitem[{Fischler \& Bolles(1981)}]{fischler_bolles_1981}
Fischler, M., \& Bolles, R. 1981, Communications of the ACM, 24, 381.
\newblock \url{/brokenurl#
  http://publication.wilsonwong.me/load.php?id=233282275}

\bibitem[{{Flewelling} {et~al.}(2016){Flewelling}, {Magnier}, {Chambers},
  {Heasley}, {Holmberg}, {Huber}, {Sweeney}, {Waters}, {Calamida}, {Casertano},
  {Chen}, {Farrow}, {Hasinger}, {Henderson}, {Long}, {Metcalfe}, {Narayan},
  {Nieto-Santisteban}, {Norberg}, {Rest}, {Saglia}, {Szalay}, {Thakar},
  {Tonry}, {Valenti}, {Werner}, {White}, {Denneau}, {Draper}, {Hodapp},
  {Jedicke}, {Kaiser}, {Kudritzki}, {Price}, {Wainscoat}, {Builders},
  {Chastel}, {McLean}, {Postman}, \& {Shiao}}]{panstarrs_dr1}
{Flewelling}, H.~A., {Magnier}, E.~A., {Chambers}, K.~C., {et~al.} 2016, arXiv
  e-prints, arXiv:1612.05243.
\newblock \doarXiv{1612.05243}

\bibitem[{{Fox} {et~al.}(2003){Fox}, {Price}, {Soderberg}, {Berger},
  {Kulkarni}, {Sari}, {Frail}, {Harrison}, {Yost}, {Matthews}, {Peterson},
  {Tanaka}, {Christiansen}, \& {Moriarty-Schieven}}]{fox_2003}
{Fox}, D.~W., {Price}, P.~A., {Soderberg}, A.~M., {et~al.} 2003, \apjl, 586,
  L5, \dodoi{10.1086/374683}

\bibitem[{{Fresneau} \& {Osborn}(2009)}]{fresneau_rosat}
{Fresneau}, A., \& {Osborn}, W.~H. 2009, \aap, 503, 1023,
  \dodoi{10.1051/0004-6361/200810798}

\bibitem[{Förster {et~al.}(2016)Förster, Maureira, Martín, Hamuy, Martínez,
  Huijse, Cabrera, Galbany, Jaeger, González-Gaitán, \&
  et~al.}]{Forster_2016}
Förster, F., Maureira, J.~C., Martín, J.~S., {et~al.} 2016, arXiv,
  \dodoi{10.3847/0004-637x/832/2/155}

\bibitem[{{Gaia Collaboration} {et~al.}(2018){Gaia Collaboration}, {Brown},
  {Vallenari}, {Prusti}, {de Bruijne}, {Babusiaux}, {Bailer-Jones}, {Biermann},
  {Evans}, {Eyer}, {Jansen}, {Jordi}, {Klioner}, {Lammers}, {Lindegren},
  {Luri}, {Mignard}, {Panem}, {Pourbaix}, {Randich}, {Sartoretti}, {Siddiqui},
  {Soubiran}, {van Leeuwen}, {Walton}, {Arenou}, {Bastian}, {Cropper},
  {Drimmel}, {Katz}, {Lattanzi}, {Bakker}, {Cacciari}, {Casta{\~n}eda},
  {Chaoul}, {Cheek}, {De Angeli}, {Fabricius}, {Guerra}, {Holl}, {Masana},
  {Messineo}, {Mowlavi}, {Nienartowicz}, {Panuzzo}, {Portell}, {Riello},
  {Seabroke}, {Tanga}, {Th{\'e}venin}, {Gracia-Abril}, {Comoretto},
  {Garcia-Reinaldos}, {Teyssier}, {Altmann}, {Andrae}, {Audard},
  {Bellas-Velidis}, {Benson}, {Berthier}, {Blomme}, {Burgess}, {Busso},
  {Carry}, {Cellino}, {Clementini}, {Clotet}, {Creevey}, {Davidson}, {De
  Ridder}, {Delchambre}, {Dell'Oro}, {Ducourant},
  {Fern{\'a}ndez-Hern{\'a}ndez}, {Fouesneau}, {Fr{\'e}mat}, {Galluccio},
  {Garc{\'\i}a-Torres}, {Gonz{\'a}lez-N{\'u}{\~n}ez}, {Gonz{\'a}lez-Vidal},
  {Gosset}, {Guy}, {Halbwachs}, {Hambly}, {Harrison}, {Hern{\'a}ndez},
  {Hestroffer}, {Hodgkin}, {Hutton}, {Jasniewicz}, {Jean-Antoine-Piccolo},
  {Jordan}, {Korn}, {Krone-Martins}, {Lanzafame}, {Lebzelter}, {L{\"o}ffler},
  {Manteiga}, {Marrese}, {Mart{\'\i}n-Fleitas}, {Moitinho}, {Mora}, {Muinonen},
  {Osinde}, {Pancino}, {Pauwels}, {Petit}, {Recio-Blanco}, {Richards},
  {Rimoldini}, {Robin}, {Sarro}, {Siopis}, {Smith}, {Sozzetti}, {S{\"u}veges},
  {Torra}, {van Reeven}, {Abbas}, {Abreu Aramburu}, {Accart}, {Aerts},
  {Altavilla}, {{\'A}lvarez}, {Alvarez}, {Alves}, {Anderson}, {Andrei},
  {Anglada Varela}, {Antiche}, {Antoja}, {Arcay}, {Astraatmadja}, {Bach},
  {Baker}, {Balaguer-N{\'u}{\~n}ez}, {Balm}, {Barache}, {Barata}, {Barbato},
  {Barblan}, {Barklem}, {Barrado}, {Barros}, {Barstow}, {Bartholom{\'e}
  Mu{\~n}oz}, {Bassilana}, {Becciani}, {Bellazzini}, {Berihuete}, {Bertone},
  {Bianchi}, {Bienaym{\'e}}, {Blanco-Cuaresma}, {Boch}, {Boeche}, {Bombrun},
  {Borrachero}, {Bossini}, {Bouquillon}, {Bourda}, {Bragaglia}, {Bramante},
  {Breddels}, {Bressan}, {Brouillet}, {Br{\"u}semeister}, {Brugaletta},
  {Bucciarelli}, {Burlacu}, {Busonero}, {Butkevich}, {Buzzi}, {Caffau},
  {Cancelliere}, {Cannizzaro}, {Cantat-Gaudin}, {Carballo}, {Carlucci},
  {Carrasco}, {Casamiquela}, {Castellani}, {Castro-Ginard}, {Charlot},
  {Chemin}, {Chiavassa}, {Cocozza}, {Costigan}, {Cowell}, {Crifo}, {Crosta},
  {Crowley}, {Cuypers}, {Dafonte}, {Damerdji}, {Dapergolas}, {David}, {David},
  {de Laverny}, {De Luise}, {De March}, {de Martino}, {de Souza}, {de Torres},
  {Debosscher}, {del Pozo}, {Delbo}, {Delgado}, {Delgado}, {Di Matteo},
  {Diakite}, {Diener}, {Distefano}, {Dolding}, {Drazinos}, {Dur{\'a}n},
  {Edvardsson}, {Enke}, {Eriksson}, {Esquej}, {Eynard Bontemps}, {Fabre},
  {Fabrizio}, {Faigler}, {Falc{\~a}o}, {Farr{\`a}s Casas}, {Federici},
  {Fedorets}, {Fernique}, {Figueras}, {Filippi}, {Findeisen}, {Fonti},
  {Fraile}, {Fraser}, {Fr{\'e}zouls}, {Gai}, {Galleti}, {Garabato},
  {Garc{\'\i}a-Sedano}, {Garofalo}, {Garralda}, {Gavel}, {Gavras}, {Gerssen},
  {Geyer}, {Giacobbe}, {Gilmore}, {Girona}, {Giuffrida}, {Glass}, {Gomes},
  {Granvik}, {Gueguen}, {Guerrier}, {Guiraud}, {Guti{\'e}rrez-S{\'a}nchez},
  {Haigron}, {Hatzidimitriou}, {Hauser}, {Haywood}, {Heiter}, {Helmi}, {Heu},
  {Hilger}, {Hobbs}, {Hofmann}, {Holland}, {Huckle}, {Hypki}, {Icardi},
  {Jan{\ss}en}, {Jevardat de Fombelle}, {Jonker}, {Juh{\'a}sz}, {Julbe},
  {Karampelas}, {Kewley}, {Klar}, {Kochoska}, {Kohley}, {Kolenberg},
  {Kontizas}, {Kontizas}, {Koposov}, {Kordopatis}, {Kostrzewa-Rutkowska},
  {Koubsky}, {Lambert}, {Lanza}, {Lasne}, {Lavigne}, {Le Fustec}, {Le
  Poncin-Lafitte}, {Lebreton}, {Leccia}, {Leclerc}, {Lecoeur-Taibi},
  {Lenhardt}, {Leroux}, {Liao}, {Licata}, {Lindstr{\o}m}, {Lister}, {Livanou},
  {Lobel}, {L{\'o}pez}, {Managau}, {Mann}, {Mantelet}, {Marchal}, {Marchant},
  {Marconi}, {Marinoni}, {Marschalk{\'o}}, {Marshall}, {Martino}, {Marton},
  {Mary}, {Massari}, {Matijevi{\v{c}}}, {Mazeh}, {McMillan}, {Messina},
  {Michalik}, {Millar}, {Molina}, {Molinaro}, {Moln{\'a}r}, {Montegriffo},
  {Mor}, {Morbidelli}, {Morel}, {Morris}, {Mulone}, {Muraveva}, {Musella},
  {Nelemans}, {Nicastro}, {Noval}, {O'Mullane}, {Ord{\'e}novic},
  {Ord{\'o}{\~n}ez-Blanco}, {Osborne}, {Pagani}, {Pagano}, {Pailler},
  {Palacin}, {Palaversa}, {Panahi}, {Pawlak}, {Piersimoni}, {Pineau}, {Plachy},
  {Plum}, {Poggio}, {Poujoulet}, {Pr{\v{s}}a}, {Pulone}, {Racero}, {Ragaini},
  {Rambaux}, {Ramos-Lerate}, {Regibo}, {Reyl{\'e}}, {Riclet}, {Ripepi}, {Riva},
  {Rivard}, {Rixon}, {Roegiers}, {Roelens}, {Romero-G{\'o}mez}, {Rowell},
  {Royer}, {Ruiz-Dern}, {Sadowski}, {Sagrist{\`a} Sell{\'e}s}, {Sahlmann},
  {Salgado}, {Salguero}, {Sanna}, {Santana-Ros}, {Sarasso}, {Savietto},
  {Schultheis}, {Sciacca}, {Segol}, {Segovia}, {S{\'e}gransan}, {Shih},
  {Siltala}, {Silva}, {Smart}, {Smith}, {Solano}, {Solitro}, {Sordo}, {Soria
  Nieto}, {Souchay}, {Spagna}, {Spoto}, {Stampa}, {Steele},
  {Steidelm{\"u}ller}, {Stephenson}, {Stoev}, {Suess}, {Surdej}, {Szabados},
  {Szegedi-Elek}, {Tapiador}, {Taris}, {Tauran}, {Taylor}, {Teixeira},
  {Terrett}, {Teyssand ier}, {Thuillot}, {Titarenko}, {Torra Clotet}, {Turon},
  {Ulla}, {Utrilla}, {Uzzi}, {Vaillant}, {Valentini}, {Valette}, {van Elteren},
  {Van Hemelryck}, {van Leeuwen}, {Vaschetto}, {Vecchiato}, {Veljanoski},
  {Viala}, {Vicente}, {Vogt}, {von Essen}, {Voss}, {Votruba}, {Voutsinas},
  {Walmsley}, {Weiler}, {Wertz}, {Wevers}, {Wyrzykowski}, {Yoldas},
  {{\v{Z}}erjal}, {Ziaeepour}, {Zorec}, {Zschocke}, {Zucker}, {Zurbach}, \&
  {Zwitter}}]{gaia_dr2}
{Gaia Collaboration}, {Brown}, A.~G.~A., {Vallenari}, A., {et~al.} 2018, \aap,
  616, A1, \dodoi{10.1051/0004-6361/201833051}

\bibitem[{{Gal} \& {Ghahramani}(2015{\natexlab{a}})}]{Gal_2015}
{Gal}, Y., \& {Ghahramani}, Z. 2015{\natexlab{a}}, arXiv e-prints,
  arXiv:1506.02142.
\newblock \doarXiv{1506.02142}

\bibitem[{{Gal} \& {Ghahramani}(2015{\natexlab{b}})}]{mcdropout}
---. 2015{\natexlab{b}}, arXiv e-prints, arXiv:1506.02158.
\newblock \doarXiv{1506.02158}

\bibitem[{{Garnavich} {et~al.}(2016){Garnavich}, {Tucker}, {Rest}, {Shaya},
  {Olling}, {Kasen}, \& {Villar}}]{garnavich_2016}
{Garnavich}, P.~M., {Tucker}, B.~E., {Rest}, A., {et~al.} 2016, \apj, 820, 23,
  \dodoi{10.3847/0004-637X/820/1/23}

\bibitem[{{Ghosh} {et~al.}(2016){Ghosh}, {Manwani}, \& {Sastry}}]{ghosh_2016}
{Ghosh}, A., {Manwani}, N., \& {Sastry}, P.~S. 2016, arXiv e-prints,
  arXiv:1605.06296.
\newblock \doarXiv{1605.06296}

\bibitem[{{Goldstein} {et~al.}(2015){Goldstein}, {D'Andrea}, {Fischer},
  {Foley}, {Gupta}, {Kessler}, {Kim}, {Nichol}, {Nugent}, {Papadopoulos},
  {Sako}, {Smith}, {Sullivan}, {Thomas}, {Wester}, {Wolf}, {Abdalla},
  {Banerji}, {Benoit-L{\'e}vy}, {Bertin}, {Brooks}, {Carnero Rosell},
  {Castander}, {da Costa}, {Covarrubias}, {DePoy}, {Desai}, {Diehl}, {Doel},
  {Eifler}, {Fausti Neto}, {Finley}, {Flaugher}, {Fosalba}, {Frieman},
  {Gerdes}, {Gruen}, {Gruendl}, {James}, {Kuehn}, {Kuropatkin}, {Lahav}, {Li},
  {Maia}, {Makler}, {March}, {Marshall}, {Martini}, {Merritt}, {Miquel},
  {Nord}, {Ogando}, {Plazas}, {Romer}, {Roodman}, {Sanchez}, {Scarpine},
  {Schubnell}, {Sevilla-Noarbe}, {Smith}, {Soares-Santos}, {Sobreira},
  {Suchyta}, {Swanson}, {Tarle}, {Thaler}, \& {Walker}}]{goldstein_2015}
{Goldstein}, D.~A., {D'Andrea}, C.~B., {Fischer}, J.~A., {et~al.} 2015, \aj,
  150, 82, \dodoi{10.1088/0004-6256/150/3/82}

\bibitem[{{G{\'o}rski} \& {Hivon}(2011)}]{gorski_2011}
{G{\'o}rski}, K.~M., \& {Hivon}, E. 2011, {HEALPix: Hierarchical Equal Area
  isoLatitude Pixelization of a sphere}, Astrophysics Source Code Library,
  record ascl:1107.018.
\newblock \doeprint{1107.018}

\bibitem[{{Hamuy} {et~al.}(1994){Hamuy}, {Suntzeff}, {Heathcote}, {Walker},
  {Gigoux}, \& {Phillips}}]{hamuy_94}
{Hamuy}, M., {Suntzeff}, N.~B., {Heathcote}, S.~R., {et~al.} 1994, \pasp, 106,
  566, \dodoi{10.1086/133417}

\bibitem[{{Hamuy} {et~al.}(1992){Hamuy}, {Walker}, {Suntzeff}, {Gigoux},
  {Heathcote}, \& {Phillips}}]{hamuy_92}
{Hamuy}, M., {Walker}, A.~R., {Suntzeff}, N.~B., {et~al.} 1992, \pasp, 104,
  533, \dodoi{10.1086/133028}

\bibitem[{{Hedges} {et~al.}(2021){Hedges}, {Luger}, {Martinez-Palomera},
  {Dotson}, \& {Barentsen}}]{hedges_2021}
{Hedges}, C., {Luger}, R., {Martinez-Palomera}, J., {Dotson}, J., \&
  {Barentsen}, G. 2021, \aj, 162, 107, \dodoi{10.3847/1538-3881/ac0825}

\bibitem[{{Heinze} {et~al.}(2018){Heinze}, {Tonry}, {Denneau}, {Flewelling},
  {Stalder}, {Rest}, {Smith}, {Smartt}, \& {Weiland}}]{atlas_variables}
{Heinze}, A.~N., {Tonry}, J.~L., {Denneau}, L., {et~al.} 2018, \aj, 156, 241,
  \dodoi{10.3847/1538-3881/aae47f}

\bibitem[{{Henden} {et~al.}(2016){Henden}, {Templeton}, {Terrell}, {Smith},
  {Levine}, \& {Welch}}]{apass_dr9}
{Henden}, A.~A., {Templeton}, M., {Terrell}, D., {et~al.} 2016, VizieR Online
  Data Catalog, II/336

\bibitem[{{Ho} {et~al.}(2018){Ho}, {Kulkarni}, {Nugent}, {Zhao}, {Rusu},
  {Cenko}, {Ravi}, {Kasliwal}, {Perley}, {Adams}, {Bellm}, {Brady}, {Fremling},
  {Gal-Yam}, {Kann}, {Kaplan}, {Laher}, {Masci}, {Ofek}, {Sollerman}, \&
  {Urban}}]{ho_2018}
{Ho}, A. Y.~Q., {Kulkarni}, S.~R., {Nugent}, P.~E., {et~al.} 2018, \apjl, 854,
  L13, \dodoi{10.3847/2041-8213/aaaa62}

\bibitem[{{Hoffleit} \& {Jaschek}(1991)}]{yale_bsc_1991}
{Hoffleit}, D., \& {Jaschek}, C. 1991, {The Bright star catalogue} (Yale
  University Observatory)

\bibitem[{{H{\o}g} {et~al.}(2000){H{\o}g}, {Fabricius}, {Makarov}, {Urban},
  {Corbin}, {Wycoff}, {Bastian}, {Schwekendiek}, \& {Wicenec}}]{tycho2}
{H{\o}g}, E., {Fabricius}, C., {Makarov}, V.~V., {et~al.} 2000, \aap, 355, L27

\bibitem[{{Howard} \& {MacGregor}(2022)}]{Howard_2022}
{Howard}, W.~S., \& {MacGregor}, M.~A. 2022, \apj, 926, 204,
  \dodoi{10.3847/1538-4357/ac426e}

\bibitem[{{Howard} {et~al.}(2018){Howard}, {Tilley}, {Corbett}, {Youngblood},
  {Loyd}, {Ratzloff}, {Law}, {Fors}, {del Ser}, {Shkolnik}, {Ziegler}, {Goeke},
  {Pietraallo}, \& {Haislip}}]{proxima_superflare}
{Howard}, W.~S., {Tilley}, M.~A., {Corbett}, H., {et~al.} 2018, \apjl, 860,
  L30, \dodoi{10.3847/2041-8213/aacaf3}

\bibitem[{{Howard} {et~al.}(2020){Howard}, {Corbett}, {Law}, {Ratzloff},
  {Galliher}, {Glazier}, {Gonzalez}, {Vasquez Soto}, {Fors}, {del Ser}, \&
  {Haislip}}]{howard_2020}
{Howard}, W.~S., {Corbett}, H., {Law}, N.~M., {et~al.} 2020, \apj, 902, 115,
  \dodoi{10.3847/1538-4357/abb5b4}

\bibitem[{{Howard} {et~al.}(2021){Howard}, {Teske}, {Corbett}, {Law}, {Wang},
  {Ratzloff}, {Galliher}, {Gonzalez}, {Soto}, {Glazier}, \&
  {Haislip}}]{Howard_2021}
{Howard}, W.~S., {Teske}, J., {Corbett}, H., {et~al.} 2021, \aj, 162, 147,
  \dodoi{10.3847/1538-3881/ac0fe3}

\bibitem[{{Hu} {et~al.}(2022){Hu}, {Wang}, {Chen}, \& {Yang}}]{Hu_2022}
{Hu}, L., {Wang}, L., {Chen}, X., \& {Yang}, J. 2022, \apj, 936, 157,
  \dodoi{10.3847/1538-4357/ac7394}

\bibitem[{{Ioffe} \& {Szegedy}(2015)}]{ioffe_2015}
{Ioffe}, S., \& {Szegedy}, C. 2015, arXiv e-prints, arXiv:1502.03167.
\newblock \doarXiv{1502.03167}

\bibitem[{{Jayasinghe} {et~al.}(2018){Jayasinghe}, {Kochanek}, {Stanek},
  {Shappee}, {Holoien}, {Thompson}, {Prieto}, {Dong}, {Pawlak}, {Shields},
  {Pojmanski}, {Otero}, {Britt}, \& {Will}}]{asassn_variables}
{Jayasinghe}, T., {Kochanek}, C.~S., {Stanek}, K.~Z., {et~al.} 2018, \mnras,
  477, 3145, \dodoi{10.1093/mnras/sty838}

\bibitem[{{Kaiser} {et~al.}(2010){Kaiser}, {Burgett}, {Chambers}, {Denneau},
  {Heasley}, {Jedicke}, {Magnier}, {Morgan}, {Onaka}, \&
  {Tonry}}]{panstarrs_instrument}
{Kaiser}, N., {Burgett}, W., {Chambers}, K., {et~al.} 2010, in Society of
  Photo-Optical Instrumentation Engineers (SPIE) Conference Series, Vol. 7733,
  \procspie, 77330E, \dodoi{10.1117/12.859188}

\bibitem[{{Keller} {et~al.}(2007){Keller}, {Schmidt}, {Bessell}, {Conroy},
  {Francis}, {Granlund}, {Kowald}, {Oates}, {Martin-Jones}, {Preston},
  {Tisserand }, {Vaccarella}, \& {Waterson}}]{skymapper_instrument}
{Keller}, S.~C., {Schmidt}, B.~P., {Bessell}, M.~S., {et~al.} 2007, \pasa, 24,
  1, \dodoi{10.1071/AS07001}

\bibitem[{{Killestein} {et~al.}(2021){Killestein}, {Lyman}, {Steeghs},
  {Ackley}, {Dyer}, {Ulaczyk}, {Cutter}, {Mong}, {Galloway}, {Dhillon},
  {O'Brien}, {Ramsay}, {Poshyachinda}, {Kotak}, {Breton}, {Nuttall},
  {Pall{\'e}}, {Pollacco}, {Thrane}, {Aukkaravittayapun}, {Awiphan},
  {Burhanudin}, {Chote}, {Chrimes}, {Daw}, {Duffy}, {Eyles-Ferris}, {Gompertz},
  {Heikkil{\"a}}, {Irawati}, {Kennedy}, {Levan}, {Littlefair}, {Makrygianni},
  {Mata S{\'a}nchez}, {Mattila}, {Maund}, {McCormac}, {Mkrtichian}, {Mullaney},
  {Rol}, {Sawangwit}, {Stanway}, {Starling}, {Str{\o}m}, {Tooke}, {Wiersema},
  \& {Williams}}]{Killestein_2021}
{Killestein}, T.~L., {Lyman}, J., {Steeghs}, D., {et~al.} 2021, \mnras, 503,
  4838, \dodoi{10.1093/mnras/stab633}

\bibitem[{{Kingma} \& {Ba}(2014)}]{kingma_2014}
{Kingma}, D.~P., \& {Ba}, J. 2014, arXiv e-prints, arXiv:1412.6980.
\newblock \doarXiv{1412.6980}

\bibitem[{{Koposov} \& {Bartunov}(2006)}]{q3c_paper}
{Koposov}, S., \& {Bartunov}, O. 2006, in Astronomical Society of the Pacific
  Conference Series, Vol. 351, Astronomical Data Analysis Software and Systems
  XV, ed. C.~{Gabriel}, C.~{Arviset}, D.~{Ponz}, \& S.~{Enrique}, 735

\bibitem[{{Koposov} \& {Bartunov}(2019)}]{q3c_ascl}
{Koposov}, S., \& {Bartunov}, O. 2019, {Q3C: A PostgreSQL package for spatial
  queries and cross-matches of large astronomical catalogs}.
\newblock \doeprint{1905.008}

\bibitem[{{Kowalski} {et~al.}(2013){Kowalski}, {Hawley}, {Wisniewski}, {Osten},
  {Hilton}, {Holtzman}, {Schmidt}, \& {Davenport}}]{kowalski_2013}
{Kowalski}, A.~F., {Hawley}, S.~L., {Wisniewski}, J.~P., {et~al.} 2013, \apjs,
  207, 15, \dodoi{10.1088/0067-0049/207/1/15}

\bibitem[{Kreps {et~al.}(2011)Kreps, Narkhede, \& Rao}]{Kreps2011KafkaA}
Kreps, J., Narkhede, N., \& Rao, J. 2011, in Proceedings of the NetDB, 1--7

\bibitem[{{Kulkarni} \& {Rau}(2006)}]{kulkarni_2006}
{Kulkarni}, S.~R., \& {Rau}, A. 2006, \apjl, 644, L63, \dodoi{10.1086/505423}

\bibitem[{Kumar(1988)}]{kumar1988world}
Kumar, M. 1988, Marine Geodesy, 12, 117

\bibitem[{{Kumar} {et~al.}(2015){Kumar}, {Gezari}, {Heinis}, {Chornock},
  {Berger}, {Rest}, {Huber}, {Foley}, {Narayan}, {Marion}, {Scolnic},
  {Soderberg}, {Lawrence}, {Stubbs}, {Kirshner}, {Riess}, {Smartt}, {Smith},
  {Wood-Vasey}, {Burgett}, {Chambers}, {Flewelling}, {Kaiser}, {Metcalfe},
  {Price}, {Tonry}, \& {Wainscoat}}]{Kumar_2015}
{Kumar}, S., {Gezari}, S., {Heinis}, S., {et~al.} 2015, \apj, 802, 27,
  \dodoi{10.1088/0004-637X/802/1/27}

\bibitem[{{Lang} {et~al.}(2010){Lang}, {Hogg}, {Mierle}, {Blanton}, \&
  {Roweis}}]{Lang_Hogg_2010}
{Lang}, D., {Hogg}, D.~W., {Mierle}, K., {Blanton}, M., \& {Roweis}, S. 2010,
  \aj, 139, 1782, \dodoi{10.1088/0004-6256/139/5/1782}

\bibitem[{{Larson} {et~al.}(2003){Larson}, {Beshore}, {Hill}, {Christensen},
  {McLean}, {Kolar}, {McNaught}, \& {Garradd}}]{catalina_instrument}
{Larson}, S., {Beshore}, E., {Hill}, R., {et~al.} 2003, in AAS/Division for
  Planetary Sciences Meeting Abstracts \#35, AAS/Division for Planetary
  Sciences Meeting Abstracts, 36.04

\bibitem[{{Law} {et~al.}(2022){Law}, {Vasquez Soto}, {Corbett}, {Galliher},
  {Gonzalez}, {Machia}, \& {Walters}}]{law_2022b}
{Law}, N., {Vasquez Soto}, A., {Corbett}, H., {et~al.} 2022, in Society of
  Photo-Optical Instrumentation Engineers (SPIE) Conference Series, Vol. 12182,
  Ground-based and Airborne Telescopes IX, ed. H.~K. {Marshall},
  J.~{Spyromilio}, \& T.~{Usuda}, 121824H, \dodoi{10.1117/12.2630037}

\bibitem[{{Law} {et~al.}(2009){Law}, {Kulkarni}, {Dekany}, {Ofek}, {Quimby},
  {Nugent}, {Surace}, {Grillmair}, {Bloom}, {Kasliwal}, {Bildsten}, {Brown},
  {Cenko}, {Ciardi}, {Croner}, {Djorgovski}, {van Eyken}, {Filippenko}, {Fox},
  {Gal-Yam}, {Hale}, {Hamam}, {Helou}, {Henning}, {Howell}, {Jacobsen},
  {Laher}, {Mattingly}, {McKenna}, {Pickles}, {Poznanski}, {Rahmer}, {Rau},
  {Rosing}, {Shara}, {Smith}, {Starr}, {Sullivan}, {Velur}, {Walters}, \&
  {Zolkower}}]{law_2009}
{Law}, N.~M., {Kulkarni}, S.~R., {Dekany}, R.~G., {et~al.} 2009, \pasp, 121,
  1395, \dodoi{10.1086/648598}

\bibitem[{{Law} {et~al.}(2015){Law}, {Fors}, {Ratzloff}, {Wulfken},
  {Kavanaugh}, {Sitar}, {Pruett}, {Birchard}, {Barlow}, {Cannon}, {Cenko},
  {Dunlap}, {Kraus}, \& {Maccarone}}]{evryscope_project}
{Law}, N.~M., {Fors}, O., {Ratzloff}, J., {et~al.} 2015, \pasp, 127, 234,
  \dodoi{10.1086/680521}

\bibitem[{{Law} {et~al.}(2021){Law}, {Corbett}, {Galliher}, {Gonzalez},
  {Vasquez}, {Walters}, {Machia}, {Ratzloff}, {Ackley}, {Bizon}, {Clemens},
  {Cox}, {Eikenberry}, {Howard}, {Glazier}, {Mann}, {Quimby}, {Reichart}, \&
  {Trilling}}]{law_2022}
{Law}, N.~M., {Corbett}, H., {Galliher}, N.~W., {et~al.} 2021, arXiv e-prints,
  arXiv:2107.00664.
\newblock \doarXiv{2107.00664}

\bibitem[{{Le Folgoc} {et~al.}(2021){Le Folgoc}, {Baltatzis}, {Desai},
  {Devaraj}, {Ellis}, {Martinez Manzanera}, {Nair}, {Qiu}, {Schnabel}, \&
  {Glocker}}]{folgoc_2021}
{Le Folgoc}, L., {Baltatzis}, V., {Desai}, S., {et~al.} 2021, arXiv e-prints,
  arXiv:2110.04286.
\newblock \doarXiv{2110.04286}

\bibitem[{LeCun \& Bengio(1995)}]{lecun_convnets}
LeCun, Y., \& Bengio, Y. 1995, in The Handbook of Brain Theory and Neural
  Networks, ed. M.~A. Arbib (MIT Press)

\bibitem[{LeCun {et~al.}(1989)LeCun, Boser, Denker, Henderson, Howard, Hubbard,
  \& Jackel}]{lecun-89}
LeCun, Y., Boser, B., Denker, J.~S., {et~al.} 1989, Neural Computation, 1, 541

\bibitem[{Li {et~al.}(2018)Li, Jamieson, DeSalvo, Rostamizadeh, \&
  Talwalkar}]{hyperband}
Li, L., Jamieson, K., DeSalvo, G., Rostamizadeh, A., \& Talwalkar, A. 2018,
  Journal of Machine Learning Research, 18, 1.
\newblock \url{http://jmlr.org/papers/v18/16-558.html}

\bibitem[{{Lipunov} {et~al.}(2004){Lipunov}, {Krylov}, {Kornilov}, {Borisov},
  {Kuvshinov}, {Belinsky}, {Kuznetsov}, {Potanin}, {Antipov}, {Tyurina},
  {Gorbovskoy}, \& {Chilingaryan}}]{master_instrument}
{Lipunov}, V.~M., {Krylov}, A.~V., {Kornilov}, V.~G., {et~al.} 2004,
  Astronomische Nachrichten, 325, 580, \dodoi{10.1002/asna.200410284}

\bibitem[{{Mahabal} {et~al.}(2019){Mahabal}, {Rebbapragada}, {Walters},
  {Masci}, {Blagorodnova}, {van Roestel}, {Ye}, {Biswas}, {Burdge}, {Chang},
  {Duev}, {Golkhou}, {Miller}, {Nordin}, {Ward}, {Adams}, {Bellm}, {Branton},
  {Bue}, {Cannella}, {Connolly}, {Dekany}, {Feindt}, {Hung}, {Fortson},
  {Frederick}, {Fremling}, {Gezari}, {Graham}, {Groom}, {Kasliwal}, {Kulkarni},
  {Kupfer}, {Lin}, {Lintott}, {Lunnan}, {Parejko}, {Prince}, {Riddle},
  {Rusholme}, {Saunders}, {Sedaghat}, {Shupe}, {Singer}, {Soumagnac}, {Szkody},
  {Tachibana}, {Tirumala}, {van Velzen}, \& {Wright}}]{mahabal_2019}
{Mahabal}, A., {Rebbapragada}, U., {Walters}, R., {et~al.} 2019, \pasp, 131,
  038002, \dodoi{10.1088/1538-3873/aaf3fa}

\bibitem[{Makhlouf {et~al.}(2021)Makhlouf, Turpin, Corre, Karpov, Kann, \&
  Klotz}]{makhlouf_2021}
Makhlouf, K., Turpin, D., Corre, D., {et~al.} 2021, arXiv

\bibitem[{Maley(1991)}]{Maley_1991}
Maley, P. 1991, Advances in Space Research, 11, 33–36,
  \dodoi{10.1016/0273-1177(91)90539-v}

\bibitem[{Maley(1987)}]{Maley_1987}
Maley, P.~D. 1987, The Astrophysical Journal, 317, L39, \dodoi{10.1086/184909}

\bibitem[{{Martin-Carrillo} {et~al.}(2014){Martin-Carrillo}, {Hanlon},
  {Topinka}, {LaCluyz{\'e}}, {Savchenko}, {Kann}, {Trotter}, {Covino},
  {Kr{\"u}hler}, {Greiner}, {McGlynn}, {Murphy}, {Tisdall}, {Meehan}, {Wade},
  {McBreen}, {Reichart}, {Fugazza}, {Haislip}, {Rossi}, {Schady}, {Elliott}, \&
  {Klose}}]{martin_carrillo_2014}
{Martin-Carrillo}, A., {Hanlon}, L., {Topinka}, M., {et~al.} 2014, \aap, 567,
  A84, \dodoi{10.1051/0004-6361/201220872}

\bibitem[{{Matheson} {et~al.}(2021){Matheson}, {Stubens}, {Wolf}, {Lee},
  {Narayan}, {Saha}, {Scott}, {Soraisam}, {Bolton}, {Hauger}, {Silva},
  {Kececioglu}, {Scheidegger}, {Snodgrass}, {Aleo}, {Evans-Jacquez}, {Singh},
  {Wang}, {Yang}, \& {Zhao}}]{antares_paper}
{Matheson}, T., {Stubens}, C., {Wolf}, N., {et~al.} 2021, \aj, 161, 107,
  \dodoi{10.3847/1538-3881/abd703}

\bibitem[{Mcculloch \& Pitts(1943)}]{mcculloch_1943}
Mcculloch, W., \& Pitts, W. 1943, Bulletin of Mathematical Biophysics, 5, 127

\bibitem[{{Nir} {et~al.}(2020){Nir}, {Ofek}, {Ben-Ami}, {Segev}, {Polishook},
  \& {Manulis}}]{nir_2020}
{Nir}, G., {Ofek}, E.~O., {Ben-Ami}, S., {et~al.} 2020, arXiv e-prints,
  arXiv:2011.03497.
\newblock \doarXiv{2011.03497}

\bibitem[{{Osten} \& {Wolk}(2015)}]{osten_2015}
{Osten}, R.~A., \& {Wolk}, S.~J. 2015, \apj, 809, 79,
  \dodoi{10.1088/0004-637X/809/1/79}

\bibitem[{{Patterson} {et~al.}(2019){Patterson}, {Bellm}, {Rusholme}, {Masci},
  {Juric}, {Krughoff}, {Golkhou}, {Graham}, {Kulkarni}, {Helou}, \& {Zwicky
  Transient Facility Collaboration}}]{zads_paper}
{Patterson}, M.~T., {Bellm}, E.~C., {Rusholme}, B., {et~al.} 2019, \pasp, 131,
  018001, \dodoi{10.1088/1538-3873/aae904}

\bibitem[{Perrett {et~al.}(2010)Perrett, Balam, Sullivan, Pritchet, Conley,
  Carlberg, Astier, Balland, Basa, Fouchez, \& et~al.}]{Perrett_2010}
Perrett, K., Balam, D., Sullivan, M., {et~al.} 2010, The Astronomical Journal,
  140, 518, \dodoi{10.1088/0004-6256/140/2/518}

\bibitem[{{Pickles} \& {Depagne}(2010)}]{tycho2_griz}
{Pickles}, A., \& {Depagne}, {\'E}. 2010, \pasp, 122, 1437,
  \dodoi{10.1086/657947}

\bibitem[{{Pietras} {et~al.}(2022){Pietras}, {Falewicz}, {Siarkowski}, {Bicz},
  \& {Pre{\'s}}}]{pietras_2022}
{Pietras}, M., {Falewicz}, R., {Siarkowski}, M., {Bicz}, K., \& {Pre{\'s}}, P.
  2022, \apj, 935, 143, \dodoi{10.3847/1538-4357/ac8352}

\bibitem[{Powell {et~al.}(2021)Powell, Kostov, Rappaport, Tokovinin, Shporer,
  Collins, Corbett, Borkovits, Gary, Chiang, Rodriguez, Law, Barclay, Gagliano,
  Vanderburg, Olmschenk, Kruse, Schlieder, Soto, Goeke, Jacobs, Kristiansen,
  LaCourse, Omohundro, Schwengeler, Terentev, \& Schmitt}]{Powell_2021}
Powell, B.~P., Kostov, V.~B., Rappaport, S.~A., {et~al.} 2021, The Astronomical
  Journal, 162, 299, \dodoi{10.3847/1538-3881/ac2c81}

\bibitem[{{Quimby} {et~al.}(2021){Quimby}, {Shafter}, \&
  {Corbett}}]{Quimby_2021}
{Quimby}, R.~M., {Shafter}, A.~W., \& {Corbett}, H. 2021, Research Notes of the
  American Astronomical Society, 5, 160, \dodoi{10.3847/2515-5172/ac14c0}

\bibitem[{{Ranjan} {et~al.}(2017){Ranjan}, {Wordsworth}, \&
  {Sasselov}}]{Ranjan_2017}
{Ranjan}, S., {Wordsworth}, R., \& {Sasselov}, D.~D. 2017, \apj, 843, 110,
  \dodoi{10.3847/1538-4357/aa773e}

\bibitem[{Rast(1991)}]{Rast_1991}
Rast, R.~H. 1991, Icarus, 90, 328–329, \dodoi{10.1016/0019-1035(91)90112-7}

\bibitem[{{Ratzloff} {et~al.}(2020){Ratzloff}, {Law}, {Corbett}, {Fors}, \&
  {del Ser}}]{ratzloff_robotilter}
{Ratzloff}, J.~K., {Law}, N.~M., {Corbett}, H.~T., {Fors}, O., \& {del Ser}, D.
  2020, arXiv e-prints, arXiv:2001.00879.
\newblock \doarXiv{2001.00879}

\bibitem[{{Ratzloff} {et~al.}(2019){Ratzloff}, {Law}, {Fors}, {Corbett},
  {Howard}, {del Ser}, \& {Haislip}}]{evryscope_instrument}
{Ratzloff}, J.~K., {Law}, N.~M., {Fors}, O., {et~al.} 2019, \pasp, 131, 075001,
  \dodoi{10.1088/1538-3873/ab19d0}

\bibitem[{{Richmond} {et~al.}(2020){Richmond}, {Tanaka}, {Morokuma}, {Sako},
  {Ohsawa}, {Arima}, {Tominaga}, {Doi}, {Aoki}, {Arimatsu}, {Ichiki}, {Ikeda},
  {Ita}, {Kasuga}, {Kawabata}, {Kawakita}, {Kobayashi}, {Kokubo}, {Konishi},
  {Maehara}, {Mito}, {Miyata}, {Mori}, {Morii}, {Motohara}, {Nakada},
  {Okumura}, {Onozato}, {Sarugaku}, {Sato}, {Shigeyama}, {Soyano}, {Takahashi},
  {Tanikawa}, {Tarusawa}, {Urakawa}, {Usui}, {Watanabe}, {Yamashita}, \&
  {Yoshikawa}}]{richmond_2020}
{Richmond}, M.~W., {Tanaka}, M., {Morokuma}, T., {et~al.} 2020, \pasj, 72, 3,
  \dodoi{10.1093/pasj/psz120}

\bibitem[{{Ricker} {et~al.}(2014){Ricker}, {Winn}, {Vanderspek}, {Latham},
  {Bakos}, {Bean}, {Berta-Thompson}, {Brown}, {Buchhave}, {Butler}, {Butler},
  {Chaplin}, {Charbonneau}, {Christensen-Dalsgaard}, {Clampin}, {Deming},
  {Doty}, {De Lee}, {Dressing}, {Dunham}, {Endl}, {Fressin}, {Ge}, {Henning},
  {Holman}, {Howard}, {Ida}, {Jenkins}, {Jernigan}, {Johnson}, {Kaltenegger},
  {Kawai}, {Kjeldsen}, {Laughlin}, {Levine}, {Lin}, {Lissauer}, {MacQueen},
  {Marcy}, {McCullough}, {Morton}, {Narita}, {Paegert}, {Palle}, {Pepe},
  {Pepper}, {Quirrenbach}, {Rinehart}, {Sasselov}, {Sato}, {Seager},
  {Sozzetti}, {Stassun}, {Sullivan}, {Szentgyorgyi}, {Torres}, {Udry}, \&
  {Villasenor}}]{tess_instrument}
{Ricker}, G.~R., {Winn}, J.~N., {Vanderspek}, R., {et~al.} 2014, in Society of
  Photo-Optical Instrumentation Engineers (SPIE) Conference Series, Vol. 9143,
  Space Telescopes and Instrumentation 2014: Optical, Infrared, and Millimeter
  Wave, ed. J.~{Oschmann}, Jacobus~M., M.~{Clampin}, G.~G. {Fazio}, \& H.~A.
  {MacEwen}, 914320, \dodoi{10.1117/12.2063489}

\bibitem[{{Rolnick} {et~al.}(2017){Rolnick}, {Veit}, {Belongie}, \&
  {Shavit}}]{rolnick_2018}
{Rolnick}, D., {Veit}, A., {Belongie}, S., \& {Shavit}, N. 2017, arXiv
  e-prints, arXiv:1705.10694.
\newblock \doarXiv{1705.10694}

\bibitem[{{Schaefer} {et~al.}(2000){Schaefer}, {King}, \&
  {Deliyannis}}]{schaefer_2000}
{Schaefer}, B.~E., {King}, J.~R., \& {Deliyannis}, C.~P. 2000, \apj, 529, 1026,
  \dodoi{10.1086/308325}

\bibitem[{{Schaefer} {et~al.}(1987){Schaefer}, {Barber}, {Brooks}, {Deforrest},
  {Maley}, {McLeod}, {McNiel}, {Noymer}, {Presnell}, {Schwartz}, \&
  {Whitney}}]{schaefer_1987}
{Schaefer}, B.~E., {Barber}, M., {Brooks}, J.~J., {et~al.} 1987, \apj, 320,
  398, \dodoi{10.1086/165552}

\bibitem[{{Segura} {et~al.}(2010){Segura}, {Walkowicz}, {Meadows}, {Kasting},
  \& {Hawley}}]{segura_2010}
{Segura}, A., {Walkowicz}, L.~M., {Meadows}, V., {Kasting}, J., \& {Hawley}, S.
  2010, Astrobiology, 10, 751, \dodoi{10.1089/ast.2009.0376}

\bibitem[{{Shamir} \& {Nemiroff}(2006)}]{shamir_2006}
{Shamir}, L., \& {Nemiroff}, R.~J. 2006, \pasp, 118, 1180,
  \dodoi{10.1086/506989}

\bibitem[{{Shappee} {et~al.}(2014){Shappee}, {Prieto}, {Grupe}, {Kochanek},
  {Stanek}, {De Rosa}, {Mathur}, {Zu}, {Peterson}, {Pogge}, {Komossa}, {Im},
  {Jencson}, {Holoien}, {Basu}, {Beacom}, {Szczygie{\l}}, {Brimacombe},
  {Adams}, {Campillay}, {Choi}, {Contreras}, {Dietrich}, {Dubberley},
  {Elphick}, {Foale}, {Giustini}, {Gonzalez}, {Hawkins}, {Howell}, {Hsiao},
  {Koss}, {Leighly}, {Morrell}, {Mudd}, {Mullins}, {Nugent}, {Parrent},
  {Phillips}, {Pojmanski}, {Rosing}, {Ross}, {Sand}, {Terndrup}, {Valenti},
  {Walker}, \& {Yoon}}]{asassn_instrument}
{Shappee}, B.~J., {Prieto}, J.~L., {Grupe}, D., {et~al.} 2014, \apj, 788, 48,
  \dodoi{10.1088/0004-637X/788/1/48}

\bibitem[{{Simonyan} \& {Zisserman}(2014)}]{vggnet}
{Simonyan}, K., \& {Zisserman}, A. 2014, arXiv e-prints, arXiv:1409.1556.
\newblock \doarXiv{1409.1556}

\bibitem[{Srivastava {et~al.}(2014)Srivastava, Hinton, Krizhevsky, Sutskever,
  \& Salakhutdinov}]{srivastava14a}
Srivastava, N., Hinton, G., Krizhevsky, A., Sutskever, I., \& Salakhutdinov, R.
  2014, Journal of Machine Learning Research, 15, 1929.
\newblock \url{http://jmlr.org/papers/v15/srivastava14a.html}

\bibitem[{{Stoppa} {et~al.}(2022){Stoppa}, {Vreeswijk}, {Bloemen},
  {Bhattacharyya}, {Caron}, {J{\'o}hannesson}, {Ruiz de Austri}, {van den
  Oetelaar}, {Zaharijas}, {Groot}, {Cator}, \& {Nelemans}}]{autosourcid-light}
{Stoppa}, F., {Vreeswijk}, P., {Bloemen}, S., {et~al.} 2022, arXiv e-prints,
  arXiv:2202.00489.
\newblock \doarXiv{2202.00489}

\bibitem[{{Tamuz} {et~al.}(2005){Tamuz}, {Mazeh}, \& {Zucker}}]{tamuz_2005}
{Tamuz}, O., {Mazeh}, T., \& {Zucker}, S. 2005, \mnras, 356, 1466,
  \dodoi{10.1111/j.1365-2966.2004.08585.x}

\bibitem[{Tonry {et~al.}(2018)Tonry, Denneau, Heinze, Stalder, Smith, Smartt,
  Stubbs, Weiland, \& Rest}]{atlas_instrument}
Tonry, J.~L., Denneau, L., Heinze, A.~N., {et~al.} 2018, Publications of the
  Astronomical Society of the Pacific, 130, 064505,
  \dodoi{10.1088/1538-3873/aabadf}

\bibitem[{{Tonry} {et~al.}(2018){Tonry}, {Denneau}, {Flewelling}, {Heinze},
  {Onken}, {Smartt}, {Stalder}, {Weiland}, \& {Wolf}}]{atlas_catalog}
{Tonry}, J.~L., {Denneau}, L., {Flewelling}, H., {et~al.} 2018, \apj, 867, 105,
  \dodoi{10.3847/1538-4357/aae386}

\bibitem[{{Troja} {et~al.}(2017){Troja}, {Lipunov}, {Mundell}, {Butler},
  {Watson}, {Kobayashi}, {Cenko}, {Marshall}, {Ricci}, {Fruchter}, {Wieringa},
  {Gorbovskoy}, {Kornilov}, {Kutyrev}, {Lee}, {Toy}, {Tyurina}, {Budnev},
  {Buckley}, {Gonz{\'a}lez}, {Gress}, {Horesh}, {Panasyuk}, {Prochaska},
  {Ramirez-Ruiz}, {Rebolo Lopez}, {Richer}, {Roman-Zuniga}, {Serra-Ricart},
  {Yurkov}, \& {Gehrels}}]{troja_2017}
{Troja}, E., {Lipunov}, V.~M., {Mundell}, C.~G., {et~al.} 2017, \nat, 547, 425,
  \dodoi{10.1038/nature23289}

\bibitem[{{van Roestel} {et~al.}(2019){van Roestel}, {Groot}, {Kupfer},
  {Verbeek}, {van Velzen}, {Bours}, {Nugent}, {Prince}, {Levitan}, {Nissanke},
  {Kulkarni}, \& {Laher}}]{van_roestel_2019}
{van Roestel}, J., {Groot}, P.~J., {Kupfer}, T., {et~al.} 2019, \mnras, 484,
  4507, \dodoi{10.1093/mnras/stz241}

\bibitem[{{Vestrand} {et~al.}(2014){Vestrand}, {Wren}, {Panaitescu}, {Wozniak},
  {Davis}, {Palmer}, {Vianello}, {Omodei}, {Xiong}, {Briggs}, {Elphick},
  {Paciesas}, \& {Rosing}}]{vestrand_2014}
{Vestrand}, W.~T., {Wren}, J.~A., {Panaitescu}, A., {et~al.} 2014, Science,
  343, 38, \dodoi{10.1126/science.1242316}

\bibitem[{{Voges} {et~al.}(1999){Voges}, {Aschenbach}, {Boller},
  {Br{\"a}uninger}, {Briel}, {Burkert}, {Dennerl}, {Englhauser}, {Gruber},
  {Haberl}, {Hartner}, {Hasinger}, {K{\"u}rster}, {Pfeffermann}, {Pietsch},
  {Predehl}, {Rosso}, {Schmitt}, {Tr{\"u}mper}, \& {Zimmermann}}]{rosat_survey}
{Voges}, W., {Aschenbach}, B., {Boller}, T., {et~al.} 1999, \aap, 349, 389.
\newblock \doarXiv{astro-ph/9909315}

\bibitem[{Walkowicz {et~al.}(2008)Walkowicz, Johns-Krull, \&
  Hawley}]{Walkowicz_2008}
Walkowicz, L.~M., Johns-Krull, C.~M., \& Hawley, S.~L. 2008, The Astrophysical
  Journal, 677, 593, \dodoi{10.1086/526421}

\bibitem[{{Wang} {et~al.}(2021){Wang}, {Xin}, {Li}, {Li}, {Sun}, {Gao}, {Han},
  {Dai}, {Liang}, {Wang}, \& {Wei}}]{wang_2021}
{Wang}, J., {Xin}, L.~P., {Li}, H.~L., {et~al.} 2021, \apj, 916, 92,
  \dodoi{10.3847/1538-4357/ac096f}

\bibitem[{{Watson} {et~al.}(2020){Watson}, {Henden}, \& {Price}}]{vsx_catalog}
{Watson}, C., {Henden}, A.~A., \& {Price}, A. 2020, VizieR Online Data Catalog,
  B/vsx

\bibitem[{{Wee} {et~al.}(2020){Wee}, {Blagorodnova}, {Penprase}, {Facey},
  {Morioka}, {Corbett}, {Barlow}, {Kupfer}, {Law}, {Ratzloff}, {Howard},
  {Gonzalez Chavez}, {Glazier}, {Soto}, \& {Horiuchi}}]{Wee_2020}
{Wee}, J., {Blagorodnova}, N., {Penprase}, B.~E., {et~al.} 2020, \apj, 899,
  162, \dodoi{10.3847/1538-4357/aba3cc}

\bibitem[{{Wolf} {et~al.}(2018){Wolf}, {Onken}, {Luvaul}, {Schmidt}, {Bessell},
  {Chang}, {Da Costa}, {Mackey}, {Martin-Jones}, {Murphy}, {Preston}, {Scalzo},
  {Shao}, {Smillie}, {Tisserand}, {White}, \& {Yuan}}]{skymapper_dr1}
{Wolf}, C., {Onken}, C.~A., {Luvaul}, L.~C., {et~al.} 2018, \pasa, 35, e010,
  \dodoi{10.1017/pasa.2018.5}

\bibitem[{{Xin} {et~al.}(2021){Xin}, {Li}, {Wang}, {Han}, {Xu}, {Meng}, {Cai},
  {Huang}, {Lu}, {Qiu}, {Wang}, {Liang}, {Dai}, {Wang}, {Wu}, {Zhang}, {Li},
  {Turpin}, {Feng}, {Deng}, {Sun}, {Zheng}, {Yang}, \& {Wei}}]{xin_2021}
{Xin}, L.~P., {Li}, H.~L., {Wang}, J., {et~al.} 2021, \apj, 909, 106,
  \dodoi{10.3847/1538-4357/abdd1b}

\bibitem[{{Zackay} {et~al.}(2016){Zackay}, {Ofek}, \& {Gal-Yam}}]{zogy}
{Zackay}, B., {Ofek}, E.~O., \& {Gal-Yam}, A. 2016, \apj, 830, 27,
  \dodoi{10.3847/0004-637X/830/1/27}

\end{thebibliography}
\bibliographystyle{aasjournal}

\end{document}